\documentclass{cernyrep}
\def\const{\mbox{const}}

\def\d{\partial}
\def\l{\left(}
\def\r{\right)}

\newcommand{\be}{\begin{equation}}
\newcommand{\ee}{\end{equation}}
\newcommand{\bea}{\begin{eqnarray}}
\newcommand{\eea}{\end{eqnarray}}

\renewcommand{\ln}{\mathop{\rm ln}\nolimits}
\newcommand{\sm}[1]{{\scriptscriptstyle \rm #1}}

\begin{document}

\title{Cosmology}


\author{V.A.~Rubakov}

\institute{Institute for Nuclear Research 
of the Russian Academy of Sciences, 
Moscow, Russia}

\maketitle

\begin{abstract}
In these lectures we first concentrate on
the cosmological problems which, hopefully, have to do with 
the 
new physics to be probed at the LHC: the nature and origin of dark 
matter and generation of matter-antimatter asymmetry.
We give several examples showing the LHC cosmological
potential. These are
WIMPs as cold dark matter, gravitinos as warm dark matter, and
electroweak baryogenesis as a mechanism  for
generating matter-antimatter
asymmetry. In the remaining part of the lectures we discuss
the cosmological perturbations
as a tool for studying the epoch preceeding the conventional
hot stage of the cosmological evolution. 
\end{abstract}

\section{Introduction}

The more we learn about our Universe, the better we understand 
that
%
it is full of mysteries. These fall into three broad classes.
One major mystery is dark energy, which deserves
a separate class. We briefly discuss dark energy
in Section~\ref{sec:de}, although, honestly speaking,
we do not have much to say about it.
The second class
most likely has to do with the 
early hot epoch of the cosmological evolution, and the third one
with an even earlier stage which preceeded the hot epoch.
Along with dark energy,
we encounter mysteries of the second class when
studying the
present composition of the Universe.
It hosts matter but not antimatter,
and after 40 years after it was understood that this is a
problem, we do not  have an established theory explaining
this asymmetry. The Universe hosts dark matter, and we do not know
what it is made of. 
In this context, one of the key players is the Large Hadron Collider.
Optimistically, the LHC experiments may discover
dark matter particles and their companions, and establish physics behind
the matter-antimatter asymmetry. Otherwise they will rule out
some very plausible scenarios; this will also have profound
impact on our understanding of the early Universe. 
Let us mention also exotic hypotheses on physics beyond the
Standard Model, like TeV scale gravity;
their support by the LHC will have an effect on the early cosmology,
which is hard to overestimate.  

In the first part of
these lectures we concentrate on a few examples showing the LHC
cosmological potential. 
Before coming to that, we briefly introduce
the basic notions of cosmology that are useful for our main
discussion. We then turn to dark matter, and present the WIMP
scenario for cold dark matter, which is currently the most popular one
---
for good reason. We also consider
light gravitino scenario for
warm dark matter. Both are probed by the LHC, as they require
rather particular new physics in the LHC energy range. 
We then discuss electroweak baryogenesis --- a mechanism for
the generation of matter-antimatter asymmetry that may have
operated at temperature of order 100\,GeV in the early Universe.
This mechanism also needs new physics at energies
$100 - 300$\,GeV, so it will be confirmed or ruled out by
the LHC. 

The third class of mysteries
is related to 
cosmological perturbations, i.e., inhomogeneities in energy
density and associated gravitational potentials and, possibly,
relic gravity waves.  
As we explain in the second part of these lectures,
the observed properties of density perturbations
show that they were generated
at some epoch that preceeded the hot stage of the
cosmological evolution. Obviously, the very fact that we are 
confident
about the existence of such an
epoch 
is a fundamental result of theoretical and observational
cosmology.
The most plausible
hypothesis on that epoch is cosmological inflation,
though the observational support of this hypothesis
is presently not particularly strong, and alternative scenarios
have not been ruled out. We will briefly discuss the potential
of future cosmological observations 
in discriminating between different options.

These lectures are meant to be self-contained, but
we necessarily omit numerous details,
while trying to make clear the basic ideas and results.
More complete accounts of particle physics aspects of cosmology
may be found in reviews~\cite{gen-cosmo}. Dark matter,
including various hypotheses about its particles,
is reviewed in Ref.~\cite{dark-rev}. Electroweak baryogenesis
is discussed in detail in reviews~\cite{ew-rev}. 
For reviews on dark energy, see, e.g.,~Ref.\cite{de-rev}.
The theory and observations of cosmological perturbations are
presented in Ref.~\cite{pert-rev}.

\section{Homogeneous isotropic Universe}

\subsection{Friedmann--Lema\^itre--Robertson--Walker metric}

Two basic facts about our visible Universe are that it is 
{\it homogeneous and isotropic} at large spatial scales,
and that it {\it expands}. 

There are three types of homogeneous and isotropic three-dimensional 
spaces.
These are\footnote{Strictly 
speaking, this statement is valid only locally: in principle, homogeneous 
and isotropic Universe may have complex global properties. As an example,
spatially flat Universe may have topology of three-torus. There is some
discussion of such a possibility in literature, and fairly strong limits
have been obtained by the analyses of cosmic microwave 
background radiation~\cite{topology}.} three-sphere, flat 
(Euclidean) space
and three-hyperboloid.
Accordingly,
one  speaks about 
closed, flat and open Universe; 
in the latter two cases the spatial size of the
Universe is
infinite, whereas in the former the Universe is compact.

The  homogeneity and isotropy of the Universe
mean that its hypersurfaces of constant time are
either three-spheres or Euclidean spaces  or three-hyperboloids.
The distances between points may (and, indeed, do) depend on
time, i.e., the interval has the form
\begin{equation}
 ds^2 = dt^2 - a^2(t) d{\bf x}^2 \; ,
\label{FRW}
\end{equation}
where $d{\bf x}^2$ is the distance on unit 
three-sphere/Euclidean space/hyperboloid. The metric~(\ref{FRW})
is usually called Friedmann--Lema\^itre--Robertson--Walker (FLRW) metric,
and $a(t)$ is called the scale factor.
In our Universe
$   \dot{a} \equiv \frac{da}{dt} > 0$,
which means that the distance between points of fixed spatial
coordinates
${\bf x}$ grows,
$  dl^2 = a^{2}(t) d{\bf x}^2$. The space stretches out;
the Universe expands.

The coordinates ${\bf x}$ are often called
comoving coordinates. It is straightforward to check that
${\bf x}=\mbox{const}$ is a time-like geodesic, so a galaxy
put at a certain ${\bf x}$ at zero velocity will stay at the same
${\bf x}$. Furthermore, as the Universe expands, non-relativistic
objects loose their velocities $\dot{\bf x}$,
i.e., they get frozen in the comoving coordinate frame. 

Observational data set strong constraints on the spatial curvature
of the Universe. They tell that to a very good approximation our
Universe is spatially flat, i.e., our {\it 3-dimensional space
is Euclidean}. In what follows $d{\bf x}^2$ is simply
the line interval in the Euclidean 3-dimensional space. 

\subsection{Redshift}

Like the distances between free particles
in the expanding Universe,
the photon wavelength increases too.
We will always label the present values of time-dependent quantities
by subscript $0$: the present wavelength of a photon
is thus denoted by $\lambda_0$, the present time is $t_0$,
the present value of the 
scale factor is $a_0 \equiv a(t_0)$, etc. If a photon was emitted
at some moment of 
time $t$ in the past, and its  wavelength
at the moment of emission was $\lambda_e$, then we receive today a photon
 whose physical wavelength  is longer,
\[
 \frac{\lambda_0}{\lambda_e}
 = \frac{a_0}{a(t)} \equiv 1 + z \; .
\]
Here we introduced the redshift $z$.
The redshift of an object
is directly measurable.  $\lambda_e$ is fixed by 
physics of the source, say, it is the wavelength of a photon
emitted by an excited hydrogen atom. So,
one identifies a series of emission 
or absorption lines, thus determining $\lambda_e$, 
and measures their actual wavelengths $\lambda_0$. 
These spectroscopic measurements give accurate values of $z$ 
even for distant sources. On the other hand, the redshift
is related to
the time of emission, and hence to the distance to the source.

Let us consider a ``nearby'' source, for which
$  z \ll 1$.
This corresponds to relatively small $(t_0 - t)$.
Expanding $a(t)$, one writes
\begin{equation}
  a(t) = a_0 - \dot{a}(t_0) (t_0 - t) \; .
\label{int1}
\end{equation}
To the leading order in $z$, the difference between the 
present time and the emission time is equal to the distance to the
source $r$ (the speed of light is set equal to 1). 
Let us define the Hubble parameter
\[
   H(t) = \frac{\dot{a}(t)}{a(t)}
\]
and denote its present value by $H_0$. Then Eq.~(\ref{int1})
takes the form
$  a(t) = a_0(1-H_0 r)$,
and we get for the redshift, again to the leading non-trivial order 
in $z$,
\[
  1+ z = \frac{1}{1-H_0r} = 1 + H_0 r \; .
\]
In this way we obtain the Hubble law,
\be
        z = H_0 r \; , \;\;\;\; z \ll 1 \; .
\label{hubble2}
\ee
Traditionally, one tends to interpret the expansion of the Universe
as runaway of galaxies from each other, and redshift as the Doppler 
effect. Then at small $z$ one writes $z=v$, where $v$ is the radial
velocity of the source with respect to the Earth, 
so $H_0$ is traditionally measured in units ``velocity per distance''.
Observational data
give~\cite{WMAP7}
\be
     H_0 = [71.0 \pm 2.5]~\frac{\mbox{km/s}}{\mbox{Mpc}} \approx
(14 \cdot 10^9~\mbox{yrs})^{-1} \; ,
\label{H00}
\ee
where $1~\mbox{Mpc} = 3\cdot 10^6~\mbox{light~yrs} = 
3\cdot 10^{24}~\mbox{cm}$ is the distance measure often used in cosmology.
Traditionally, the present value of the Hubble parameter is written as
\begin{equation}
   H_0 = h \cdot 100 ~\frac{\mbox{km}}{\mbox{s}\cdot \mbox{Mpc}} \; .
\label{H0}
\end{equation}
Thus $   h \approx 0.71$.
We will use this value in further estimates.

Let us point out that the interpretation of redshift in terms of
the Doppler shift is actually 
not adequate, especially for large enough $z$.
In fact, there is no need in this interpretation at all: the
``radial velocity'' enters neither theory nor observations, so this
notion may be safely dropped. Physically meaningful quantity is
redshift $z$ itself.

A final comment is that $H_0^{-1}$ has dimension of time, or length,
as indicated in Eq.~(\ref{H00}).
Clearly, this quantity sets the cosmological
scales of time and distance at the present epoch.

\subsection{Hot Universe}

Our Universe is filled with cosmic microwave background (CMB).
Cosmic microwave background as observed today
consists of photons with excellent black-body spectrum of temperature
\be
   T_0 = 2.726 \pm 0.001~\mbox{K} \; .
\label{temperature}
\ee
The spectrum has been precisely measured by various instruments
and does not show any deviation from the Planck 
spectrum~\cite{black-body}.

Thus, the present Universe is ``warm''. Earlier Universe was
warmer; it cooled down because of the expansion. While the CMB 
photons freely propagate today, it was not so at early stage.
When the Universe was hot, the usual matter (electrons and protons 
with rather small admixture of light nuclei) was in the plasma phase.
At that time photons strongly interacted with electrons 
due to the Thomson scattering
and protons interacted with electrons via Coulomb force,
so all these particles were in thermal equilibrium. 
As the Universe cooled down, electrons ``recombined'' with protons into
neutral hydrogen atoms, and the Universe became transparent to photons.
The temperature scale of recombination is, very crudely speaking, 
determined by the ionisation energy of hydrogen, which is of order 
10\,eV. In fact,  recombination occured at lower 
temperature\footnote{The reason is that the number density of electrons
and protons is small compared to the number density  of photons.
At temperature above 3000~K, a hydrogen atom formed in 
an electron-proton encounter
is quickly
destroyed by absorbing a photon from the high energy
tail of the Planck distribution, and after that
the electron/proton
lives long time before it meets proton/electron
and forms a hydrogen atom again. In thermodynamical terms,
 at 
temperatures above 3000~K
there is large entropy per electron/proton, and
recombination 
is
not thermodynamically favourable
because of entropy considerations.},
$       T_{rec} \approx 3000~{\mbox K}$.
An important point is that the duration of the period of
recombination was considerably shorter
than the Hubble time at that epoch; to a reasonable approximation,
recombination occured instantaneously. 

The importance of the recombination epoch (more precisely,
the epoch of photon last scattering; we will use the term recombination
for brevity) is that the CMB photons travel freely after it:
the density of hydrogen atoms was so small
(about $250~\mbox{cm}^{-3}$ right after recombination)
that the gas was transparent to photons. So, CMB photons
give the photographic picture of the Universe at recombination, i.e.,
at redshift and age
\be
z_{rec} =1090 \; , \;\;\;\;\;\;\;\
t_{rec} = 370~000~\mbox{years} \; .
\label{sep12-11-5}
\ee

It is worth noting  that even though after recombination photons
no longer were in thermal equilibrium with anything, the shape 
of the photon distribution function has not changed, 
except for overall redshift. Indeed, the thermal distribution function
for {\it ultra-relativistic} particles, the Planck distribution, depends 
only on the ratio of frequency to temperature, 
$f_{Planck}(p,T) = f\left(\omega_p/T\right)$, $\omega_p =|p|$.
As the Universe expands, the photon momentum
gets redshifted, $p (t) = p(t_{rec}) \cdot \frac{a (t_{rec})}{a(t)}$,
 the frequency is redshifted in the same way,
but the shape of the spectrum remains
Planckian, with redshifted
temperature.
Hence, the Planckian form of the
observed spectrum is no surprise. Generaly speaking, this property 
does not hold for massive particles.

At even earlier times, the temperature of the Universe was even
higher.
The earliest time at the hot stage
which has been observationally probed so far
is the Big Bang Nucleosynthesis epoch; that epoch began
at temperature of order 1\,MeV, when the lifetime of the Universe
was about 1~s. At that time the weak processes like
\[
e^- + p \longleftrightarrow n + \nu_e
\]
switched off, and the comoving number density of neutrons
freezed out. Somewhat later, 
these neutrons combined with protons into
light elements in thermonuclear reactions 
\begin{eqnarray}
  p + n &\to& {^2H} + \gamma\; ,
\nonumber\\
  ^2H + p &\to& {^3He} + \gamma \; ,
\nonumber\\
   ^3He + ^2H &\to&  {^4He} + p  \; ,
\end{eqnarray}
etc., up to $^7Li$. Comparison of the Big Bang Nucleosynthesis
theory with the observational determination of the composition
of cosmic medium gives us confidence that we understand the
Universe at that epoch. Notably, we are convinced that
the cosmological expansion was
governed by General Relativity.

\subsection{Properties of components of cosmic medium}

Let us come back to photons. Their effective temperature 
after recombination
 scales as
\begin{equation}
  T(t) \propto a^{-1}(t) \; .
\label{T}
\end{equation}
This behaviour is characteristic of {\it ultra-relativistic} free species
(at zero chemical potential).
The same formula is valid for ultra-relativistic particles
(at zero chemical potential) which are in thermal equilibrium. 
Thermal equilibrium means adiabatic expansion; during adiabatic 
expansion, the temperature of ultra-relativistic gas scales
as the inverse size of the system, according to
usual thermodynamics. The energy density of
 ultra-relativistic gas scales as $\rho \propto T^4$,
and pressure is $p= \rho/3$. 

Both for free photons, and for photons in thermal equilibrium,
the number density behaves as follows,
\[
  n_{\gamma} = \mbox{const} \cdot T^3 \propto a^{-3} \; ,
\]
and the energy density is given by the Stefan--Boltzmann law,
\begin{equation}
  \rho_{\gamma} = \frac{\pi^2}{30} \cdot 2 \cdot T^4 
\propto a^{-4} \; ,
\label{rel}
\end{equation}
where the factor $2$ accounts for  two photon polarizations.
The present number density of relic photons is 
\be 
n_{\gamma, 0} = 410~~\mbox{cm}^{-3} \; ,
\label{nga} 
\ee 
 and their energy density is 
\begin{equation}
  \rho_{\gamma, 0} = 2.7 \cdot 10^{-10}\,\frac{\mbox{GeV}}{\mbox{cm}^3} \; .
\label{rhogamma}
\end{equation}

An important characteristic of the early Universe is the entropy
density of cosmic plasma in thermal equilibrium. It is given by
\be
s =  \frac{2 \pi^2}{45} g_* T^3 \; ,
\label{edensi}
\ee
where
$g_*$ is the number of degrees of freedom 
 with $m \lesssim T$, that is,
the degrees of freedom which are relativistic at temperature $T$;
each spin state counts as an
independent degree of freedom, and 
fermions
contribute to $g_*$ with a factor of $7/8$.
The point is that the entropy density scales {\it exactly} as $a^{-3}$,
\be
sa^3 = \mbox{const} \; ,
\label{sep13-11-1}
\ee
while 
temperature scales {\it approximately} as $a^{-1}$.
The property \eqref{sep13-11-1} is nothing but 
the reflection of the fact that the Universe expands
relatively slowly, and the evolution is adiabatic
(barrig fairly exotic scenarios with strong entropy generation
at some early cosmological epoch).
The temperature would scale as $a^{-1}$ if the number
of relativistic degrees of freedom would be independent of
time. This is not the case, however.  
Indeed, the value of $g_*$ depends on temperature:
at $T \sim 10$\,MeV relativistic species are
photons, neutrinos, electrons and positrons, while
at $T \sim 1$\,GeV four flavors of quarks, gluons, muons
and $\tau$-leptons
are relativistic too. The number of degrees of freedom
in the Standard Model at $T \gtrsim 100$\,GeV
is 
\[
g_* (100\,\mbox{GeV}) \approx 100 \; .
\]
The present value of the entropy density (taking into account 
neutrinos as if they were massless) is
\be
s_0 \approx 3000~\mbox{cm}^{-3} \; .
\label{e-density}
\ee

The parameter $g_*$ determines not only the entropy density but also
the energy density of the cosmic plasma in thermal equilibrium.
The Stefan--Boltzmann law gives 
\be
\rho_{rad} = \frac{\pi^2}{30} g_* T^4 \; ,
\label{sep13-11-2}
\ee
where subscript {\it rad} indicates that we are talking about
the relativistic component (radiation in broad sense).

Let us now turn to non-relativistic particles: baryons, massive
neutrinos,
dark matter particles, etc. If they are not
destroyed during the evolution of the Universe (that is, they are
stable and do not annihilate), their number density merely
gets diluted,
\be
      n \propto a^{-3} \; .
\label{non}
\ee
This means, in particular, that the baryon-to-photon ratio stays
constant in time (we consider for definiteness
the late Universe, $T\lesssim 100$\,keV),
  \be
   \eta_B \equiv \frac{n_B}{n_\gamma} =\mbox{const} \approx 6.1\cdot 10^{-10}
\; .
\label{etadef} 
\ee
The numerical value here is determined by two independent methods:
one is Big Bang Nucleosynthesis theory and measurements  
of the light element abundances, and another is the measurements of
the CMB temperature anisotropy. It is reassuring that these methods give
consistent results (with comparable precision).

The energy density of non-relativistic particles
scales as
\be
   \rho (t) = m \cdot n(t) \propto a^{-3} (t) \; , 
\label{nonrel}
\ee
in contrast to more rapid fall off~(\ref{rel}) characteristic of
relativistic species.

Finally,
dark energy density does not decrease in
time as fast as in Eqs.~(\ref{rel}) or~(\ref{nonrel}). 
In fact, to a reasonable approximation dark energy
density does not depend on time at all,
\begin{equation}
  \rho_\Lambda= \mbox{const} \; .
\label{vac}
\end{equation}
Dark energy with exactly time-independent energy density
is the same thing as the
cosmological constant, or $\Lambda$-term.

\subsection{Composition of the present Universe}

The basic equation governing the expansion rate of the Universe
is the Friedmann equation, which we write for the case of spatially
flat Universe,
\begin{equation}
    H^2 \equiv \left( \frac{\dot{a}}{a} \right)^2 
= \frac{8\pi}{3} G \rho \; ,
\label{Friedmann}
\end{equation}
where dot denotes derivative with respect to time $t$, $\rho$ is the
{\it total} energy density in the Universe
and $G$ is Newton's gravity constant; in
natural units
  $G = M_{Pl}^{-2}$
where
  $M_{Pl} = 1.2\cdot 10^{19}\,\mbox{GeV}$
is the Planck mass. The Friedmann equation is nothing but the $(00)$-component
of the Einstein equations of General Relativity,
\[
R_{00} - \frac{1}{2}g_{00}R = 8\pi T_{00} \; ,
\]
specified to FLRW metric.

Let us introduce the parameter
\be
\rho_c = \frac{3}{8\pi G} H_0^2 \approx 5\cdot 10^{-6}\,
\frac{\mbox{GeV}}{\mbox{cm}^3} \; .
\label{rhoc-new}
\ee
According to Eq.~(\ref{Friedmann}), it is equal to the sum of all forms
of energy density
in the present Universe. As a side remark, we note that
the latter statement would not be true if our
Universe were not spatially flat. However, according to observations,
spatial flatness holds to a very good precision, corresponding
 to less than 1 per cent deviation of the total energy density from
$\rho_c$~\cite{WMAP7a}.

As we now discuss, the cosmological data correspond to
a very weird composition of the Universe.

Before proceeding, let us introduce
a notion traditional in the analysis of the composition
of the present Universe. For every type of matter $i$ with
the present energy density $\rho_{i,0}$, one defines the parameter
\be
   \Omega_i = \frac{\rho_{i,0}}{\rho_{c}} \; .
\nonumber
\ee
Then Eq.~(\ref{Friedmann}) tells that
$  \sum_i \Omega_i = 1 $
where the sum runs over all forms of energy.
Let us now discuss contributions of different species to this sum.

We begin with {\bf baryons}.  The result~(\ref{etadef})
gives
\be 
 \rho_{B, 0} = m_B \cdot n_{B, 0}   
\approx 2.4\cdot 10^{-7}\,
\frac{\mbox{GeV}}{\mbox{cm}^3} \; .
\label{rhoB}
\ee
Comparing this result with the value of $\rho_c$ given in Eq.~(\ref{rhoc-new}), one finds
\[
   \Omega_B = 0.045 \; .
\]
Thus, baryons constiute rather small fraction of the present energy 
density in the Universe.

{\bf Photons} 
contribute even smaller fraction, as is clear from Eq.~(\ref{rhogamma}), namely
$    \Omega_{\gamma} \approx  5\cdot 10^{-5}$.
From electric neutrality, the number density of {\bf electrons}
is about
the same as  that of baryons, 
so electrons contribute negligible fraction
to the total mass density. The remaining known
stable particles are {\bf neutrinos}. Their number density is
calculable in Hot Big Bang theory and these calculations are
nicely confirmed by Big Bang Nucleosynthesis. The present
number density
of each type of neutrinos is 
\[
  n_{\nu_{\alpha}, 0} = 110 ~\mbox{cm}^{-3} \; ,
\]
where $\nu_{\alpha} = \nu_e, \nu_{\mu}, \nu_{\tau}$ (more appropriately,
$\nu_\alpha$ are neutrino mass eigenstates). Direct limit
on the mass of electron neutrino, $m_{\nu_e} < 2$\,eV, together
with the observations of neutrino
oscillations  suggest that every
type of neutrino has mass smaller than 2\,eV (neutrinos with masses
above 0.05\,eV must be degenerate, according to neutrino oscillation
data). 
The energy density
of all types of neutrinos is thus smaller than $\rho_c$:
\[
  \rho_{\nu, total} = \sum_{\alpha} m_{\nu_{\alpha}} n_{\nu_{\alpha}}
 < 3\cdot 2\,\mbox{eV} \cdot 110 ~\frac{1}{\mbox{cm}^3}
\sim 6\cdot 10^{-7}\,\frac{\mbox{GeV}}{\mbox{cm}^3} \; ,
\]
which means that
$  \Omega_{\nu, total} < 0.12$.
This estimate does not make use of any cosmological data.
In fact, cosmological observations give stronger bound
\be
\Omega_{\nu, total} \lesssim 0.014 \; .
\label{omeganu}
\ee
This bound is mostly due to the analysis of the structures at
relatively 
small length scales, and has to do with streaming of neutrinos from 
the gravitational potential wells at  early times when neutrinos
were moving fast. In terms of the neutrino masses the bound~(\ref{omeganu}) 
reads~\cite{Seljak:2004xh,WMAP5}
\[
   \sum m_{\nu_\alpha} < 0.6\,\mbox{eV} \; ,
\]
so every neutrino must be lighter than 0.2\,eV. 
It is worth noting that the
atmospheric neutrino data, as well as K2K,
Minos and T2K experiments tell us that the mass of at
least one neutrino must be larger than about 0.05\,eV. Comparing these
numbers, one sees that it may be feasible to measure neutrino
masses by cosmological observations (!) in the future.

Coming back to our main topic here, we  conclude that
most of the energy density in the present Universe is not in the
form of known particles; most energy in the present Universe
must be in ``something unknown''. Furthermore, 
this ``something unknown'' has two
components:
clustered (dark matter) and unclustered (dark energy).

{\bf Clustered dark matter} consists presumably of new stable massive
particles. These make clumps of energy (mass) 
which constitute 
most of the mass of galaxies and  
clusters of galaxies. 
There are various ways of estimating the contribution of
non-baryonic dark matter into the total energy density of the
Universe (see Ref.~\cite{dark-rev} for details):

-- Composition of the Universe affects the angular anisotropy of
cosmic microwave background. Quite accurate measurements of the
CMB anisotropy, available today, enable one to estimate the total mass
density
of  dark matter. 

-- Composition of the Universe, and especially the density of
 non-baryonic dark matter, is crucial for structure
formation of the Universe. Comparison of the results of numerical
simulations of structure formation with observational data gives
reliable estimate of the mass density of non-baryonic clustered dark
matter.

The bottom line is that the non-relativistic component constitutes
about 27 per cent of the total present energy density, which means
that non-baryonic dark matter has
\be
   \Omega_{DM} \approx 0.22 \; ,
\label{omegacdm}
\ee
the rest is due to baryons.

There is direct evidence that dark matter exists in the largest 
gravitationally bound objects -- clusters of
galaxies. There are various methods to determine the gravitating mass
of a cluster, and even mass distribution in a cluster, which give
consistent results. To name a few:

-- One measures velocities of galaxies in 
galactic clusters, and makes use of
the gravitational virial theorem,
\[
\mbox{Kinetic~~energy~~of~~a~~galaxy} = 
\frac{1}{2}~\mbox{Potential~~energy} \; .
\]
In this way one obtains the gravitational potential, and thus the
distribution of the total mass in a cluster. 

-- Another measurement of masses of clusters makes use of intracluster
gas. Its temperature obtained from X-ray measurements is also
related to the gravitational potential.

-- Fairly accurate reconstruction of mass distributions in clusters is
obtained from the observations of gravitational lensing of background
galaxies by  clusters. 

These methods enable one to measure mass-to-light ratio in clusters of
galaxies. Assuming that this ratio applies to all matter in the
Universe\footnote{This is a fairly
 strong assumption, since only about 10 per
cent of galaxies are in clusters.}, one arrives at the estimate
for the mass density of clumped matter in the present
Universe. Remarkably, this estimate agrees with Eq.~(\ref{omegacdm}).

Finally, dark matter exists also in galaxies. Its distribution is
measured by the observations of rotation velocities of distant stars
and gas clouds around a galaxy.

Thus, cosmologists are confident that much of the energy density in
our Universe consists of new stable particles. We will see
that there is good chance for the LHC to produce these particles.

{\bf Unclustered dark energy.} 
Non-baryonic clustered dark matter is not the whole story.
Making use of the above estimates, one obtains an
estimate for the energy density of all 
particles,
$  \Omega_{\gamma} + \Omega_B + \Omega_{\nu, total} +
\Omega_{DM} \approx 0.27$.
This implies  that 73 per cent of the energy
density is unclustered. This component is called dark energy;
it has the properties similar to those of vacuum. We will briefly 
discuss
dark energy in Section~\ref{sec:de}.

All this fits nicely all cosmological observations, but does not
fit to the Standard Model of particle physics. It is our hope that
the LHC will shed light at least on some of the properties of the
Universe. 

\subsection{Regimes of cosmological expansion}
\label{sec:regimes}

The cosmological expansion at the  present epoch is
determined mostly by dark energy, since its contribution to the
right hand side of the Friedmann equation \eqref{Friedmann} is the 
largest,
\[
\Omega_\Lambda = 0.73 \; .
\]
Non-relativistic matter (dark matter and baryons) is also
non-negligible,
\be
\Omega_M = 0.27 \; ,
\label{sep12-11-2}
\ee
while the energy density of relativistic matter 
(photons and neutrinos, if one of the neutrino species
is massless or very light) is negligible today.
This was not always the case. Making use of Eq.~\eqref{rel}
for photons and relativistic neutrinos,
Eq.~\eqref{non} for non-relativistic matter, and assuming 
for definiteness that dark energy density is constant in time,
we can rewrite the Friedmann equation \eqref{Friedmann} in the 
following form
\begin{align}
{ H^2(t)}  &= \frac{8\pi}{3 M_{Pl}^2} [\rho_\Lambda + \rho_M (t)
+ \rho_{rad}(t)]
\nonumber \\
&= H_0^2 \left[{ \Omega_\Lambda} + { \Omega_M}
 \left(\frac{a_0}{a(t)}\right)^3
+ \Omega_{rad} \left(\frac{a_0}{a(t)}\right)^4 \right] \; .
\label{sep12-11}
\end{align}
It is appropriate for our purposes to treat neutrinos as
massless particles; including their contribution to
$\Omega_{rad}$ one has
\be
\Omega_{rad} = 8.4 \cdot 10^{-5} \; .
\label{sep12-11-3}
\ee
Equation \eqref{sep12-11} tells that at early times, when the scale
factor $a(t)$ was small, the expansion was dominated by relativistic matter
(``radiation''), later on 
there was long period of domination of the
non-relativistic matter, and in future the expansion will be dominated
by dark energy, 
\[
\dots \Longrightarrow \mbox{Radiation~domination}
\Longrightarrow \mbox{Matter~domination}\Longrightarrow 
\Lambda\mbox{--domination} \; .
\]
Dots here denote some cosmological
epoch preceding the hot stage of evolution; as we discuss in 
Section~\ref{sec:pertu}, we are confident that such an epoch
existed, but do not quite know what it was. Making use of
\eqref{sep12-11-2} and \eqref{sep12-11-3}, it is straightforward
to find the redshift at radiation--matter equality, when the first
two terms in \eqref{sep12-11} are equal,
\[
1+ z_{eq} = \frac{a_0}{a(t_{eq})} = \frac{\Omega_M}{\Omega_{rad}}
\approx 3000 \; ,
\]
and using the Friedmann equation one finds the age of the Universe at
equality
\[
t_{eq} \approx 60~000~\mbox{years} \; .
\]
Note that recombination occured at matter domination,
but rather soon after equality, see \eqref{sep12-11-5}.

It is useful for what follows to find the evolution of the scale factor
at the radiation domination epoch. At that time the energy density
is given by Eq.~\eqref{sep13-11-2}, so that the Friedmann 
equation can be written as follows 
\be
   H = \frac{T^2}{M_{Pl}^*} \; ,
\label{sep16-11-1}
\ee
where $M_{Pl}^* = M_{Pl}/(1.66 \sqrt{g_*})$.
Now, we neglect for simplicity
the dependence of $g_*$ on temperature, and hence
on time, and recall that in this case the temperature
scales as $a^{-1}$, see Eq.~\eqref{sep13-11-1}. Hence, we
obtain
\[
\frac{\dot{a}}{a} = \frac{\mbox{const}}{a^2} \; .
\]
This gives the desired evolution law
\be
a (t) = \mbox{const} \cdot \sqrt{t} \; .
\label{sep13-11-5}
\ee
The constant here does not have physical significance,
as one can rescale the coordinates ${\bf x}$ at some fixed moment
of time, thus changing the normailzation of $a$.

There are several points to note regarding the result
\eqref{sep13-11-5}. First, the expansion {\it decelerates}:
\[
\ddot{a} < 0 \; .
\]
This property holds also for the matter dominated epoch, but, as we see
momentarily, it does not hold for domination of the dark
energy.

Second, time $t=0$ is the Big Bang singularity
(assuming erroneously
that the Universe starts being radiation dominated).
The expansion rate
\[
H(t) = \frac{1}{2t}
\]
diverges as $t \to 0$, and so does the energy
density $\rho(t) \propto H^2 (t)$ and temperature
$T \propto \rho^{1/4}$. Of course, the classical
General Relativity and usual notions of statistical
mechanics (e.g., temperature itself) are not applicable
very near the singularity, but our result
suggests that in the picture we discuss
(hot epoch right after the Big Bang), the Universe
starts its classical evolution in a very hot and dense state, and its expansion
rate is very high in the beginning. It is customary to
assume for illustrational purposes
that the relevant quantities in the beginning of
the classical expansion take the Planck values,
$\rho \sim M_{Pl}^4$, $H \sim M_{Pl}$, etc.

Third, at a given moment of time the size of a causally connected
region is finite. Consider signals emitted right after the Big Bang
and travelling with the speed of light. These signals travel along the
light cone with
$ds=0$, and hence $a(t) dx = dt$. So, the coordinate distance
that a signal travels from the Big Bang to time $t$ is
\be
x = \int_0^{t} \frac{dt}{a(t)} \equiv \eta \; .
\label{sep13-11-6}
\ee
In the radiation dominated Universe 
\[
\eta = \mbox{\const} \cdot \sqrt{t} \; .
\]
The physical distance from the emission point to
the position of the signal is
\[
l_{H, t} = a(t) x = a(t) \int_0^{t} \frac{dt}{a(t)} =2t \; .
\]
As expected, this physical distance is finite, and it 
gives the size of a causally connected region at time $t$.
It is called the horizon size (more precisely, the
size of particle horizon). A related property is that
an observer at time $t$ can see only the part
of the Universe whose current physical size is $l_{H,t}$.
Both at radiation and matter
domination one has, modulo numerical constant of order 1,
\[
l_{H,t} \sim H^{-1}(t) \; .
\]
To give an idea of numbers, the horizon size at 
the present epoch is
\[
l_{H,t_0} \approx 15~\mbox{Gpc} \simeq 4.5 \cdot 10^{28}~\mbox{cm}
\; .
\]

\begin{figure}[htb!]
\begin{center}
\includegraphics[width=0.5\textwidth,angle=-90]{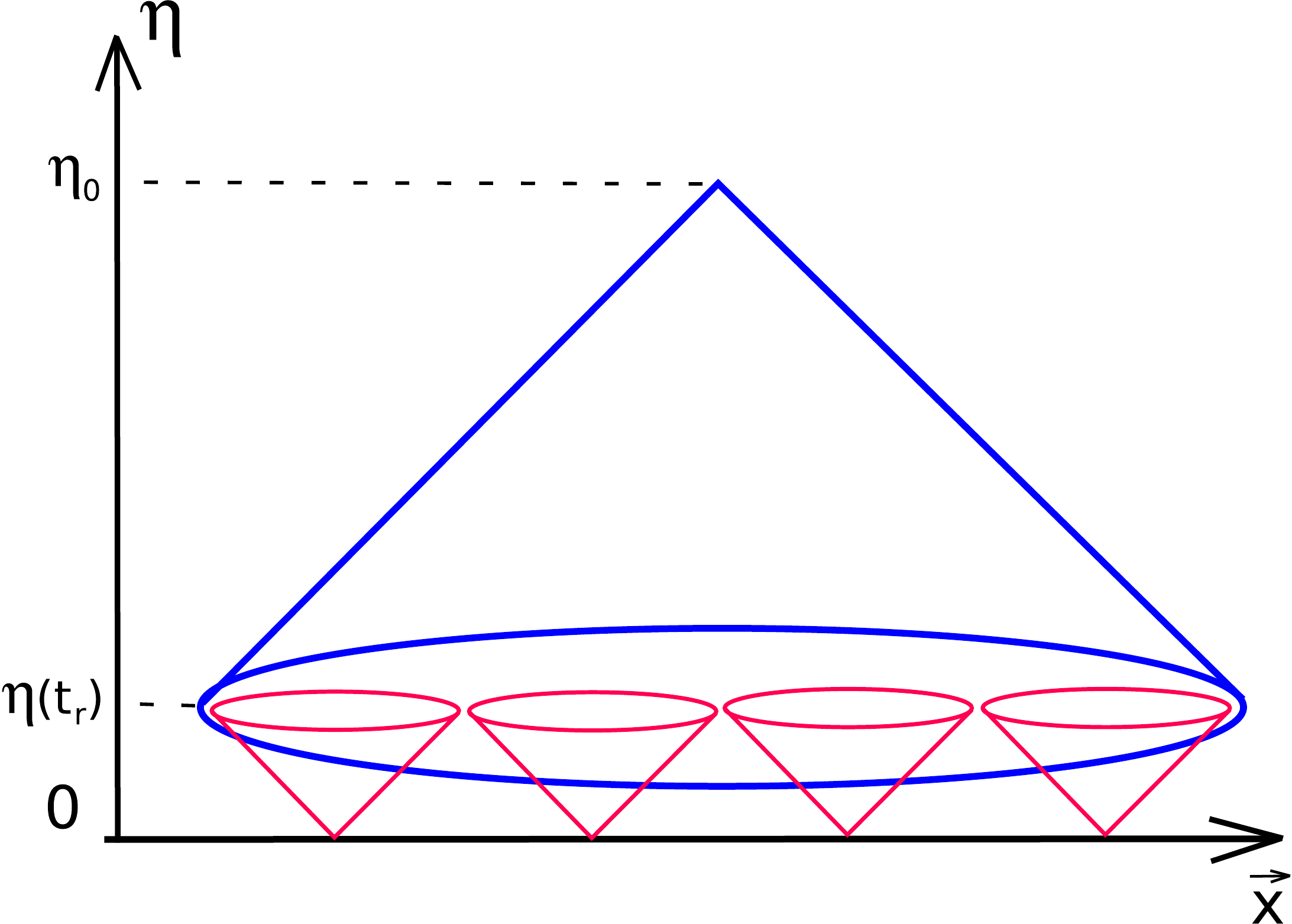}
\end{center}
\caption{ Causal structure of space-time in hot Big Bang theory
 \label{noinfl-horizon}
 }
\end{figure}

One property of the Universe  that starts its expansion from
rafiation domination is puzzling. Using 
Eq.~\eqref{sep13-11-6} one sees that the size of the observable Universe
increases in time. For example, the coordinate size of
the present horizon is about 50 times larger that the coordinate
size of the horizon at recombination. Hence, when performing
CMB observations we see $50^2$ regions on the sphere of last scattering
which were causally disconnected at the recombination epoch,
see Fig.~\ref{noinfl-horizon}.
Yet they look exactly the same!
Clearly, this is a problem for the hot Big Bang theory,
which is called horizon problem. We will see in 
Section~\ref{sec:pertu} that this problem has 
a somewhat different side, which unambiguously shows
that the hot Big Bang theory is not the whole story:
the hot epoch was preceeded by some other, very
different epoch of the cosmological evolution.

To end up this section, let us note that the properties of
the Universe dominated by dark energy are quite different.
Assuming for definiteness that $\rho_\Lambda$ is
independent of time, we immediately find the solution to
the Friedmann equation for the $\Lambda$-dominated Universe:
\be
a(t) = \mbox{const} \cdot \mbox{e}^{H_\Lambda t} \; ,
\label{sep20-11-1}
\ee
where $H_\Lambda = \sqrt{8\pi \rho_\Lambda/3 M_{Pl}^2}$.
The cosmological expansion {\it accelerates},
\[
\ddot{a} > 0 \; .
\]
The dark energy was introduced precisely for explaining the
accelerated expansion of the Universe at the present epoch.

\section{Dark matter}
\label{sec:dm}

Dark matter  is characterized by the mass-to-entropy ratio,
\be
   \left( \frac{\rho_{DM}}{s} \right)_0 = \frac{\Omega_{DM} \rho_c}{s_0}
\approx 
\frac{0.22 \cdot 5 \cdot 10^{-6}\,\mbox{GeV} \cdot \mbox{cm}^{-3}}{3000 
~\mbox{cm}^{-3}} = 4 \cdot 10^{-10}\,\mbox{GeV} \; .
\label{10p*}
\ee
This ratio is constant in time since the freeze out of
dark matter density: both number density of dark matter particles
$n_{DM}$ (and hence their mass density 
$\rho_{DM}=m_{DM} n_{DM}$) and entropy density
dilute exactly as $a^{-3}$. 

Dark matter is crucial for our existence, for the following reason.
Density perturbations in baryon-electron-photon plasma before recombination
do not grow because of high pressure, which is mostly due to photons;
instead,  perturbations are sound waves 
propagating in plasma with time-independent
amplitudes.
Hence, in a Universe without dark matter, density perturbations
in baryonic component would start to grow only after baryons decouple from
photons, i.e., after recombination. The mechanism of the growth is
pretty simple: an overdense region gravitationally attracts
surrounding matter; this matter falls into the overdense region,
and the density contrast increases. In the expanding matter dominated
Universe this gravitational instability results in the density contrast
growing like $(\delta \rho/\rho) (t) \propto a(t)$. Hence, in a Universe
without dark matter, the growth factor for baryon density perturbations
would be at most\footnote{Because of the presence of dark energy,
the growth factor is even somewhat smaller.} 
\be
   \frac{a(t_0)}{a(t_{rec})} = 1+z_{rec} = \frac{T_{rec}}{T_0} \approx 10^3 \; .
\label{bgrow}
\ee
The initial amplitude of density perturbations is very well known from
the CMB anisotropy measurements, $(\delta \rho /\rho)_i = 5 \cdot 10^{-5}$.
Hence, a Universe without dark matter would still be pretty homogeneous:
the density contrast would be in the range of a few per cent. No structure
would have been formed, no galaxies, no life. No structure would be formed
in future either, as the accelerated expansion
 due to dark energy will soon terminate
the growth of perturbations.

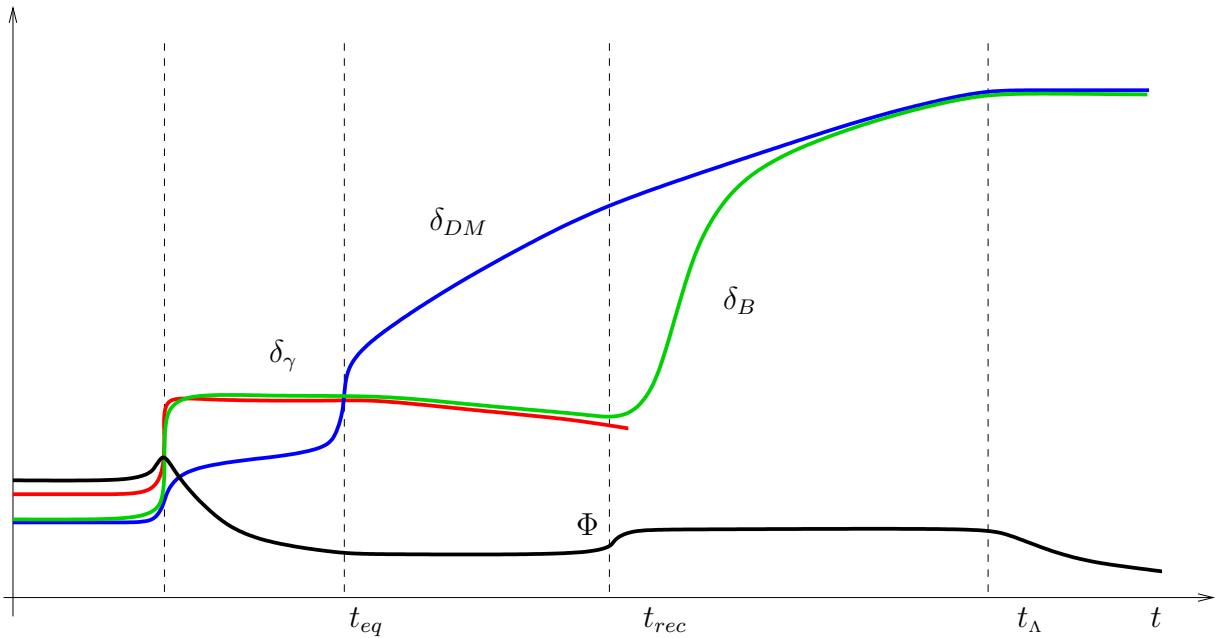
\begin{figure}[htb!]
\includegraphics[width=\textwidth]{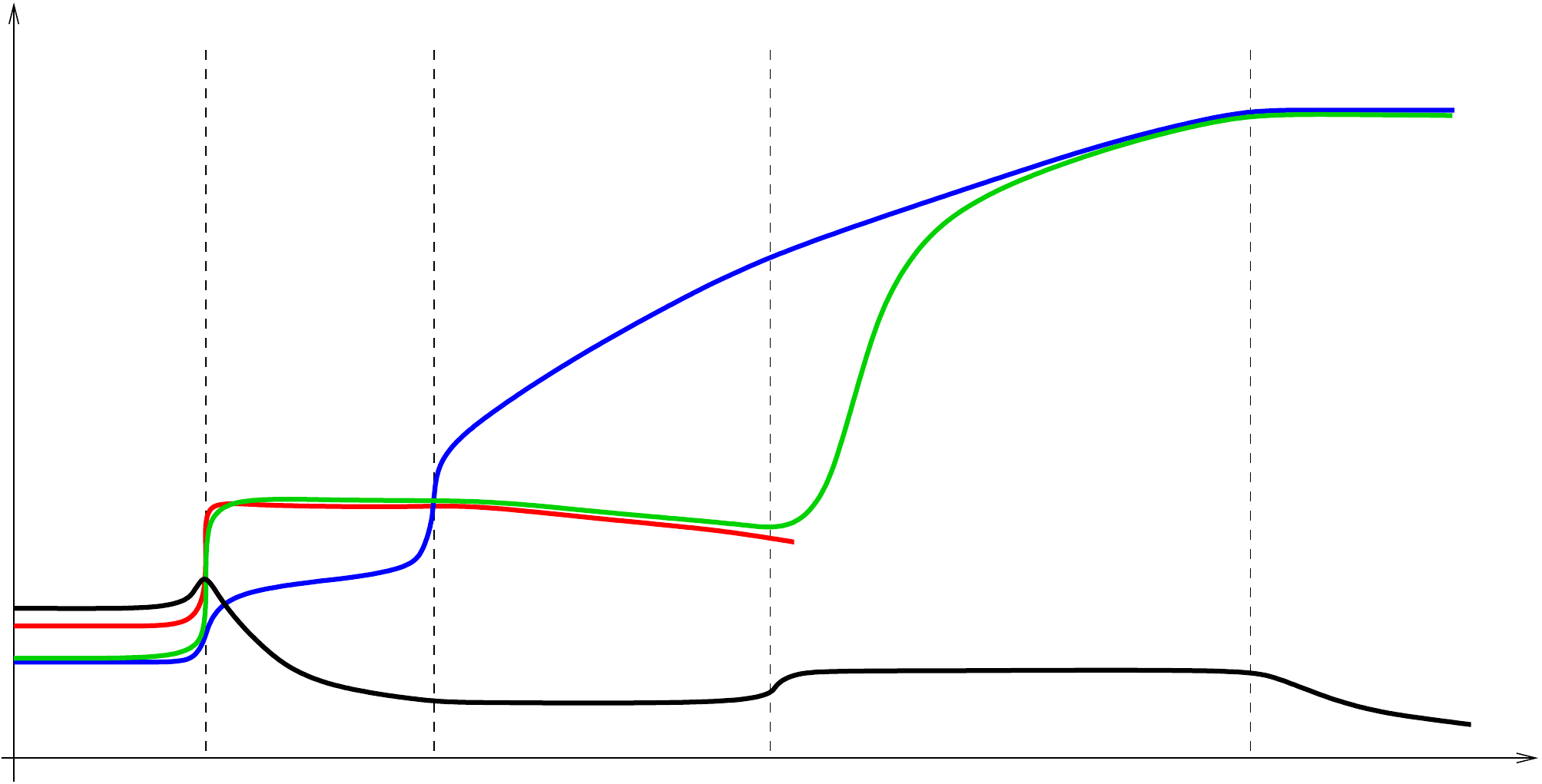}
\begin{picture}(10,10)(0,0)
{\large 
\put(90,120){{}}
\put(380,10){$t_{\sm \Lambda}$}
\put(240,10){$t_{rec}$}
\put(130,10){$t_{eq}$}
\put(430,10){$t$}
\put(215,45){$\Phi$}
\put(270,130){$\delta_B$}
\put(160,160){$\delta_{DM}$}
\put(100,110){$\delta_\gamma$}
}
\end{picture}
\caption{A sketch of the time dependence,  
in the linear regime, of density contrasts of dark matter,
baryons and photons, $\delta_{DM}\equiv \delta \rho_{DM}/\rho_{DM}$,
$\delta_B$ and $\delta_\gamma$, respectively, as well as the
Newtonian potential $\Phi$.
 $t_{eq}$  and $t_\Lambda$ 
correspond to the transitions from radiation
domination to matter domination, and from decelerated expansion to
accelerated expansion,  $t_{rec}$ refers to the recombination
epoch. 
 \label{pert}}
\end{figure}

Since dark matter particles decoupled from plasma much earlier
than baryons,
perturbations in dark matter started to grow much earlier.
The corresponding growth factor is larger than Eq.~(\ref{bgrow}),
so that the dark matter density contrast at galactic and 
sub-galactic scales
becomes of order one, perturbations  enter non-linear regime and form
dense dark matter clumps at $z = 5~-~10$. Baryons fall into 
potential wells formed by dark matter, so dark matter and baryon
perturbations develop together soon after recombination. Galaxies get 
formed in the regions where dark matter was overdense originally.
The development of perturbations in our Universe is shown in 
Fig.~\ref{pert}. For this picture to hold, dark matter particles
must be non-relativistic early enough, as relativistic particles
fly through gravitational wells instead of being trapped there.
This means, in particular, that neutrinos cannot 
constitute a considerable part
of dark matter, hence the bound~(\ref{omeganu}).

Depending on the mass of the dark matter particles and mechanism
of their production in the early Universe, dark matter may be {\it cold}
(CDM) and {\it warm} (WDM). Roughly speaking, CDM consists of heavy
particles, while the masses of WDM particles are smaller,
\begin{subequations}
\bea
&& \mbox{CDM}\; : \;\;\;\; m_{DM} \gtrsim 100\,\mbox{keV} \; ,
\label{cdmmass} \\
&& \mbox{WDM}\; : \;\;\;\; m_{DM} = 3~-~ 30\,\mbox{keV} \; .
\label{wdmmass}
\eea
\end{subequations}
This assumes that the dark matter particles were in thermal
(kinteic) equilibrium at some early times, or, more generally,
that their kinetic energy was comparable to temperature.
This need not be the case for very weakly interacting particles;
a well known example is axions which are {\it cold} dark matter
candidates despite their very small mass. Likewise, very weakly
interacting warm dark matter
particles may be much heavier than Eq. \eqref{wdmmass} suggests.

We will discuss warm dark matter option later on, and now we
move on to CDM.

\subsection{WIMPS: best guess for cold dark
matter}

There is a simple mechanism of the dark matter generation in the early
Universe. It applies to {\it cold} dark matter. Because of its simplicity
and robustness, it is considered by many as a very likely one,
and the corresponding dark matter candidates --- weakly interacting massive
particles, WIMPs 
--- as the best candidates. Let us describe this mechanism
in some detail.

Let us assume that there exists a heavy stable neutral particle $Y$, and
that 
$Y$-particles can only be destroyed or created via their
pair-annihilation or creation, with annihilation products being
the particles of the Standard Model. 
We will see that the overall 
cosmological behaviour of $Y$-particles is as follows.
At high temperatures, $T \gg m_Y$, the
$Y$-particles are in thermal equilibrium with
the rest of cosmic plasma; there are lots of $Y$-particles
in the plasma, which are continuously created and annihilate.  
As the temperature drops below $m_Y$, the equilibrium number
density decreases.
 At some ``freeze-out'' temperature $T_f$ the 
number density becomes so small, that $Y$-particles can no longer
meet each other during the Hubble time, and their annihilation
terminates. After that the number density of  survived
$Y$'s decreases like $a^{-3}$, and these relic particles
contribute to the mass density in the present Universe. Our purpose is
to estimate the range of properties of $Y$-particles, in which they
serve as dark matter.

Assuming thermal equilibrium,
elementary considerations of mean free path of a particle in gas
give for the lifetime of a non-relativistic
 $Y$-particle in cosmic plasma, $\tau_{ann}$,
\[
      \sigma_{ann}\cdot v \cdot \tau_{ann} \cdot n_{Y} \sim 1 \; ,
\]
where $v$ is the velocity of $Y$-particle,
$\sigma_{ann}$ is the annihilation cross section at velocity $v$
and $n_Y$ is the equilibrium number density given by the Boltzmann
law at zero chemical potential,
\[
   n_Y = g_Y \cdot \left(\frac{m_Y T}{2\pi} \right)^{3/2} 
\mbox{e}^{-\frac{m_Y}{T}}\; ,
\]
where $g_Y$ is the number of spin states of $Y$-particle.
Note that we consider non-relativistic regime, $m_Y \ll T$.
Let us introduce the notation 
\[
      \sigma_{ann} v = \sigma_0
\]
(in fact, the left hand side is to be understood as thermal
average). If the
annihilation occurs in $s$-wave, then $\sigma_0$ is a constant
independent of temperature, for $p$-wave it is somewhat suppressed
at $T \ll m_Y$.
One should compare the lifetime with the Hubble
time,
or annihilation rate $\Gamma_{ann} \equiv \tau^{-1}_{ann}$ 
with the expansion rate $H$.  
At $T \sim m_Y$, the equilibrium density is of order
$n_Y \sim T^3$, and $\Gamma_{ann} \gg H$ for not too small $\sigma_0$.
This means that annihilation (and, by reciprocity, creation) of
$Y$-pairs is indeed rapid, and $Y$-particles are indeed in
thermal equilibrium with the plasma. At very low temperature, on the
other hand, the equilibrium 
number density $n_Y^{(eq)}$ is exponentially small, and the
equilibrium rate is small,
$\Gamma_{ann}^{(eq)} \ll H$. At low temperatures we cannot, of course, 
make
use of the equilibrium formulas: $Y$-particles no longer annihilate
(and, by reciprocity, are no longer created), there is no thermal
equilibrium with respect to creation--annihilation processes, 
and the number density $n_Y$ gets diluted only because of the
cosmological expansion.

The freeze-out temperature $T_f$ is determined by the relation
\[
  \tau_{ann}^{-1} \equiv \Gamma_{ann} \sim H \; ,
\]
where we can still use the equilibrium formulas, as $Y$-particles
are in thermal equilibrium (with respect to annihilation and creation)
just before freeze-out. 
Making use of the relation \eqref{sep16-11-1} between the Hubble
parameter and temperature at radiation domination, we obtain
\begin{equation}
      \sigma_0 (T_f) \cdot n_Y (T_f) \sim  \frac{T_f^2}{M_{Pl}^{*}} \; ,
\label{dec1}
\end{equation}
or
\[
   \sigma_0 (T_f)
\cdot g_Y \cdot \left(\frac{m_Y T_f}{2\pi} \right)^{3/2}
\mbox{e}^{-\frac{m_Y}{T_f}} \sim \frac{T_f^2}{M_{Pl}^{*}} \; .
\]
The latter equation gives the freeze-out temperature, which, up to
loglog corrections, is
\be
   T_f \approx \frac{m_Y}{\ln (M_{Pl}^{*} m_Y \sigma_0)} \; 
\label{may10-1}
\ee
(the possible depndence of $\sigma_0$ on temperature
is irrelevant in the right hand side: we are doing the
calculation in the leading-log approximation anyway). 
Note that this temperature is somewhat lower than $m_Y$, if the
relevant microscopic mass scale is much below $M_{Pl}$. This means
that $Y$-particles freeze out when they are indeed non-relativistic, hence
the term ``cold dark matter''. The fact that the annihilation and
creation of $Y$-particles terminate at relatively low temperature has
to do with rather slow expansion of the Universe, which should be
compensated for by the smallness of the number density $n_Y$. 

At the freeze-out temperature, we make use of Eq.~(\ref{dec1})
and obtain
\[
   n_Y (T_f) = \frac{T_f^2}{M_{Pl}^{*} \sigma_0 (T_f)} \; .
\]
Note that this density is inversely proportional to the annihilation
cross section (modulo logarithm). The reason is that
for higher annihilation cross section, the creation--annihilation
processes are longer in equilibrium, and less $Y$-particles survive.

Up to a numerical factor of order 1, the number-to-entropy ratio 
at freeze-out
is
\be
   \frac{n_Y}{s} \simeq \frac{1}{g_* (T_f) M_{Pl}^{*}T_f \sigma_0(T_f)}  \; .
\label{nYs}
\ee
This ratio stays constant until the present time, so 
 the present number density of
$Y$-particles is
$   n_{Y, 0} = s_0 \cdot  
\left(n_Y/s \right)_{freeze-out}$,
and the mass-to-entropy ratio is
\be
  \frac{\rho_{Y, 0}}{s_0} = \frac{m_Y n_{Y,0}}{s_0} 
   \simeq   
\frac{\ln (M_{Pl}^{*} m_Y \sigma_0)}{g_*(T_f) M_{Pl}^{*} \sigma_0(T_f)}
\simeq \frac{\ln (M_{Pl}^{*} m_Y \sigma_0)}{\sqrt{g_*(T_f)} M_{Pl} \sigma_0 (T_f)} \; ,
\nonumber
\ee
where we made use of Eq.~(\ref{may10-1}).
This formula is remarkable. The mass density depends mostly on
one parameter, the annihilation cross section $\sigma_0$. The
dependence on the mass
of $Y$-particle is through the logarithm and through
$g_* (T_f)$; it is very
mild. 
The value of the logarithm here is between 30 and 40, depending on
parameters
(this means, in particular, that freeze-out occurs when the
temperature drops 30 to 40 times below the mass of $Y$-particle).
Plugging in other numerical values 
($g_* (T_f) \sim 100$,
$M_{Pl}^* \sim 10^{18}\,\mbox{GeV}$), as well as numerical factor omitted
in Eq.~(\ref{nYs}), and comparing with Eq.~(\ref{10p*}) we obtain the estimate
\be
   \sigma_0 (T_f) \equiv \langle \sigma v \rangle (T_f)
= (1 \div 2) \cdot 10^{-36}~\mbox{cm}^2 \; .
\label{estim}
\ee
This is a weak scale cross section, 
which tells us that the relevant energy scale
is TeV. We note in passing that the estimate~(\ref{estim}) is quite precise and robust.

If the annihilation occurs in $s$-wave,
the annihilation cross section may be parametrized as
$   \sigma_0 = \alpha^2/ M^2$
where $\alpha$ is some coupling constant, and $M$ is
a mass scale (which may be higher than 
$m_Y$). This parametrization is suggested by the picture of
$Y$ pair-annihilation via the exchange by another particle of mass 
$M$. With $\alpha \sim
10^{-2}$, the estimate for the mass scale is roughly
$ M \sim 1\,\mbox{TeV}$.
Thus, with very mild assumptions, we find that the
non-baryonic dark matter may naturally originate from
the TeV-scale physics. In fact, what we have found can be understood as
an approximate equality between the cosmological parameter, mass-to-entropy
ratio of dark matter, and the particle physics parameters,
\[
\mbox{mass-to-entropy} \simeq \frac{1}{M_{Pl}}
\l \frac{\mbox{TeV}}{\alpha_W} \r^2 \; .
\]
Both are of order $10^{-10}\,\mbox{GeV}$, and it is very tempting
to think that this is not a mere coincidence.
If it is not, the dark matter particle should be found
at the LHC.

Of course, the most prominent candidate
for WIMP is neutralino of the supersymmetric extensions of the Standard
Model. The situation with neutralino is somewhat tense, however.
The point is that the pair-annihilation of neutralinos often occurs
in $p$-wave, rather than $s$-wave. This gives the suppression factor
in $\sigma_0 \equiv \langle \sigma_{ann} v\rangle$, 
proportional to $v^2 \sim T_f / m_Y \sim 1/30$.
Hence, neutralinos tend to be overproduced in most of the parameter
space of MSSM and other models. Yet neutralino remains a good candidate,
especially at high $\tan \beta$.

\subsection{Warm dark matter: light gravitinos}

The cold dark matter scenario successfully describes the bulk
of the cosmological data. Yet, there are clouds above it.
First, according to numerical simulations, CDM scenario tends to
overproduce small objects --- dwarf galaxies:  
it predicts hundreds of satellite
dwarf galaxies
in the vicinity of a
large galaxy like Milky Way, 
whereas only dozens of satellites have been observed so far. Second,
again according to simulations,
CDM tends to produce too high densities in galactic centers (cusps
in density profiles); this feature is not confirmed by observations
either.
There is no strong discrepancy 
yet, but one may be motivated to analyse a possibility
that dark matter is not that cold.

An alternative to CDM is warm dark matter whose particles decouple
being relativistic. Let us assume for definiteness
that they are in kinetic equilibrium
with cosmic plasma when their number density freezes out
(thermal relic). After
kinetic equilibrium breaks down, and WDM particles decouple completely,
their spatial momenta decrease as $a^{-1}$,
i.e., the momenta are  of order $T$ all the time after decoupling.
WDM particles become non-relativistic at $T\sim m$, where $m$ is their mass.
Only after that the WDM perturbations start to 
grow\footnote{The situation in fact is somewhat more complicated,
but this simplified picture will be sufficient
 for our  estimates.}: as we mentioned above,
relativistic particles escape from gravitational potentials, so
the gravitational potentials get smeared out instead of getting deeper.
Before becoming non-relativistic, WDM particles travel the distance of the
order of the horizon size; the WDM perturbations therefore are suppressed
at those scales.
The horizon size at the time $t_{nr}$ when $T\sim m$ is of order
\[
   l(t_{nr}) \simeq
H^{-1} (T\sim m) = \frac{M_{Pl}^*}{T^2} \sim  \frac{M_{Pl}^*}{m^2} 
\; .
\]
Due to the expansion of the Universe,
the corresponding length at present is
\be
  l_0 =  l(t_{nr}) \frac{a_0}{a(t_{nr})} \sim 
l(t_{nr}) \frac{T}{T_0} \sim \frac{M_{Pl}}{m T_0} \; ,
\label{l0dwarf}
\ee
where we neglected  (rather weak) dependence on $g_*$.
Hence, in WDM scenario,
 structures of sizes smaller than $l_0$ are less abundant
as compared to CDM. Let us point out that $l_0$ refers to the
size of the perturbation as if it were in the linear regime; in other 
words, this is the size of the region from which matter collapses into a
compact object.

The present size of a dwarf galaxy is a few kpc, and the density 
is about $10^6$ of the average density in the Universe.
Hence, the size $l_0$ for these objects is of order $100~\mbox{kpc}
\simeq 3 \cdot 10^{23}~\mbox{cm}$. Requiring that perturbations of this
size, 
but not much larger, are suppressed, we obtain from Eq.~(\ref{l0dwarf})
the estimate~(\ref{wdmmass}) for the mass of WDM particles.

Among WDM candidates, light gravitino is probably the best
motivated. The gravitino mass is of order
\[
m_{3/2} \simeq \frac{F}{M_{Pl}} \; ,
\]
where $\sqrt{F}$ is the supersymmetry breaking scale. Hence, gravitino
masses are in the right ballpark for rather low supersymmetry breaking
scales,  $\sqrt{F} \sim 10^{6} - 10^{7}\,\mbox{GeV}$. This is the case, 
e.g., in gauge mediation scenario. With so low mass, gravitino
is the lightest supersymmetric particle (LSP), so it is stable in
many supersymmetric extensions of the Standard Model.
From this viewpoint
gravitinos can indeed serve as dark matter particles.
For what follows, important parameters are the
widths of decays of other superpartners into  gravitino and the Standard
Model particles. These are of order
\be
\Gamma_{\tilde S} \simeq \frac{M_{\tilde S}^5}{F^2} \simeq 
\frac{ M_{\tilde S}^5}{ m_{3/2}^2 M_{Pl}^2} \; ,
\label{swidth}
\ee
where $M_{\tilde S}$ is the mass of the superpartner.

One mechanism of the gravitino production in the early Universe
is decays of other superpartners. Gravitino interacts with everything
else so weakly, that once produced, it moves freely, without
interacting with cosmic plasma. At production, gravitinos are
relativistic, hence they are indeed {\it warm} dark matter candidates.
Let  us assume that production in decays is the dominant mechanism
and consider
under what circumstances the present mass density of gravitinos 
coincides with that of dark matter.

The rate of gravitino production in decays of superpartners of the type
$\tilde S$ in the
early Universe is
\[
 \frac{d (n_{3/2}/s)}{dt} = \frac{n_{\tilde S}}{s} \Gamma_{\tilde S} \; ,
\]
where $n_{3/2}$ and $n_{\tilde S}$ are number densities of gravitinos
and superpartners, respectively,
and $s$ is the entropy density. 
For superpartners in thermal equilibrium,
one has  $n_{\tilde S}/s =\mbox{const} \sim g_*^{-1}$ 
for $T \gtrsim M_{\tilde S}$,
and $n_{\tilde S}/s \propto \mbox{exp} (- M_{\tilde S}/T)$
at  $T \ll M_{\tilde S}$. Hence, the production is most efficient
at $T \sim M_{\tilde S}$, when the number density of superpartners
is still large, while the Universe expands most slowly. The density of
gravitinos produced in decays of ${\tilde S}$'s is thus given by
\[
 \frac{n_{3/2}}{s}  \simeq \left( \frac{d (n_{3/2}/s)}{dt} \cdot H^{-1}
\right)_{T \sim M_{\tilde S}} \simeq 
\frac{\Gamma_{\tilde S}}{g_*} H^{-1}(T \sim M_{\tilde S})
\simeq \frac{1}{g_*} \cdot
\frac{ M_{\tilde S}^5}{ m_{3/2}^2 M_{Pl}^2} \cdot 
\frac{M_{Pl}^*}{M_{\tilde S}^2} \; .
\]
This gives the mass-to-entropy ratio today:
\be
   \frac{m_{3/2} n_{3/2}}{s} \simeq \sum_{\tilde S}
\frac{M_{\tilde S}^3}{g_*^{3/2} M_{Pl} m_{3/2}} \; ,
\label{sum-ino}
\ee
where the sum runs over all  superpartner species {\it
which have ever been relativistic
in thermal equilibrium}. The correct value~(\ref{10p*}) is obtained for gravitino masses in the 
range~(\ref{wdmmass}) at
\be
M_{\tilde S} = 100 - 300\,\mbox{GeV} \; .
\label{msup}
\ee
Thus, the scenario with gravitino as warm dark
matter particle requires light superpartners, which are to be
discovered at the LHC.

A few comments are in order. First, decays of superpartners is not
the only mechanism of gravitino production: gravitinos may also be
produced in scattering of superpartners. To avoid overproduction
of gravitinos in the latter processes, one has to assume that the
maximum temperature in the Universe (reached after post-inflationary
reheating stage) is quite low, $T_{max} \sim 1 - 10$\,TeV.
This is not a particularly
 plausible assumption, but it is
consistent with everything else in cosmology
and can indeed be realized
in some models of inflation.
Second,
existing constraints on masses of
strongly interacting superpartners (gluinos and squarks)
suggest that their masses exceed Eq.~(\ref{msup}). Hence, these particles 
should not contribute to the sum in Eq.~(\ref{sum-ino}), otherwise WDM gravitinos would be overproduced. 
This is possible, if masses of
squarks and gluinos are larger
than $T_{max}$, so that they were never abundant in the
early Universe. 
Third, gravitino produced in decays of superpartners is {\it not}
a thermal relic, as it was never in thermal equilibrium
with the rest of cosmic plasma. Nevertheless, since gravitinos are
produced at $T \sim M_{\tilde{S}}$ and at that time have energy
$E  \sim M_{\tilde{S}} \sim T$, our estimate \eqref{l0dwarf}
does apply.
Finally, the decay into gravitino and the Standard
Model particles is the only decay channel for the next-to-lightest
superpartner (NLSP). Hence, the estimate for
the total width of
NLSP is given by Eq.~(\ref{swidth}), so that
\[
c\tau_{NLSP} = \mbox{a~few} \cdot \mbox{mm} - \mbox{a~few} \cdot 100~{\mbox m}
\]
for $m_{2/3} = 3 - 30$\,keV and $M_{NLSP} = 100 - 300$\,GeV.
Thus, NLSP should either
be visible in a detector, or  fly it through.

Needless to say, the warm gravitino scenario is a lot more contrived
than the WIMP option. It is reassuring, however, that it can be ruled out
or confirmed at the LHC.

\subsection{Discussion}

If dark matter particles are indeed WIMPs, and the relevant
energy scale is of order 1\,TeV, then the Hot Big Bang theory will be
probed experimentally up to temperature of $(\mbox{a~few})\cdot
(10 - 100)$\,GeV
and down to age $10^{-9} - 10^{-11}~$s in relatively near future (compare
to 1\,MeV and 1~s accessible today through Big Bang Nucleosynthesis). 
With microscopic physics to be
known from collider experiments, the WIMP density will be reliably
calculated and checked against the data from observational cosmology.
Thus, WIMP scenario offers a window to
a very early stage of the evolution of the Universe.

If dark matter particles are gravitinos, the prospect of 
probing quantitatively so early stage of the cosmological
evolution is not so bright: it would be very hard, if at all possible,
to get an experimental handle on the gravitino mass;
furthermore, the present gravitino mass
density depends on an unknown reheat temperature
$T_{max}$. On the other hand, if this scenario is realized in Nature,
then the whole picture of the early Universe will be quite different from
our best guess on the
early cosmology.   Indeed, gravitino
scenario requires low reheat temperature, which in turn calls for
rather exotic mechanism of inflation.

The mechanisms discussed here are by no means the only ones
capable of producing dark matter, and WIMPs and gravitinos  are by no means 
the only candidates for dark matter particles. Other dark matter
candidates include 
axions, sterile neutrinos, Q-balls, 
very heavy relics produced 
towards the end of inflation, etc. Hence, even though there are grounds to hope
that the dark matter problem will be solved by the LHC, there is no guarantee
at all.

\section{Baryon asymmetry of the Universe}
\label{sec:bau}

In the present Universe, there are baryons and almost no antibaryons.
The number density of baryons today is characterized by the ratio
$\eta_B$, see Eq.~(\ref{etadef}).
In the early
Universe,
the appropriate quantity is
\[
       \Delta_B = \frac{n_B - n_{\bar{B}}}{s} \; ,
\]
where $n_{\bar{B}}$ is the number density of antibaryons, and $s$ is
the entropy density. If the baryon number is conserved, and the
Universe expands adiabatically, $\Delta_B$ is constant, and its value, 
up to a numerical factor, is equal to $\eta$ (cf. Eqs.~(\ref{nga}) and~(\ref{e-density})). More precisely, 
\[
  \Delta_B \approx 0.8 \cdot 10^{-10} \; .
\]
At early times, at temperatures well above 100\,MeV, cosmic plasma
contained many quark-antiquark pairs, whose number density
was of the order of the entropy density,
\[
    n_q + n_{\bar{q}} \sim s \; ,
\]
while the baryon number density was related to densities of quarks and
antiquarks as follows (baryon number of a quark equals $1/3$),
\[
  n_B = \frac{1}{3} (n_q - n_{\bar{q}}) \; .
\]
Hence, in terms of quantities characterizing the very early epoch, the
baryon asymmetry may be expressed as
\[
  \Delta_B \sim \frac{n_q - n_{\bar{q}}}{n_q + n_{\bar{q}}} \; .
\]
We see that there was one extra quark per about 10 billion
quark-antiquark pairs! It is this tiny excess that is responsible for
the entire baryonic matter in the present Universe: as the Universe
expanded and cooled down, antiquarks annihilated with quarks,
and only the excessive quarks remained and formed baryons.

There is no logical contradiction to suppose that the tiny excess of
quarks over antiquarks was built in as an initial condition. This is
not at all satisfactory for a physicist, however. Furthermore,
inflationary scenario
 does not provide such an initial condition for the hot Big Bang epoch;
rather, inflation theory
predicts that the Universe
was baryon-symmetric just after inflation.
Hence, one would like to explain the baryon asymmetry dynamically.

The baryon asymmetry may be generated from initially symmetric state
only if three necessary conditions, dubbed Sakharov's conditions, are
satisfied. These are 

(i) baryon number non-conservation;

(ii) C- and CP-violation;

(iii) deviation from thermal equilibrium.

All three conditions are easily understood.
(i) If baryon number were conserved, and initial net baryon number in
the Universe was zero, the Universe today would still be symmetric.
(ii) If C or CP were conserved, then the rate of
reactions with particles would be the same as the rate of reactions
with antiparticles, and no asymmetry would be generated. 
(iii) Thermal equilibrium means that the system is stationary (no time
dependence at all). Hence, if the initial baryon number is zero, it is
zero forever, unless there are deviations from thermal equilibrium.

There are two well understood mechanisms of baryon number
non-conservation. One of them emerges in Grand Unified Theories and is
due to the exchange of super-massive particles. It is similar,
say, to the mechanism of charm non-conservation in weak interactions,
which occurs via the exchange of heavy $W$-bosons. The scale of these
new, baryon number violating interactions is the Grand Unification
scale, presumably of order $M_{GUT} \simeq
10^{16}\,\mbox{GeV}$. It is rather unlikely
that the baryon asymmetry was generated due to this mechanism:
the relevant temperature would be of order $M_{GUT}$, 
while so high reheat
temperature after inflation is difficult to obtain.

Another mechanism is non-perturbative~\cite{'tHooft:1976up} 
and is related to the triangle
anomaly in the baryonic current (a keyword here is 
``sphaleron''~\cite{Klinkhamer-Manton,Kuzmin:1985mm}).
It exists already in the Standard
Model, and, possibly with slight modifications, operates in all its
extensions. The two main features of this mechanism, as applied to the
early Universe, is that it is effective over a wide range of
temperatures, 
$100\,\mbox{GeV} < T < 10^{11}\,\mbox{GeV}$, and that it conserves
$(B-L)$.

Let us pause here to discuss the physics behind electroweak baryon 
and lepton number non-conservation in little more detail,
though still at a qualitative level. A detailed analysis can be found in
the book~\cite{Rubakov:2002fi} and in references therein.

The first object to consider is
the
baryonic current,
\[
B^\mu = \frac{1}{3} \cdot \sum_{i} \bar{q}_{i} \gamma^\mu q_{i} \; ,
\]
where the sum runs over quark flavors. Naively, the baryonic current is
conserved, but at the quantum level its divergence is non-zero,
the effect called triangle anomaly (similar effect
goes under the name of axial anomaly in the context
of QED and QCD),
 \[
    \partial_\mu B^\mu =  
\frac{1}{3} \cdot 3_{colors}
\cdot 3_{generations} \cdot  \frac{g^2}{32 \pi^2}
\epsilon^{\mu \nu \lambda \rho}
F^a_{\mu \nu} F^a_{\lambda \rho} \; ,
\]
where $F^a_{\mu \nu}$ and $g$ are the field strength of the $SU(2)_W$
gauge field and the $SU(2)_W$ gauge coupling, respectively. 
Likewise, each leptonic current ($n = e, \mu, \tau$) is anomalous,
\[
    \partial_\mu L^\mu_n = \frac{g^2}{32 \pi^2} 
\cdot \epsilon^{\mu \nu \lambda \rho}
F^a_{\mu \nu} F^a_{\lambda \rho} \; .
\]
A non-trivial fact is that there exist 
large field fluctuations,  $F^a_{\mu \nu} ({\bf x}, t) \propto g^{-1}$ 
which
have
\[
Q \equiv \int~d^3x dt~   \frac{g^2}{32 \pi^2} 
\cdot \epsilon^{\mu \nu \lambda \rho}
F^a_{\mu \nu} F^a_{\lambda \rho} \neq 0 \; .
\]
Furthermore, for any such fluctuation the value of $Q$ is integer.
Suppose now that a fluctuation with non-vanishing $Q$ has occured.
Then the baryon numbers in the end and beginning of the process are different,
\be
 B_{fin} - B_{in} =
 \int~d^3x dt~  \partial_\mu B^\mu = 3 Q \; .
\label{may7-2}
\ee
Likewise
\be
   L_{n,~fin} - L_{n,~in} = Q \; .
\label{may7-3}
\ee
This explains the selection rule mentioned above:
 $B$ is violated, $(B-L)$ is not.

 At zero temprature, the large field fluctuations that induce baryon
and lepton number violation are vacuum fluctuations, called instantons,
which 
to a certain 
extent are
similar to virtual fields that emerge and disappear in 
vacuum of quantum field theory at the perturbative level. The 
peculiarity is
that instantons are {\it large} field fluctuations. The latter property
results in the exponential suppression of the probability of
their emergence, and hence
the rate of baryon number violating processes, by a factor
\[
\mbox{e}^{- { \frac{16\pi^2}{g^2}}} \sim 10^{-165} \; .
\]
On the other hand, at high temperatures
there are large {\it thermal} fluctuations
(``sphalerons'') whose rate is not necessarily small.
And, indeed, 
$B$-violation in the early Universe
is rapid as compared to the cosmological expansion at
sufficiently high temperatures, provided that
\be
\langle\phi \rangle_T < T \; ,
\label{may7-4}
\ee
where
$ \langle\phi \rangle_T$ is the
 Higgs expectation value at temperature $T$.

\begin{figure}[htb!]
\begin{center}
\includegraphics[width=0.2\textwidth]{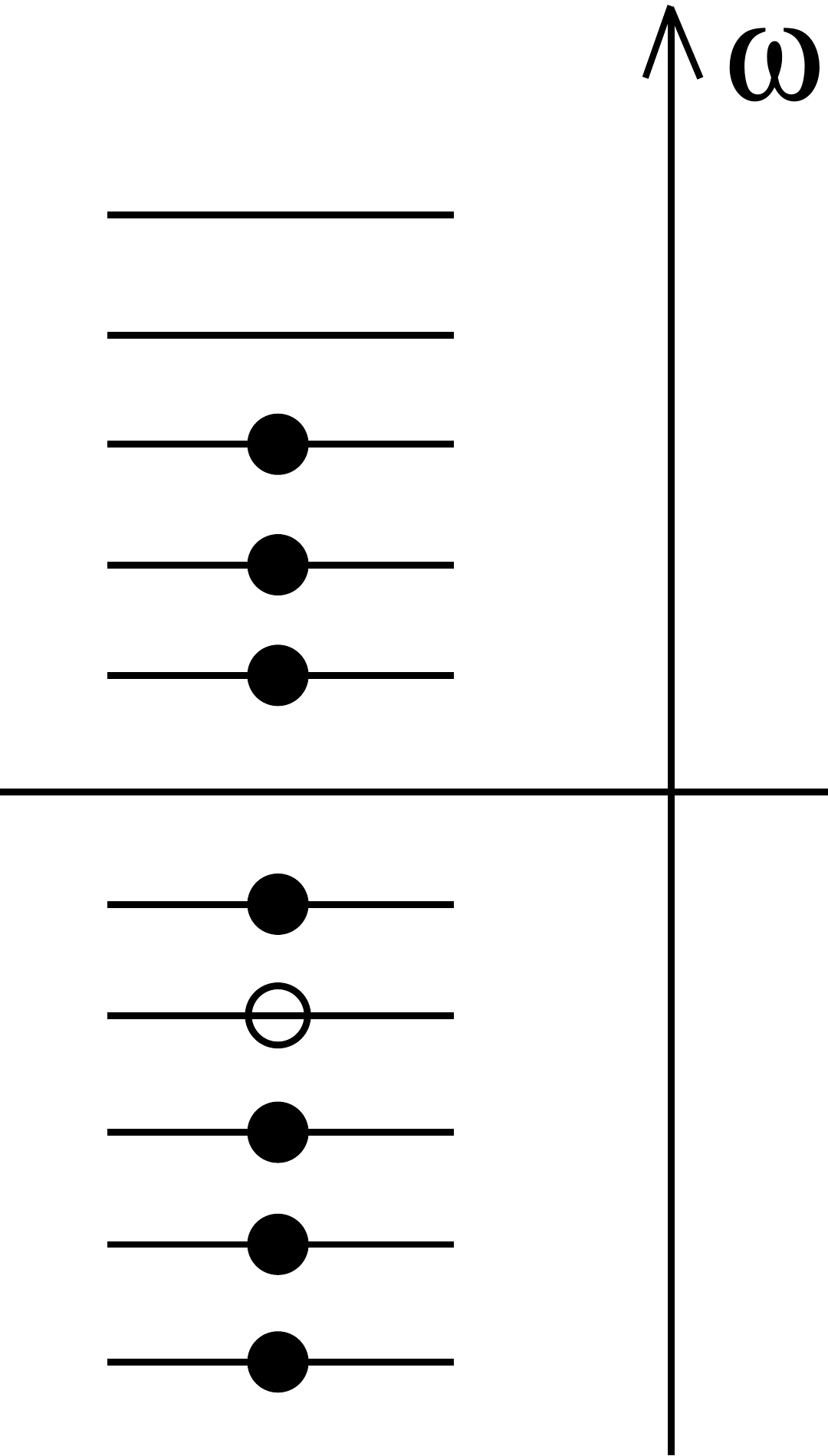}
\end{center}
\caption{Fermion energy levels at zero background gauge field.
 \label{a=0}
 }
\end{figure}

One may wonder how baryon number may be not conserved even though
there are no baryon number violating terms in the Lagrangian of
the Standard Model. To understand what is going on, let us consider
a massless {\it left handed} fermion field in the background of the $SU(2)$
gauge field
${\bf A}({\bf x}, t)$, which depends on space-time coordinates in a
non-trivial way. As a technicality, 
 we set the temporal component of the gauge field equal to zero,
$A_0 =0$, by the choice of gauge.
One way to understand the behavior of the fermion field in
the gauge field background is to study the system of
eigenvalues of the
Dirac Hamiltonian $\{ \omega (t) \}$. The Hamiltonian is defined in the
standard way
\[
  H_{Dirac} (t)
= i\alpha^i \left(\d_i - ig A_i ({\bf x}, t)\right) \frac{1- \gamma_5}{2} \; ,
\]
where $\alpha^i = \gamma^0 \gamma^i$, so that the Dirac equation has the
Schr\"odinger form,
\[
 i \frac{\d \psi}{\d t} = H_{Dirac} \psi \; .
\]
We are going to discuss the eigenvalues $\omega_n (t)$
of the operator $H_{Dirac}(t)$,
treating $t$ as a parameter. These eigenvalues are found from
\[
H_{Dirac} (t) \psi_n = \omega_n (t) \psi_n \; .
\]
At ${\bf A} = 0$ the system
of levels is shown schematically in Fig.~\ref{a=0}. Importantly,
there are both  positive- and negative-energy levels. According to Dirac,
the lowest energy state (Dirac vacuum) has all negative energy levels
occupied, and all positive energy levels empty. Occupied positive energy
levels (three of them in Fig.~\ref{a=0}) correspond to real fermions,
while empty negative energy levels describe antifermions (one in 
Fig.~\ref{a=0}). Fermion-antifermion annihilation in
this picture is a jump 
of a fermion from a positive
energy level to an unoccupied negative energy level.

As a side remark, this original Dirac picture is, in fact, equivalent
to the more conventional (by now) procedure of the
quantization of  
fermion field, which does not make use of the notion of
negative energy levels. The discussion that follows can be translated into
the conventional language; however, the original Dirac picture turnd out
to be a
lot more transparent in our context. This is a nice example of the
complementarity of various approaches in quantum field theory.

Let us proceed with the discussion of the fermion energy levels
in gauge field backgrounds.
In weak background fields, the energy levels depend on time
(move), but nothing dramatic happens. For adiabatically varying background
fields, the fermions merely sit on their levels, while fast changing fields
generically give rise to jumps from, say, negative- to  positive-energy
levels, that is, creation of fermion-antifermion pairs. Needless to
say, fermion number $(N_f - N_{\bar f})$ is conserved.

\begin{figure}[htb!]
\begin{center}
\includegraphics[width=0.85\textwidth]{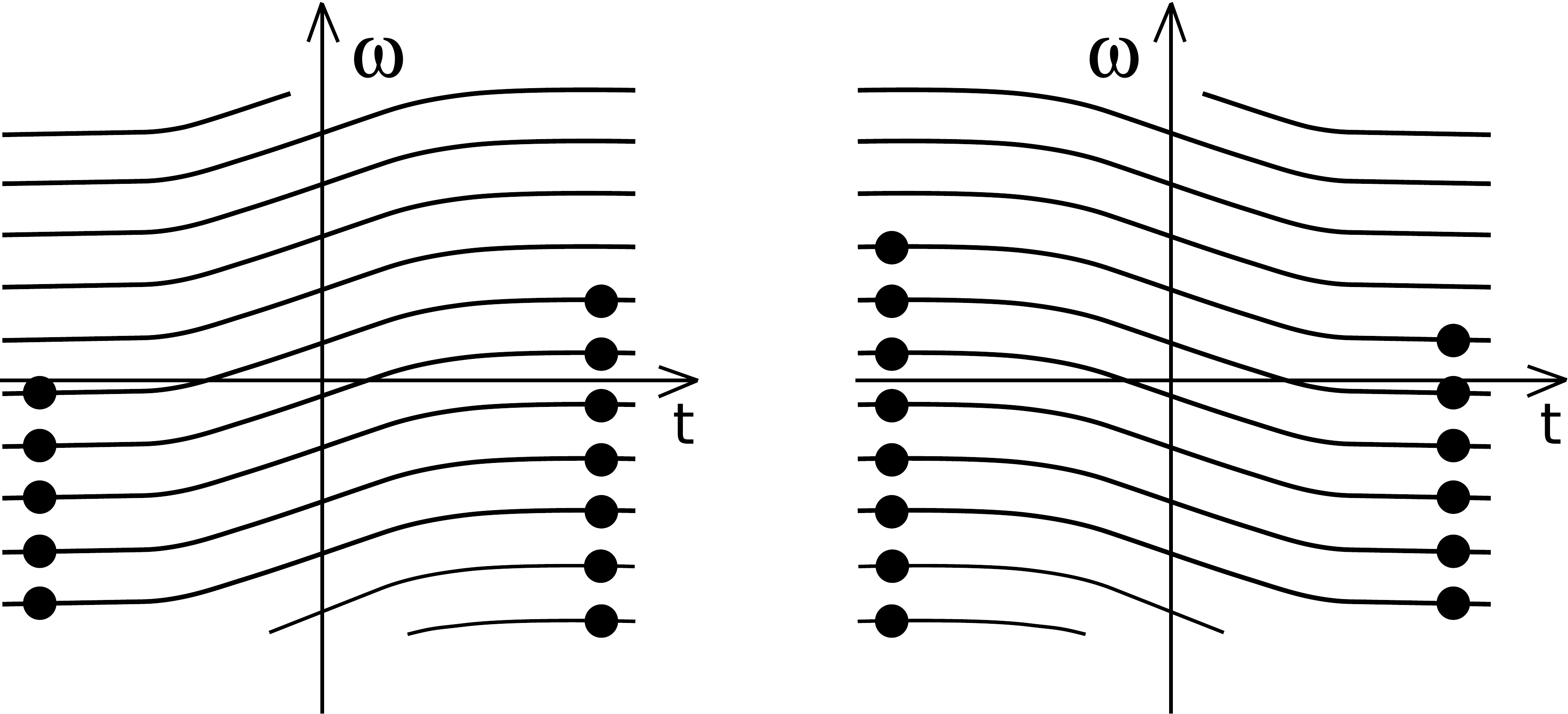}
\end{center}
\caption{Motion of fermion levels in background gauge fields
with non-vanishing $Q$ (shown is the case $Q=2$). Left panel:
left-handed fermions. Right panel: right-handed fermions.
 \label{ane0}}
\end{figure}

The situation is entirely different for the background fields with
non-zero $Q$. The levels of left-handed fermions move as shown
in the left panel of Fig.~\ref{ane0}. Some levels necessarily cross zero,
and the net number of levels crossing zero from below equals
$Q$. This means that the number of left-handed fermions is not
conserved: for adiabatically
varying gauge field ${\bf A}({\bf x}, t)$ the motion of levels
shown
in the left panel of Fig.~\ref{ane0} corresponds to the case
in which
the initial state of the
fermionic system is vacuum (no fermions at all) whereas the final state
contains $Q$ real fermions (two in the particular case shown).
If the evolution of the gauge field is not adiabatic,
the result for the fermion number non-conservation is the same:
there may be jumps from negative energy levels to positive
energy levels, or vice versa. These correspond to creation or
annihilation of fermion-antifermion pairs, but the net change
of the fermion number (number of fermions minus number of antifermions)
remains equal to $Q$.
Importantly, the initial and final field configurations of the gauge
field may be trivial, ${\bf A} = 0$
(up to gauge transformation), so that fermion number non-conservation
may occur due to a fluctuation that begins and ends in the 
gauge field vacuum\footnote{A subtlety here is that in four-dimensional
gauge theories, this is impossible for {\it Abelian}
gauge fields, so fermion number non-conservation is inherent
in {\it non-Abelian} gauge theories only.}. 
This is precisely an instanton-like vacuum fluctuation.
At finite temperatures, processes of this type occur due to
thermal fluctuations, sphalerons.


If the same gauge field interacts also with right-handed fermions,
the motion of the levels of the latter is opposite to
that of left-handed fermions. This is shown in the right
 panel of Fig.~\ref{ane0}. The change in the number of right-handed
fermions is equal to $(-Q)$. So, if the gauge interaction is vector-like,
the total fermion number $N_{left} + N_{right}$ is conserved, while
chirality $N_{left} - N_{right}$ is violated even for massless fermions.
This explains why there is no baryon number  violation in QCD.
On the other hand, non-perturbative violation of chirality in QCD
in the limit of massless quarks has non-trivial consequences, which are
indeed confirmed by phenomenology. In this sense anomalous
non-conservation of fermion quantum numbers is an experimentally
established fact.

In electroweak theory, right-handed fermions do not interact with
$SU(2)_W$ gauge field, while left-handed fermions do. Therefore, fermion 
number is not conserved. Since fermions of each $SU (2)_W$-doublet
interact with the   $SU (2)_W$ gauge bosons (essentially $W$ and $Z$)
in one and the same way, they are equally created in a process involving
a gauge field fluctuation with non-zero $Q$. This again leads to the
relations~(\ref{may7-2}) and~(\ref{may7-3}), i.e., to the selection rules
$\Delta B = \Delta L$,
$\Delta L_e = \Delta L_\mu = \Delta L_{\tau}$.

It is tempting to use this mechanism of baryon number non-conservation
for explaining the baryon asymmetry of the Universe. There are
two problems, however. One is that CP-violation in the Standard Model
is too weak:  the CKM mechanism
alone is insufficient to generate the realistic value of the
baryon asymmetry. Hence, one needs extra sources of CP-violation.
Another problem has to do with departure from thermal equilibrium
that is necessary for the generation of the baryon asymmetry.
At temperatures well above 100\,GeV electroweak symmetry is restored,
the expectation value of the Higgs field 
$\phi$ is zero\footnote{There are subtleties at this point,
see below.}, the relation~(\ref{may7-4}) is valid, and the baryon number non-conservation is rapid
as compared to the cosmological expansion. At temperatures of order
100\,GeV the  relation~(\ref{may7-4}) may be violated, but the Universe expands very slowly:
the cosmological time scale at these temperatures is
\be
H^{-1} = \frac{M_{Pl}^*}{T^2}
\sim 10^{-10}
~\mbox{s} \; ,
\label{may8-1}
\ee
which is very large by the electroweak physics standards. The only
way in which
strong departure from thermal equilibrium at these
temperatures may occur is through  the
first order phase transition.

\begin{figure}[htb!]
\includegraphics[width=\textwidth]{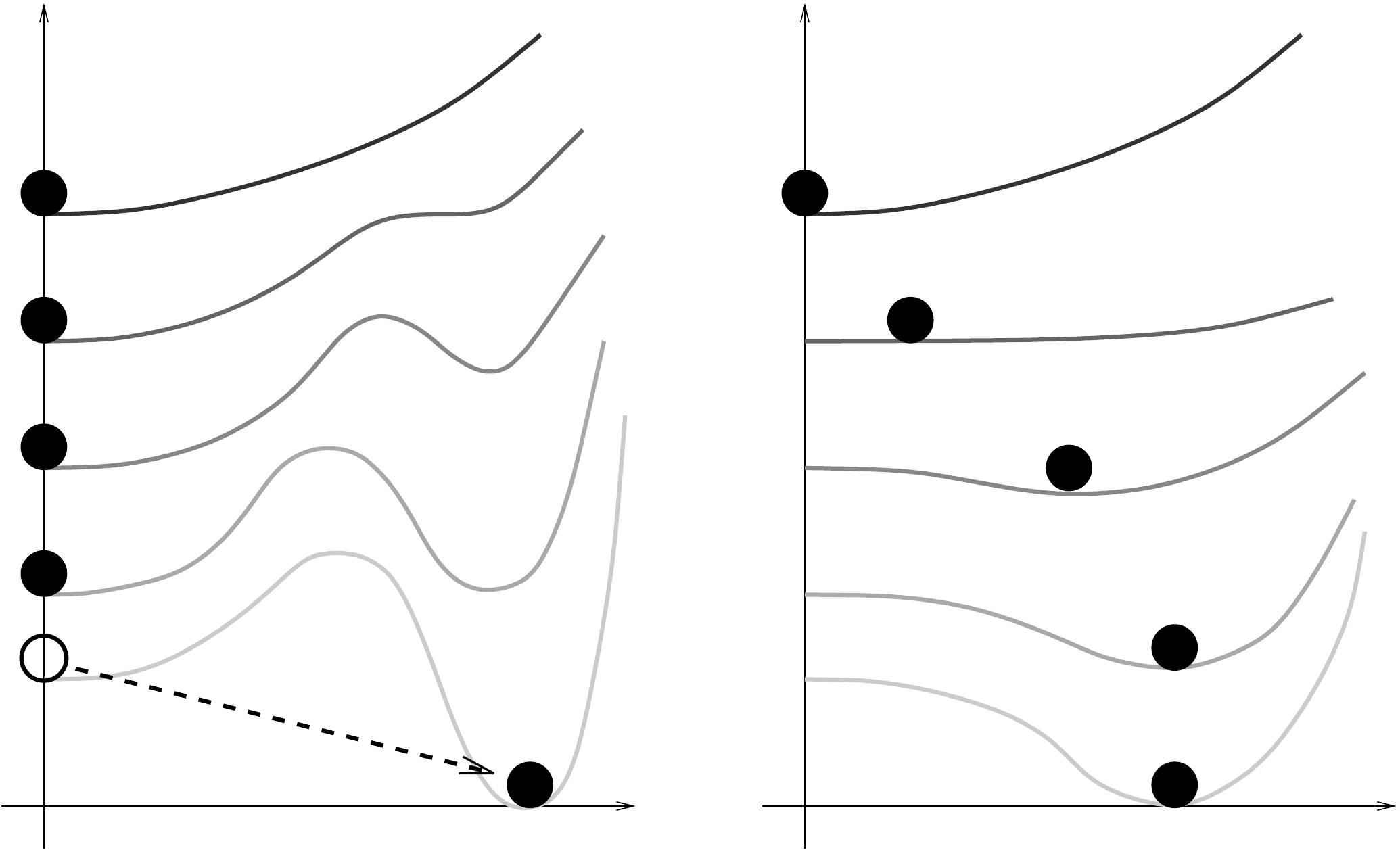}
\begin{picture}(10,10)(0,0)
{\large 
\put(90,120){{}}
\put(22,275){$V_{eff}(\phi)$}
\put(280,275){$V_{eff}(\phi)$}
\put(200,10){$\phi$}
\put(445,10){$\phi$}
}
\end{picture}
\caption{Effective potential as function of $\phi$ at different temperatures.
Left: first order phase transition. Right: second order phase transition.
Upper curves correspond to higher temperatures.
 \label{1-2order}}
\end{figure}
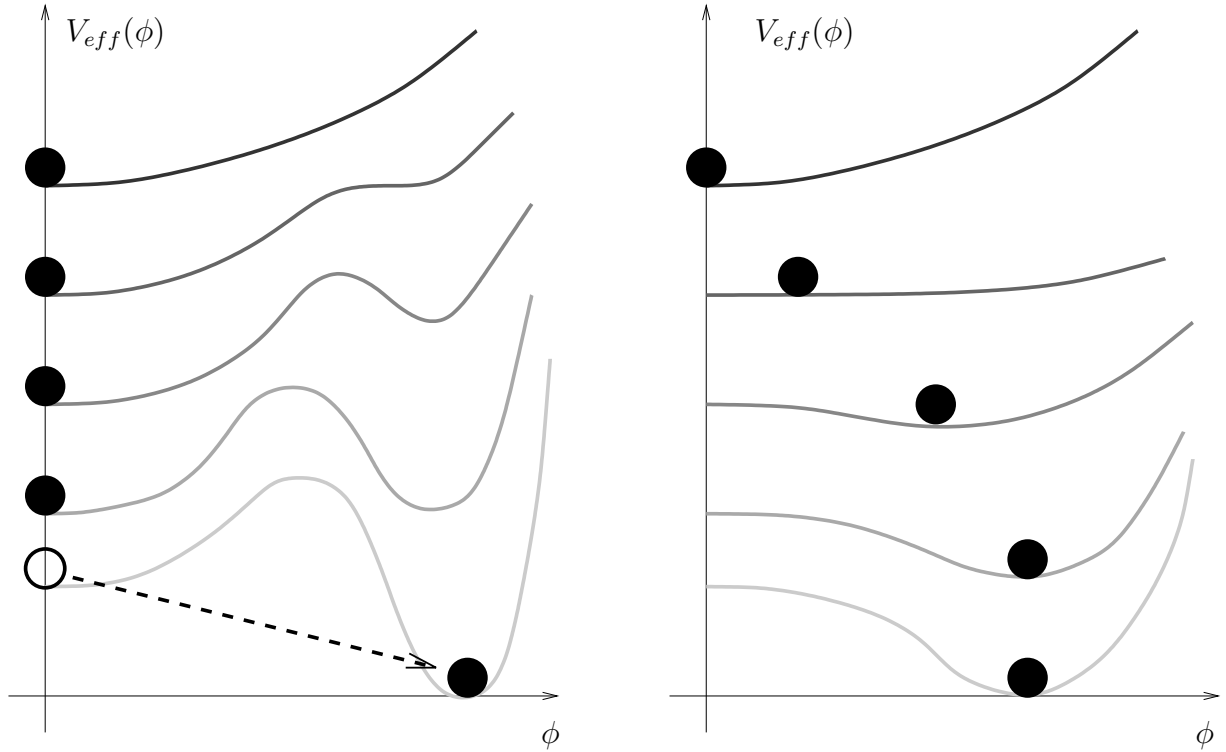

The property that
at temperatures well above 100\,GeV the expectation value of
the Higgs field is zero, while it is non-zero in vacuo,
suggests that there may be a phase transition
from the phase with $\langle \phi \rangle = 0$ to the phase
 with $\langle \phi \rangle \neq 0$. The situation is pretty
subtle here, as $\phi$ is not gauge invariant, and hence cannot serve
as an order parameter, so the notion of phases 
with $\langle \phi \rangle = 0$ and
$\langle \phi \rangle \neq 0$
is vague. In fact,  neither electroweak theory 
nor most of its extensions have a gauge-invariant order parameter,
so there is no real distinction between these ``phases''.
This situation is similar to that in liquid-vapor
system, which does not have an order parameter and may or may
not experience vapor-liquid phase transition as temperature decreases,
depending on other parameters characterizing this system, e.g.,  pressure.  
In the Standard Model the role of such a parameter is played by
the Higgs self-coupling $\lambda$ or, in other words, the Higgs
boson mass.

Continuing to use somewhat sloppy terminology, we observe that
the interesting
case for us is the first order phase transition. In this case
the effective potential (free energy density as function of
$\phi$) behaves as shown in the left panel of Fig.~\ref{1-2order}.
At high temperatures, there exists one minimum of $V_{eff}$ at
$\phi =0$, and the expectation value of the Higgs field is zero.
As the temperature decreases, another minimum appears at finite
$\phi$, and then becomes lower than the minimum at $\phi =0$.
However, the probability of the transition from the phase
$\phi=0$ to the phase $\phi \neq 0$ is very small for some time,
so the system gets overcooled. The transition occurs when the
temperature becomes sufficiently low, as shown schematically
by an arrow in Fig.~\ref{1-2order}. This is to be contrasted
to the case, e.g., of the second order phase transition with the
behavior of the effective potential shown in the right panel
of  Fig.~\ref{1-2order}. In the latter case, the field slowly evolves,
as the temperature decreases, from zero to non-zero vacuum
value, and the system remains very close to the thermal equilibrium
at all times.

\begin{figure}[htb!]
\includegraphics[width=\textwidth]{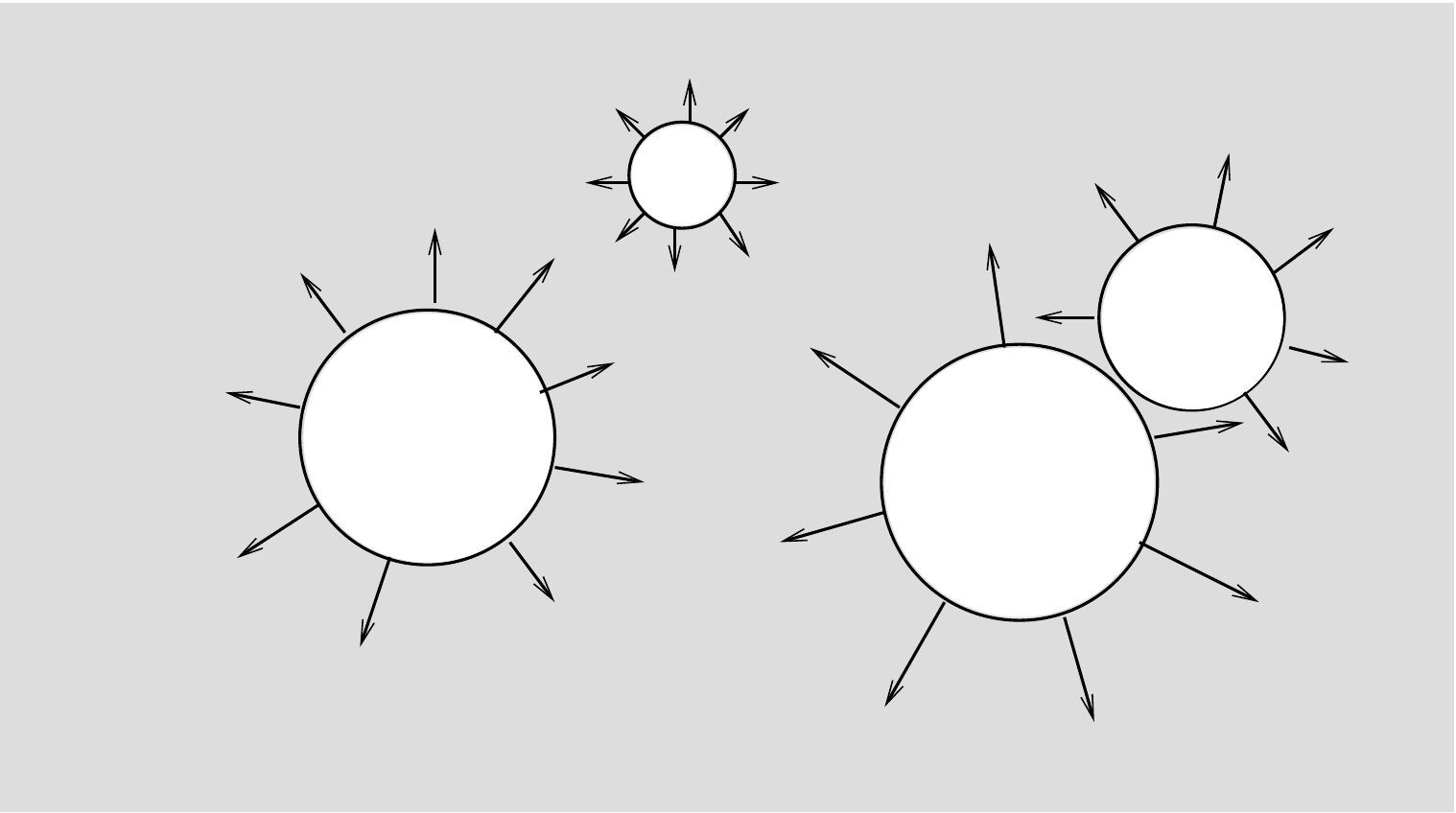}
\begin{picture}(10,10)(0,0)
{\large 
\put(90,120){{}}
\put(25,160){$\phi=0$}
\put(120,125){$\phi \neq 0$}
\put(306,115){$\phi \neq 0$}
\put(150,30){$\phi=0$}
\put(368,172){$\phi \neq 0$}
\put(320,230){$\phi=0$}
}
\end{picture}
\caption{
First order phase transition: boiling Universe.
 \label{boiling}}
\end{figure}
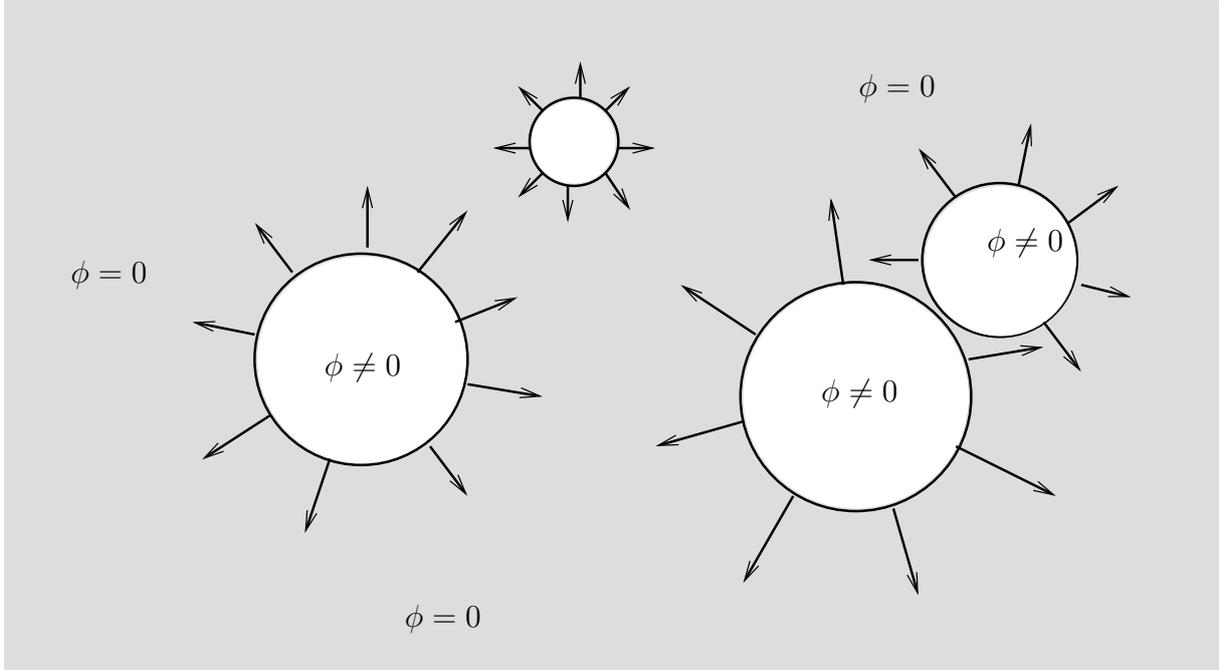

The first order phase transition occurs via spontaneous creation
of bubbles of the new phase inside the old phase. These bubbles
then grow, their walls eventually collide, and the new phase finally
occupies entire space. The Universe boils, as shown schematically 
in Fig.~\ref{boiling}. In the cosmological context, this process
happens when the bubble nucleation rate  per Hubble time
per Hubble volume is of order 1,
$  \Gamma_{nucl} \sim H^{-4}$.
The velocity of the bubble wall in the relativistic
cosmic plasma is roughly of the order of the speed of light
(in fact, it is somewhat smaller, from $0.1~c$ to $0.01~c$),
simply because there are no relevant dimensionless parameters
characterizing the system. Hence, the bubbles grow large before their
walls collide: their size at collision is roughly of order of the
Hubble size.
While at nucleation the bubble is microscopic --- its size is
dictated by the elecroweak scale and is roughly of order
$(100\,\mbox{GeV})^{-1} \sim 10^{-16}\,\mbox{cm}$ --- its size at collision
of walls is macroscopic, 
$H^{-1} \sim \mbox{a~few}~\mbox{cm}$, as follows from Eq.~(\ref{may8-1}). Clearly,  boiling  is a highly
inequilibrium process, and one may hope that the baryon asymmetry
may be generated at that time. And, indeed, there exist mechanisms
of the generation of the baryon asymmetry, which have to do with
interactions of quarks and leptons with moving bubble walls.
The value of the resulting baryon asymmetry may well be of order
$10^{-10}$, as required by observations, provided that there is enough
CP-violation in the theory.

A necessary condition for the electroweak generation of the
baryon asymmetry is that the inequality~(\ref{may7-4}) must be violated
{\it just after} the phase transition. Indeed, in the opposite case
the electroweak baryon number violating processes are fast after
the transition, and the baryon asymmetry, generated during the
transition, is washed out
afterwards. Hence, the phase transition must be of strong enough
first order. This is {\it not} the case in the Standard Model.
To see why this is so, and to get an idea in which extensions of
the Standard Model the phase
transition may be of strong enough first order, let us consider
the effective potential in some detail. At zero temperature, the
Higgs potential has the standard fom,
\[
V(\phi) = - \frac{m^2}{2} |\phi|^2 + \frac{\lambda}{4} |\phi|^4 \; .
\]
Here 
\be
   |\phi| \equiv \l \phi^\dagger \phi \r^{1/2}
\label{may8-3}
\ee
is the length of the Higgs doublet $\phi$, 
$m^2 = \lambda v^2$ and
$v = 247$\,GeV is the Higgs expectation value in vacuo.
The Higgs boson mass is related to the latter
as follows,
\be
m_H = \sqrt{2 \lambda} v \; .
\label{may8-5}
\ee
Now, to the leading order of perturbation theory, the finite temperature
effects modify the effective potential into
\be
V_{eff} (\phi, T) = \frac{\alpha}{2} |\phi|^2 - \frac{\beta}{3} T |\phi|^3
+ \frac{\lambda}{4} |\phi|^4 \; ,
\label{may8-2}
\ee
with
$\alpha (T) = -m^2 + \hat{g}^2 T^2$,
where $\hat{g}^2$ is a positive linear combination of squares of
coupling constants of all fields to the Higgs field (in the
Standard Model, a
linear combination
of $g^2$, $g^{\prime \, 2}$ and $y_i^2$, where $g$ and $g^\prime$
are gauge couplings and $y_i$ are Yukawa couplings), while
$\beta$ is a positive linear combination of cubes of 
 coupling constants of all {\it bosonic}
fields to the Higgs field.  In the
Standard Model, $\beta$ is a
linear combination
of $g^3$ and  $g^{\prime \, 3}$, i.e., a linear combination
of $M_W^3/v^3$ and $M_Z^3/v^3$,
\be
\beta =  \frac{1}{2 \pi} \frac{2 M_W^3 + M_Z^3}{v^3} \; .
\label{may8-7}
\ee
The cubic term in Eq.~(\ref{may8-2})
is rather peculiar: 
in view of Eq.~(\ref{may8-3}) it is not analytic in the original Higgs
field
$\phi$. Yet this term is crucial for the first order phase transition:
for $\beta=0$ the phase transition would be of the second order.
The origin of the non-analytic cubic term can be traced back
to the enhancement of the Bose--Einstein thermal
distribution  at low momenta,  $p, m \ll T$,
\[
f_{Bose}(p) = \frac{1}{\mbox{e}^{\frac{\sqrt{p^2 + m_a^2}}{T}} - 1}
\simeq \frac{T}{\sqrt{p^2 + m_a^2}} \; ,
\]
where $m_a \simeq g_a |\phi|$ is the mass of the boson
$a$ that is generated due to the non-vanishing Higgs field, and
$g_a$ is the coupling constant of the field $a$ to the Higgs field.
Clearly, at $p \ll g_a |\phi|$ the distribution function is
non-analytic in $\phi$,
\[
f_{Bose}(p) \simeq \frac{T}{g_a |\phi|} \; .
\]
It is this non-analyticity that gives rise to the non-analytic cubic term in
the effective potential. Importantly, the Fermi--Dirac distribution,
\[
f_{Fermi}(p) = \frac{1}{\mbox{e}^{\frac{\sqrt{p^2 + m_a^2}}{T}} + 1} \; ,
\]
is analytic in $m_a^2$, and hence $\phi^\dagger \phi$, so fermions do
not contribute to the cubic term.

With the cubic term in the effective potential, the phase transition
is indeed
of the first order: at high temperatures the coefficient
$\alpha$ is positive and 
large, and there is one minimum of the effective potential
at $\phi =0$, while for $\alpha$ small but still positive
there are two minima. The phase transition ocurs at $\alpha \approx 0$;
at that moment 
\[
V_{eff} (\phi, T) \approx - \frac{\beta T}{3} |\phi|^3
+ \frac{\lambda}{4} |\phi|^4 \; .
\]
We find from this expression that immediately after the phase
transition the minimum of $V_{eff}$ is at
\[
    \phi \simeq \frac{\beta T}{\lambda} \; .
\]
Hence, the necessary condition for successfull electroweak
baryogenesis, $\phi > T$, translates into
\be
   \beta > \lambda \; .
\label{may8-6}
\ee
According to Eq.~(\ref{may8-5}), $\lambda$ is proportional to
$m_H^2$, whereas in the Standard Model
$\beta$ is proportional to $(2 M_W^3 + M_Z^3)$.
Therefore, the relation~(\ref{may8-6}) holds for small
Higgs boson masses only; in the Standard Model one makes 
use of Eqs.~(\ref{may8-5})
and~(\ref{may8-7}) and finds that
this happens
for $m_H < 50$\,GeV, which is ruled out\footnote{In 
fact, in the Standard Model with $m_H > 114$\,GeV,
there is no phase transition at all; the electroweak transition
is  smooth crossover instead. The latter fact is not visible
from the expression~(\ref{may8-2}), but that expression
is the lowest order perturbative result, while the
perturbation theory is not applicable for describing the transition
in the Standard Model with large $m_H$.}.  

This discussion indicates a possible way to make the
electroweak phase transition strong. What one needs 
is the existence of new bosonic fields that have
large enough couplings to the Higgs field(s), and hence provide large
contriutions to $\beta$. To have an effect on the dynamics of
the transition, the new bosons must be present in the cosmic plasma
at the transition temperature, $T \sim 100$\,GeV, so their masses
should not be too high, $M \lesssim 300$\,GeV. In supersymmetric
extensions of the Standard Model, the  natural candidate for long time
has been stop (superpartner of top-quark)
whose
Yukawa coupling to the Higgs field is the same as that of top,
that is, large. The light
stop scenario for electroweak baryogenesis would indeed work, as has been
shown by the detailed analysis in Ref.~\cite{stop}.

Yet another issue is CP-violation, which has to be strong enough
for successfull electroweak baryogenesis. As the asymmetry is generated
in the interactions of quarks and leptons (and their superpartners
in supersymmetric extensions) with the bubble walls, CP-violation
must occur at the walls. Recall now that the walls are made of the
Higgs field(s). This points towards the necessity of CP-violation
in the Higgs sector, which may only be the case in a theory
with more than one Higgs fields. 

To summarize, electroweak baryogenesis requires a considerable
extension of the Standard Model, with masses of new particles in
the range $100 - 300$\,GeV. Hence, this mechanism will definitely
be ruled out or confirmed by the LHC. We stress, however, that electroweak
baryogenesis is not the only option at all: an elegant and well motivated
competitor is leptogenesis~\cite{Kayser}; several other mechanisms
have been proposed
that may be responsible for the baryon asymmetry of the Universe.

\section{Dark energy}
\label{sec:de}

Dark energy, the famous ``substance'', does not clump,
unlike dark matter.  It gives rise to the
accelerated expansion of the Universe.  As we see from
Eq.~\eqref{sep20-11-1}, the Universe with constant energy 
density should expand exponentially; if the energy density
is almost constant, the expansion
is almost exponential.  Let us make use of the first law of
thermodynamics, which for the adiabatic expansion reads
\[
dE = - pdV \; ,
\]
and apply it to comoving volume, $E = \rho V$, $V=a^3$.
We obtain for dark energy
\[
d\rho_\Lambda = -3 \frac{da}{a} (\rho_\Lambda + p_\Lambda) \; ,
\]
or
\[
\frac{d\rho_\Lambda}{\rho_\Lambda} =  -3 \frac{da}{a} (1 + w) \; ,
\]
where we introduced the equation of state parameter $w$ such that
\[
p_\Lambda = w\rho_\Lambda \; .
\]
Thus, (almost) time-independent dark energy density corresponds to
$w\approx -1$, i.e., effective pressure of dark energy is negative.
We emphasize that pressure is by definition a spatial component of
the energy-momentum tensor, which in the homogeneous and isotropic
situation has the general form
\[
T_{\mu \nu} = \mbox{diag} ~(\rho, p,p,p) \; .
\]
Dark energy density does not depend on time at all, if 
$p_\Lambda =- \rho_\Lambda$, i.e.,
\[
T_{\mu \nu} = \rho_\Lambda \eta_{\mu \nu} \; ,
\]
where $\eta_{\mu \nu}$ is the Minkowski tensor.
This is characteristic of vacuum, whose energy-momentum tensor must
be Lorentz-covariant. 
 Observationally, 
$w$ is close to $-1$ to reasonably good precision. The most accurate 
detrmination, which, however, does not include systematic errors in
supernovae data and possible time-dependence of $w$, is~\cite{WMAP7a}
\begin{eqnarray}
w=-0.98\pm 0.05 \; .
\end{eqnarray}
So, the dark energy density is almost time-independent, indeed.

The problem with dark energy is that 
its present value is extremely small by particle physics standards, 
\begin{eqnarray*}
\rho_{DE}\approx 4\,\mbox{GeV/m$^3$}=(2\times10^{-3}\,\mbox{eV})^4\,.
\end{eqnarray*}
In fact, there are two hard problems.
One is that particle physics scales are much larger 
than the scale relevant to the dark energy density,
so the dark energy density is zero to an excellent approximation. 
Another is that
it is non-zero nevertheless, and one has to understand its energy scale.
To quantify the first problem, we recall the known scales of 
particle physics and gravity,
\begin{eqnarray*}
\mbox{Strong interactions}:&\quad&\Lambda_{QCD}\sim 1\,\mbox{GeV}\,,
\cr
\mbox{Electroweak}:&&M_W\sim 100\,\mbox{GeV}\,,
\cr
\mbox{Gravitational}:&&M_{pl}\sim 10^{19}\, \mbox{GeV}\,.
\end{eqnarray*}
In principle, vacuum should contribute to $\rho_\Lambda$, and
there is absolutely no reason for vacuum to be as light as it is.  
The discrepancy here is huge, as one sees from the above numbers.

To elaborate on this point, let us note that the action
of gravity plus, say, the Standard Model has the general form
\[
S = S_{EH} + S_{SM} - \rho_{\Lambda, 0} \int~\sqrt{-g}~d^4x \; ,
\]
where $S_{EH} = -(16 \pi G_N)^{-1} \int ~R~\sqrt{-g}~d^4x$ is the
Einstein--Hilbert action of General Relativity, $S_{SM}$ is the action
of the Standard Model and   $\rho_{\Lambda, 0}$ is the bare 
cosmological constant. In order that the vacuum energy density be
almost zero, one needs fantastic
cancellations between the contributions of the Sandard Model fields
into the vacuum energy density, on the one hand, and  $\rho_{\Lambda, 0}$
on the other.  
For example, we know that QCD has a complicated vacuum structure, and
one would expect that 
the energy density of QCD combined with
 $\rho_{\Lambda, 0}$
should be of order $(1\,\mbox{GeV})^4$.
 Nevertheless, it is not, 
so at least for QCD, one needs a cancellation on the order of $10^{-44}$.
 If one goes further and considers other interactions,
the numbers get even worse.
 
What are the hints from this ``first'' cosmological constant problem?  
There are several options, though not many.  
 One is that the Universe could have a very long prehistory. 
Extremely long.  This option has to do with relaxation mechanisms.  
Suppose that the original vacuum energy density is indeed large, say, 
comparable to the particle physics scales. Then there must be a
mechanism which can  relax this value down to an 
acceptably small number.  It is easy to
convince oneself that this relaxation could not happen in the
history of the Universe we know of.  Instead,
the Universe should have a very long prehistory 
during which this relaxation process might occur.  At that 
prehistoric time, the vacuum in the 
Universe must have been exactly the same 
as our vacuum, so the Universe in its prehistory 
must have been exactly like ours, or almost exactly like ours.  
Only in that case could a relaxation mechanism  work.  
There are concrete scenarios of this sort~\cite{prehistory}.  
However, at the moment it seems that these scenarios are hardly testable,
since this is prehistory.  

Another possible hint is towards anthropic selection.  
The argument that goes back to Weinberg 
and Linde~\cite{Weinberg:1987dv,Linde:1986dq}
is that if the 
cosmological constant were larger, say, by a factor of 100, we simply would
not exist: the stars would not have formed because of the 
fast expansion of the Universe. 
So, the vacuum energy density may be selected anthropically.  
The picture is that the Universe may be much, much larger
than what we can see, and different large regions of the Universe
may have different properties. In particular, vacuum energy
density may be different in different regions. 
Now, we are somewhere in the place where one can live.  
All the rest is empty of human beings, because  
there the parameters such as vacum
energy density are not suitable for their existence.
This is disappointing for a theorist, as this point of view allows for 
arbitrary tuning of fundamental parameters.  It is 
hard to disprove this option,
on the other hand. 
We do exist, and this is an experimental fact.  
The anthropic viewpoint may, though hopefully  will not, 
get more support from the LHC, if no or insufficient 
new physics is found there.  Indeed, another candidate for 
an environmental quantity is the electroweak scale.  

Let us recall in this regard the gauge hierarchy problem:
the electroweak scale $M_W \sim 100$\,GeV
is much lower than the natural scale in 
gravitational physics, the Planck mass, $M_{Pl}\sim 10^{19}$\,GeV.
The electroweak scale in the Standard Model is unprotected from large 
contributions due to high energy physics, and in this sense
it is very similar to 
the cosmological constant.  There are various anthropic arguments showing
that the electroweak scale must
be small.  
A simple example is that if one
makes 
it larger without touching other parameters, then quarks would be too heavy.  
Neutron would be the lightest baryon, and proton would be unstable.
There would be no stable hydrogen, and that is presumably 
inconsistent with our existence.  Hence, one of the ``solutions''
to the gauge hierarchy problem is anthropic.

An interesting part
of the story is that unlike the cosmological constant, 
there are natural ways to make the electroweak scale small and render it small 
in extensions of the Standard Model, like low energy
supersymmetry.  
All these extensions 
require new physics at TeV energies. 
So we are in a situation where the experiment has to say its word.
If it says that none of these extensions is there in Nature, 
then we will have to take the 
anthropic viewpoint much more seriously than before.

Turning to the ``second'' cosmological constant problem,
we note that the scale $10^{-3}$\,eV may be associated with some new
light field(s), rather than with vacuum. This implies, in general,
that $\rho_\Lambda$ depends on time, i.e., $w \neq -1$
and $w$ may well depend on time itself. 
``Normal'' field (called
quitessence in this context) has $w>-1$, but there are examples
(rather contrived) of fields with $w < -1$ (called phantom fields).
Current data are compatible with time-independent $w$ equal to
$-1$,  but their precision is not particularly high.
We conclude that
 future cosmological
observations may shed new light on the field content of
fundamental theory.

\section{Cosmological perturbations and the very early Universe}
\label{sec:pertu}

 With Big Bang nucleosynthesis theory and observations,
 we are confident of the theory of the early Universe
at temperatures up to 
$T\simeq 1$\,MeV, that corresponds to age of $t\simeq 1$~second.
With the LHC, we hope to be able to go 
up to temperatures $T\sim 100$\,GeV and age $t \sim 10^{-10}$~second.
The question is: are we going to have a handle on even earlier epoch?

The key issue in this regard is cosmological perturbations.
These are inhomogeneities in the energy density and assoiated
gravitational potentials, in the first place.
This type of inhomogeneities is called scalar perturbations,
as they are described by 3-scalars. There may exist perturbations
of another type, called tensor; these are primordial gravity
waves. We will mostly concentrate on scalar perturbations, since
they are observed; tensor perturbations are important too,
and we comment on them later on. It is worth pointing out 
that perturbations of the present size below ten  Megaparsec
have large amplitudes today and are non-linear, but
in the past their amplitudes were small, and they can be described
within the linearized theory. Indeed, CMB temperature anisotropy
tells us that the perturbations at recombination epoch
were roughly at the level 
\[
\delta \equiv \frac{\delta \rho}{\rho} = 10^{-4} - 10^{-5} \; .
\]
Thus, the linearized theory works very well before recombination
and somewhat later. 

\begin{figure}[htb!]
\begin{center}
\includegraphics[width=0.8\textwidth]{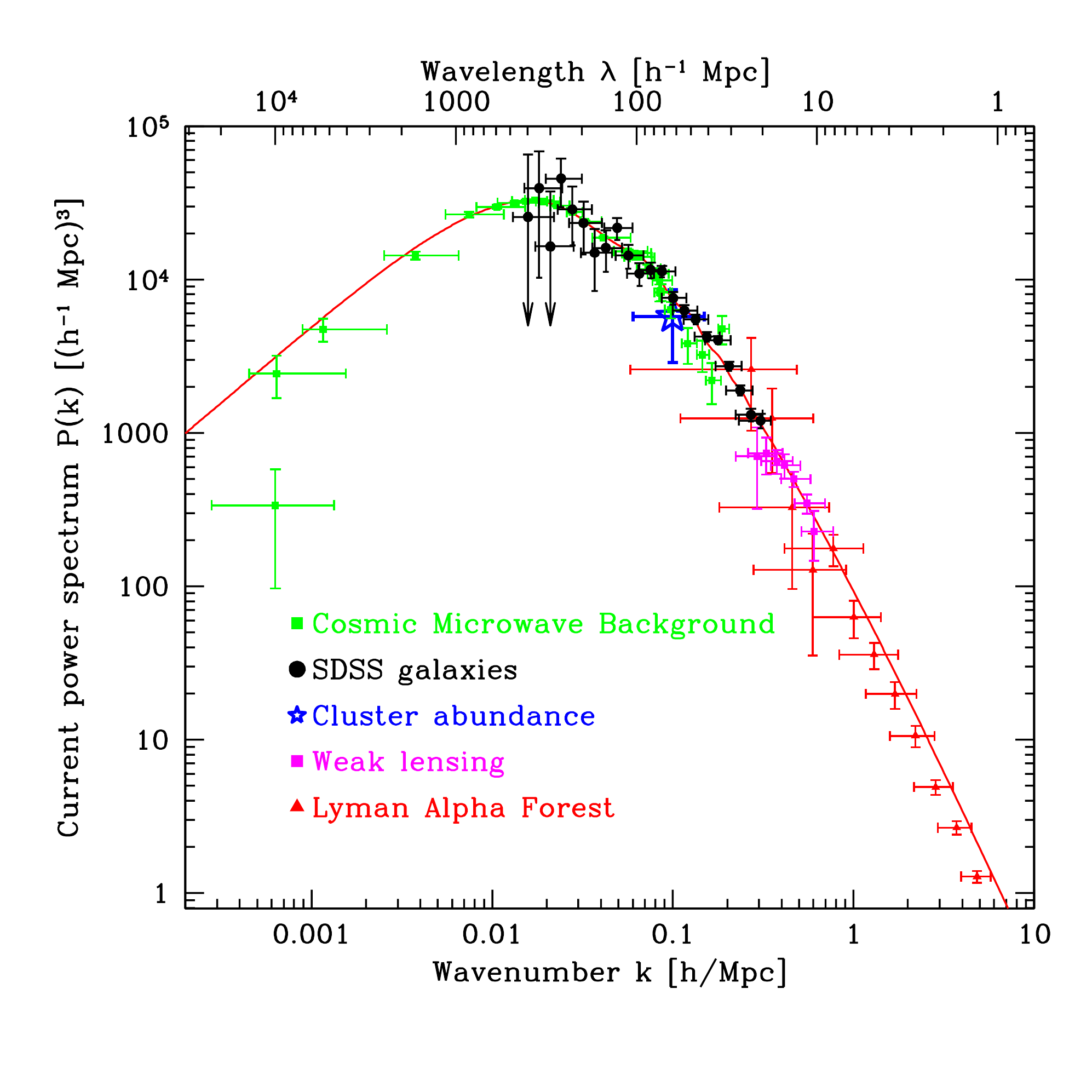}
\end{center}
\vspace{-1.2cm}
\caption{Power spectrum of density perturbations, measured by
various methods and translated to the
present epoch by using the linearized theory~\cite{Tegmark}. 
$k$ is the present wavenumber, and $h \approx 0.7$ is the
dimensionless Hubble parameter at the present epoch.
 \label{spec}
 }
\end{figure}

Properties of scalar perturbations are mesured in various ways.
Perturbations of large spatial scales leave their imprint in
CMB temperature anisotropy and polarization, 
so we have very detailed knowledge of them.
Shorter wavelength perturbations are studied by analysing
distributions of galaxies and quasars at present and in
relatively near past. There are several other methods, some of which
can probe even shorter wavelengths. As we discuss in more detail
below, scalar perturbations in the linear regime are actually
Gaussian random field, and the first thing to measure is its
power spectrum.  
Overall, independent methods
give consistent results, see Fig.~\ref{spec}.

Cosmic medium in our Universe has several components that
interact only gravitationally: baryons, photons, neutrinos,
dark matter. Hence, there may be and, in fact, there are
perturbations in each of these components. As we pointed out
in the beginning of
Section~\ref{sec:dm}, electromagnetic interactions between
baryons, electrons and photons were strong 
before recombination,
so these species made single fluid, and it is appropriate to talk
about perturbations in this fluid. After recombination, baryons and 
photons evolved independently.

The main point of this part of lectures is that
by analysing the density perturbations, we have already learned
a number of very important things. To appreciate what they are,
it is instructive to consider first the
baryon-electron-photon fluid before 
recombination. Perturbations in this fluid are nothing but sound waves;
they obey a wave equation. So, let us turn to the wave equation in
the expanding Universe.

\subsection{Wave equation in expanding Universe.
Subhorizon and superhorizon regimes.}

The actual system of equations for density perturbations in the
baryon-electron-photon fluid and associated gravitational
potentials is fairly cumbersome. So, let us simplify things.
Instead of writing and 
then solving the equations for sound waves, let us consider 
a toy example, the case of massless scalar field.
The general properties of density perturbations
are similar to this case, although there are a few places in which
they differ; we comment on the differences in due course.  
 
The action for the massless scalar field is
\[
S=\frac12\int~d^4x~\sqrt{-g}~g^{\mu\nu}\d_\mu\phi\d_\nu\phi=\int~
 d^3xdt~a^3~\left[\frac12\dot{\phi}^2-\frac{1}{2a^2}(\d_i\phi)^2\right] \; ,
\]
where we specified to FLRW metric in the second expression.
The field equation thus reads:
\[
-\frac{d}{dt}(a^3\dot{\phi})+a\d_i\d_i\phi=0,
\]
i.e.,
\be
\ddot{\phi}+3H\dot{\phi}-\frac{1}{a^2}\d_i^2\phi=0 \; , 
\label{scalarfield}
\ee
where $H\equiv\dot{a}/a$ is again the Hubble parameter. 
This equation is linear in $\phi$ and homogeneous in space, 
so it is natural to represent $\phi$ in terms of the Fourier 
harmonics,
\[
\phi({\bf x},t)=\int~e^{i{\bf k}{\bf x}}\phi_{\bf k}(t)~d^3k.
\]
Clearly, the value of ${\bf k}$ 
for a given Fourier mode is constant in time. 
However, $k$ is not  the physical wavenumber 
(physical momentum), since $x$ is not the  physical distance. 
$k$ is called conformal momentum, while physical momentum equals 
$q\equiv 2\pi/\lambda=2\pi/(a(t)\Delta x)=k/a(t)$. 
$\Delta x$ here is time-independent comoving wavelength of
perturbation, and
$\lambda$ is the physical wavelength; the latter grows 
due to the expansion of the Universe. 
Accordingly, as the Universe expands, the physical 
momentum of a given mode decreases (gets redshifted), 
$q(t)\propto a^{-1}(t)$. 
For a mode of given conformal momentum $k$, Eq.~(\ref{scalarfield}) gives:
\be
\ddot{\phi}+3H\dot{\phi}+\frac{k^2}{a^2}\phi=0.
\label{sep20-11-10}
\ee
Besides the redshift of momentum, the cosmological expansion
has the effect of inducing the second term, ``Hubble friction''.

\begin{figure}[htb!]
\begin{center}
\includegraphics[width=0.8\textwidth]{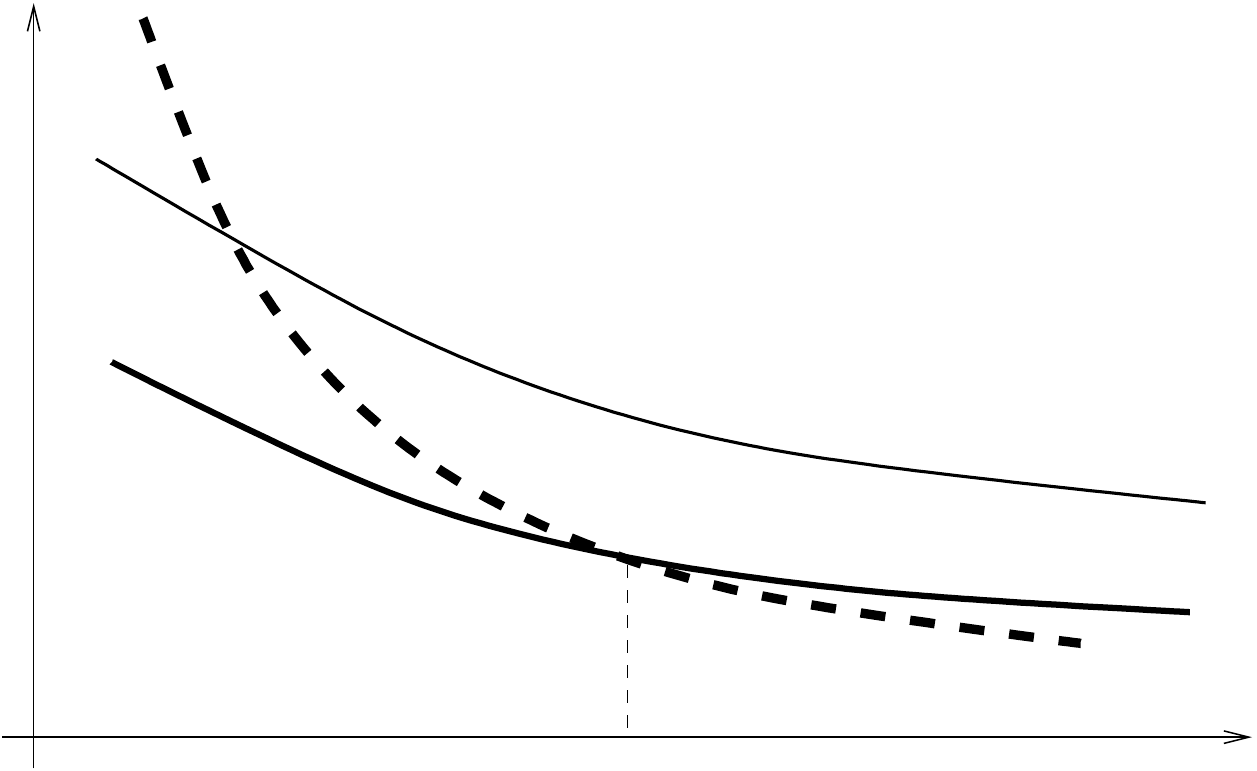}
\begin{picture}(10,10)(0,0)
{
\put(0,0){$t$}
\put(-320,80){$q_2(t)$}
\put(-100,100){$q_1(t) = k_1/a(t)$}
\put(-320,190){$H(t)$}
\put(-200,0){$t_\times$}
}
\end{picture}
\end{center}
\caption{Physical momenta (solid lines, $k_2 < k_1$)
and Hubble parameter (dashed line) at radiation-
and matter-dominated epochs. $t_\times$ is the horizon entry time.
 \label{length-hubble}
 }
\end{figure}
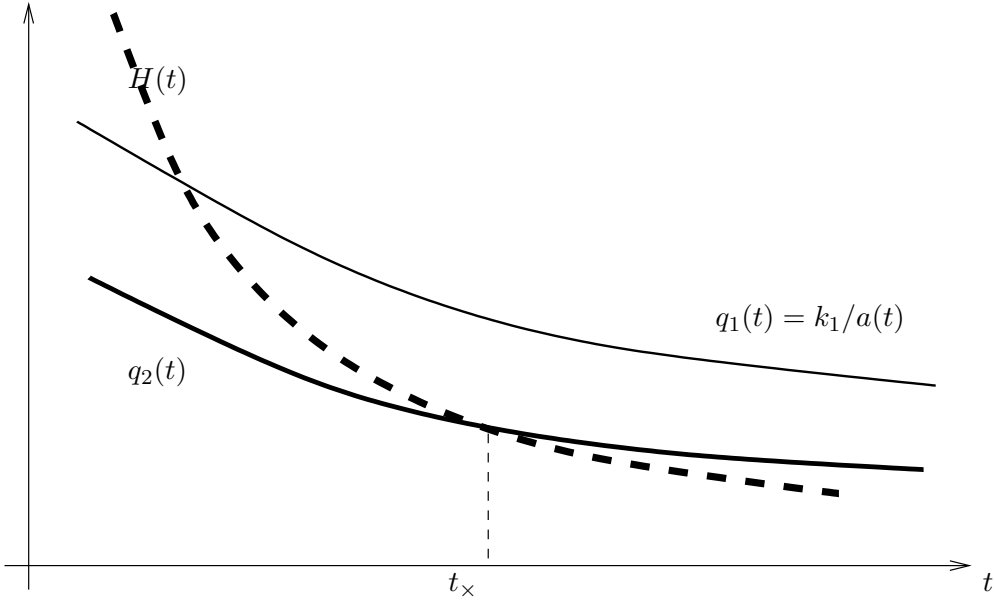

Equation~\eqref{sep20-11-10} has two time-dependent
parameters of the same dimension: $k/a$ and $H$. 
Let us consider two limiting cases: $k/a\ll H$ and $k/a\gg H$. In 
cosmological models
with conventional equation of state of the dominant component (e.g., 
matter-dominated or radiation-dominated Universe),  $H^{-1}$ is of
the order of the size of the 
cosmological horizon, see 
Section~\ref{sec:regimes}. So, the regime $k/a \ll H$ is the 
regime in which the physical wavelength $\lambda=2\pi a/k$
is greater than the horizon size (this is called
superhorizon regime), 
while for $k/a \gg H$ the physical wavelength is smaller 
than the horizon size (subhorizon regime). 
The time when the wavelength of the mode coincides 
with the horizon size is called horizon crossing. In what follows we
denote this
time by the
symbol $\times$. Both at radiation- and matter-dominated epochs, 
the ratio $k/(aH)$ grows. Indeed, in the radiation-dominated epoch
$a \propto \sqrt{t}$, while $H \propto t^{-1}$, so
$k/(aH) \propto \sqrt{t}$. 
This means that every mode was at some early time superhorizon, and later on
it becomes subhorizon, see Fig.~\ref{length-hubble}.
It is straightforward to see that for all cosmlogically interesting wavelengths,
horizon crossing occurs much later than 1~s after the Big Bang, i.e.,
at the time we are confident about. So, there is no guesswork at
this point.

Now we can address the question of the origin of density
perturbations.
By causality, any  mechanism of their generation 
that operates at the radiation- and/or matter-dominated epoch,
can only work after the horizon entry time $t_\times$.
Indeed, no physical process can create a perturbation whose
wavelength exceeds the size of an entire causally connected region. 
So, in that case
the perturbation modes were never superhorizon.
On the other hand, if modes were ever superhorizon, they have to
exist already in the beginning of the hot epoch. Hence, in the latter
situation one has to conclude that there existed another
epoch {\it before} the hot stage: that was the epoch of 
the generation of primordial density perturbations.

Observational data, notably (but not only) on CMB temperature
anisotropy and polarization, disentangle these two possibilities.
They unambiguously show  that density perturbations
{\it were} superhorizon at radiation and matter domination!

To understand how this comes about, let us see what is special
about a perturbation which was superhorizon at the hot stage.
For a superhorizon mode, 
we can neglect the term $\phi\cdot k^2/a^2$ in 
Eq.~(\ref{scalarfield}). Then the field equation, 
e.g., in
the radiation--dominated Universe ($a\propto t^{1/2},H=1/2t$), becomes
\be
\ddot{\phi}+\frac{3}{2t}\dot{\phi}=0 \; .
\ee
The general solution to this equation is
\be
\phi(t)=A+\frac{B}{\sqrt{t}} \; ,
\ee
where A and B are constants.
This behavior is generic for all cosmological perturbations
at the hot stage: there is 
a constant mode ($A$ in our case) and a mode that decays in time.
If we extrapolate the decaying mode $B/\sqrt{t}$ back in time, we 
get very strong (infinite in the limit $t\rightarrow 0$) perturbation. 
For density perturbations (and also tensor perturbations)
this means that this mode corresponds to
strongly inhomogeneous early Universe. 
Therefore, the consistency  of the cosmological model dictates that
the decaying mode has to be absent for actual 
perturbations. 
Hence, for given ${\bf k}$, the solution is determined by
a single parameter, the initial amplitude $A$ of the mode
$\phi_{{\bf k}}$. 

After entering the 
subhorizon regime, the modes oscillate --- these  are 
the analogs of conventional sound waves.  
In the subhorizon regime one makes use of the WKB
aproximation to solve the complete equation
\be
\ddot{\phi}+\frac{3}{2t}\phi+\frac{k^2}{a^2(t)}\phi=0 \; .
\label{sep21-11-1}
\ee
The general solution in the WKB approximation reads
\be
\phi(t)=\frac{A'}{a(t)}\cos\left(\int_0^t\frac{k}{a(t')}dt'+\psi_0\right)
\; ,
\label{may29-2}
\ee
with the two constants being the amplitude $A'$ and the phase $\psi_0$. 
The amplitude $A'$ of these oscillations is determined by the amplitude 
$A$ of the superhorizon initial perturbation, while {\it
the phase $\psi_0$ of these oscillations 
is uniquely determined by the condition of the absence of the decaying mode}, 
$B=0$. Imposing this condition yields
\be
\phi(t)=cA\frac{a_{\times}}{a(t)}\sin\left(\int_0^t\frac{k}{a(t')}dt'\right)\; ,
\label{may31-7}
\ee
where the constant $c$ is of order 1 and can be evaluated by solving
the complete equation \eqref{sep21-11-1}.
The decreasing amplitude of oscillations $\phi(t) \propto 1/a(t)$ and 
the particular phase $\psi_0 = -\pi/2$ in Eq.~(\ref{may29-2}) are
peculiar properties of the wave equation~(\ref{scalarfield}), as well as
the radiation-dominated cosmological expansion. However,
{\it the fact that the phase of oscillations
is uniquely determined by the requirement
of the absence of the superhorizon decaying mode is generic}.

The perturbations in the baryon--photon medium before recombination 
--- sound waves --- behave in a rather 
similar way.  Their evolution is as follows:
\be 
\delta_\gamma\equiv
\frac{\delta\rho_\gamma}{\rho_\gamma}=
\left\{\begin{array}{cc} {\rm const}, & ~~~~~{\rm outside \quad horizon}, \\
{\rm const}\cdot \cos\left(\int\limits_0^tv_s\frac{k}{a(t')}dt'\right), & 
~~~~~{\rm inside \quad horizon}, \end{array} \right.
\label{sep21-11-2}
\ee
where $v_s\equiv \sqrt{d p/d\rho}$ is the sound speed. 
The  baryon--photon medium before recombination is almost 
relativistic\footnote{This does not contradict the statement 
that the Universe is in matter-dominated regime at recombination. 
The dominant component at this stage is dark matter.},
since $\rho_{\rm B}<\rho_\gamma$. Therefore, 
$v_s\approx 1/\sqrt{3}$. Let us reiterate that the phase of the
oscillating solution in \eqref{sep21-11-2} is uniquely defined.

\subsection{Oscillations in CMB angular spectrum}

\begin{figure}[htb!]
\begin{center}
\includegraphics[width=0.7\textwidth]{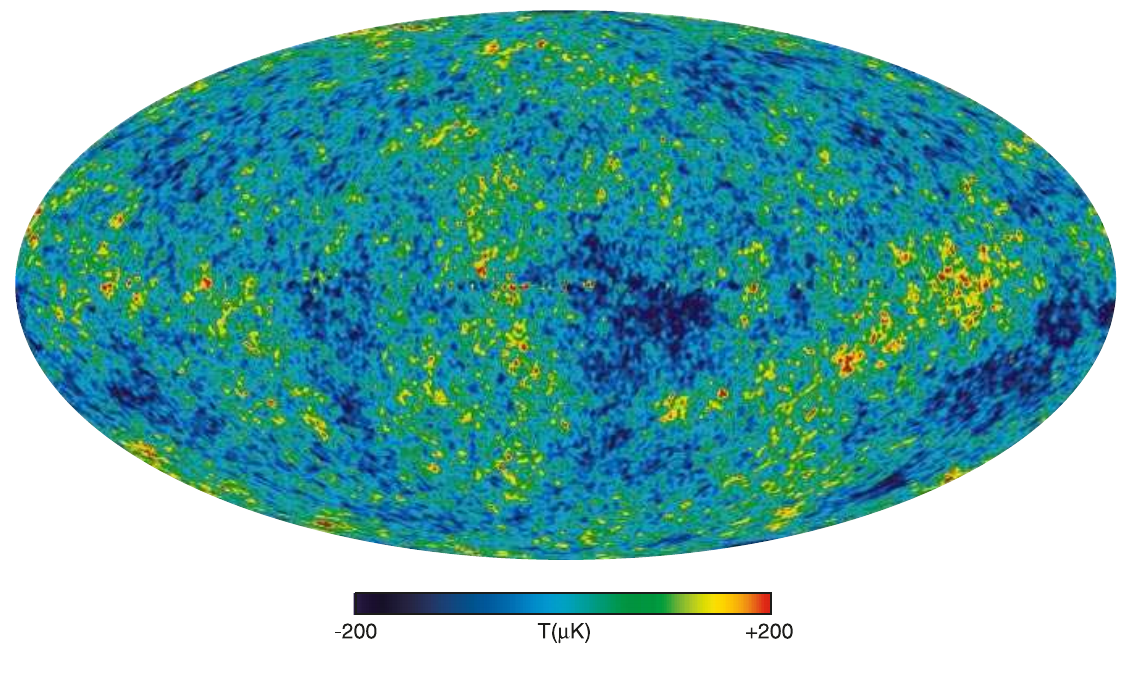}
\end{center}
\caption{ CMB sky as seen by WMAP experiment.
 \label{sky}
 }
\end{figure}

CMB gives us the photographic picture of the Universe
at recombination (photon last scattering), see Fig.~\ref{sky}.
Waves of different momenta $k$ are at different phases
at recombination. At that epoch, oscillations in time
in Eq.~\eqref{sep21-11-2} show up as oscillations in momentum.
This in turn gives rise to the observed oscillations in the
CMB angular spectrum. 

In more detail,  
at the time of last scattering $t_{rec}$ we have
\be
\delta_\gamma \equiv \frac{\delta \rho_\gamma}{\rho_\gamma}
= A(k)
\cdot \cos\left(\int_0^{t_{rec}}v_s\frac{k}{a(t')}dt'\right)= A(k)
\cdot \cos kr_s\; ,
\label{may30-5}
\ee
where $A(k)$ is linearly related to the initial amplitude
of the superhorizon perturbation and is a non-oscillatory
function of $k$, and
\[
r_s=\int_0^{t_{rec}}v_s\frac{dt'}{a(t')}
\]
 is the comoving size of the sound horizon at recombination, 
while its physical size equals $a(t_{rec})r_s$. So, we see that 
the
density perturbation at recombination indeed
oscillates as a function of wavenumber.
The period of this oscillation is determined by $r_s$, which is
a straighforwardly calculable quantity. 

Omitting details, the 
fluctuation of the
CMB temperature is partially due to 
the density perturbation in the
baryon-photon medium at recombination. The relevant place is 
the point where the photons last scatter before coming to us.
This means that the temperature fluctuation
of photons coming from the direction
${\bf n}$ in the sky is, to a reasonable accuracy,
\[
\delta T({\bf n}) \propto \delta_\gamma ({\bf x_n}, \eta_{rec})
+ \delta T_{smooth} ({\bf n}) \; ,
\]
where $  T_{smooth} ({\bf n})$ corresponds to the non-oscillatory
part of the CMB angular spectrum, and
\[
{\bf x_n}=-{\bf n} (\eta_0 - \eta_{rec}) \; .
\]
 Here
the variable $\eta$ is defined in \eqref{sep13-11-6},
and $\eta_0$ is its present value, so that $(\eta_0 - \eta_{rec})$
is the coordinate distance to the sphere of photon last scattering,
and ${\bf x_n}$ is the coordinate of the place where the photons 
coming from the direction ${\bf n}$ scatter last time. 
 $  T_{smooth} ({\bf n})$ originates from 
the gravitational
potential generated by the dark matter perturbation; 
dark matter has zero pressure
at all times, so there are no sound waves in this component,
and there are no 
oscillations at recombination as a function of momentum.

\begin{figure}[htb!]
\begin{center}
\includegraphics[width=0.8\textwidth]{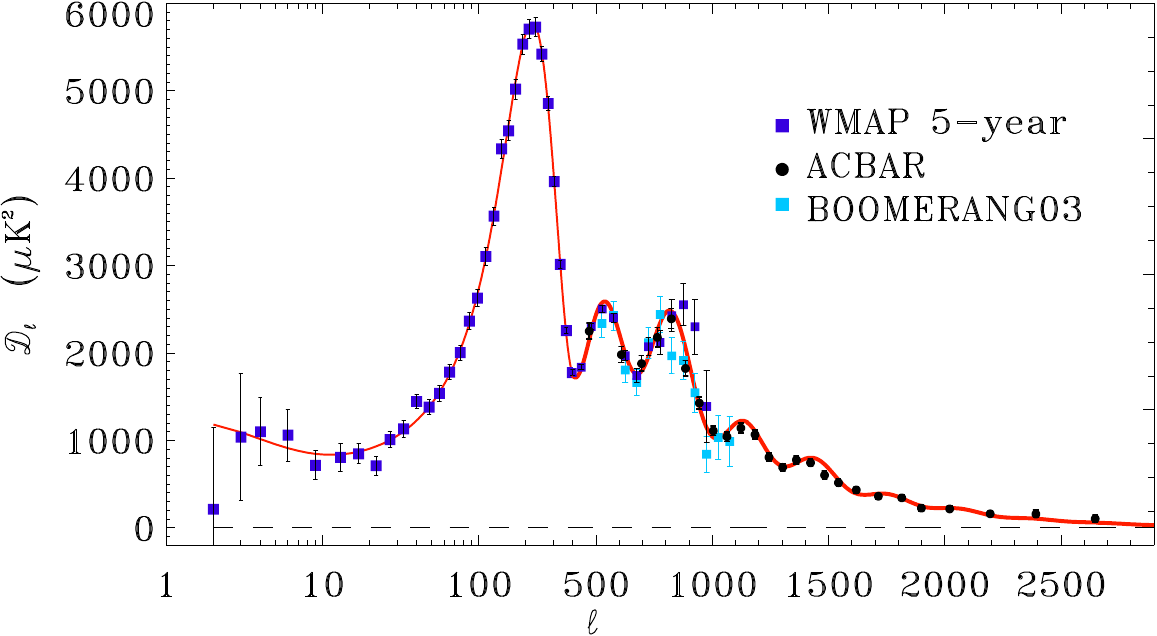}
\caption{The angular spectrum of the CMB temperature
anisotropy~\cite{cmbang}. The quantity in vertical axis  is 
$D_l$ defined in Eq.~(\ref{defdl}).
\label{anisotropyspectrum}}
\end{center}
\end{figure}

One expands
the temperature variation on celestial sphere
in spherical harmonics:
\[
\delta T({\bf n})=\sum_{lm}a_{lm}Y_{lm}(\theta,\phi).
\]
The multipole number $l$ characterizes the temperature fluctuations
at the angular scale $\Delta \theta = \pi/l$. The sound waves of momentum
$k$ are seen roughly at an angle $\Delta \theta = \Delta x/(\eta_0 - \eta_{rec})$,
where $\Delta x = \pi/k$ is coordinate half-wavelength.
 Hence, there is the correspondence
\[
  l \longleftrightarrow k (\eta_0 - \eta_{rec}) \; .
\]
Oscillations in momenta in \eqref{may30-5} thus translate into
oscillations in $l$, and these are indeed observed, see
Fig.~\ref{anisotropyspectrum}.

To understand what is shown in Fig.~\ref{anisotropyspectrum}, we note that
all
observations today support the hypothesis that 
$a_{lm}$ {\it are independent  Gaussian random variables}. 
Gaussianity means that
\be
P(a_{lm})da_{lm}=\frac{1}{\sqrt{2\pi C_l}} e^{-\frac{a_{lm}^2}{2C_l}}da_{lm},
\ee
where $P(a_{lm})$ is the probability density for the random variable
$a_{lm}$. For a hypothetical ensemble of Universes like ours, the 
average values
of products of the coefficients $a_{lm}$ would obey
\be
\langle a_{lm} a_{l^\prime m^\prime} \rangle = C_l \delta_{l l^\prime}
\delta_{m m^\prime}.
\label{may29-6}
\ee
This gives the expression for
the
temperature fluctuation:
\[
 \langle[\delta T({\bf n})]^2\rangle
=\sum_l\frac{2l+1}{4\pi}C_l\approx 
\int~ \frac{dl}{l} D_l \; ,
\]
where 
\be
D_l = \frac{l(l+1)}{2\pi} C_l \; . 
\label{defdl}
\ee
It is the latter quantity that is usually shown in plots,
in particular, in  Fig.~\ref{anisotropyspectrum}. Note the
unconventional scale on the horizontal axis, aimed at showing both
small $l$ region (large angular scales) and 
large $l$ region.

\begin{figure}[htb!]
\begin{center}
\includegraphics[width=0.8\textwidth]{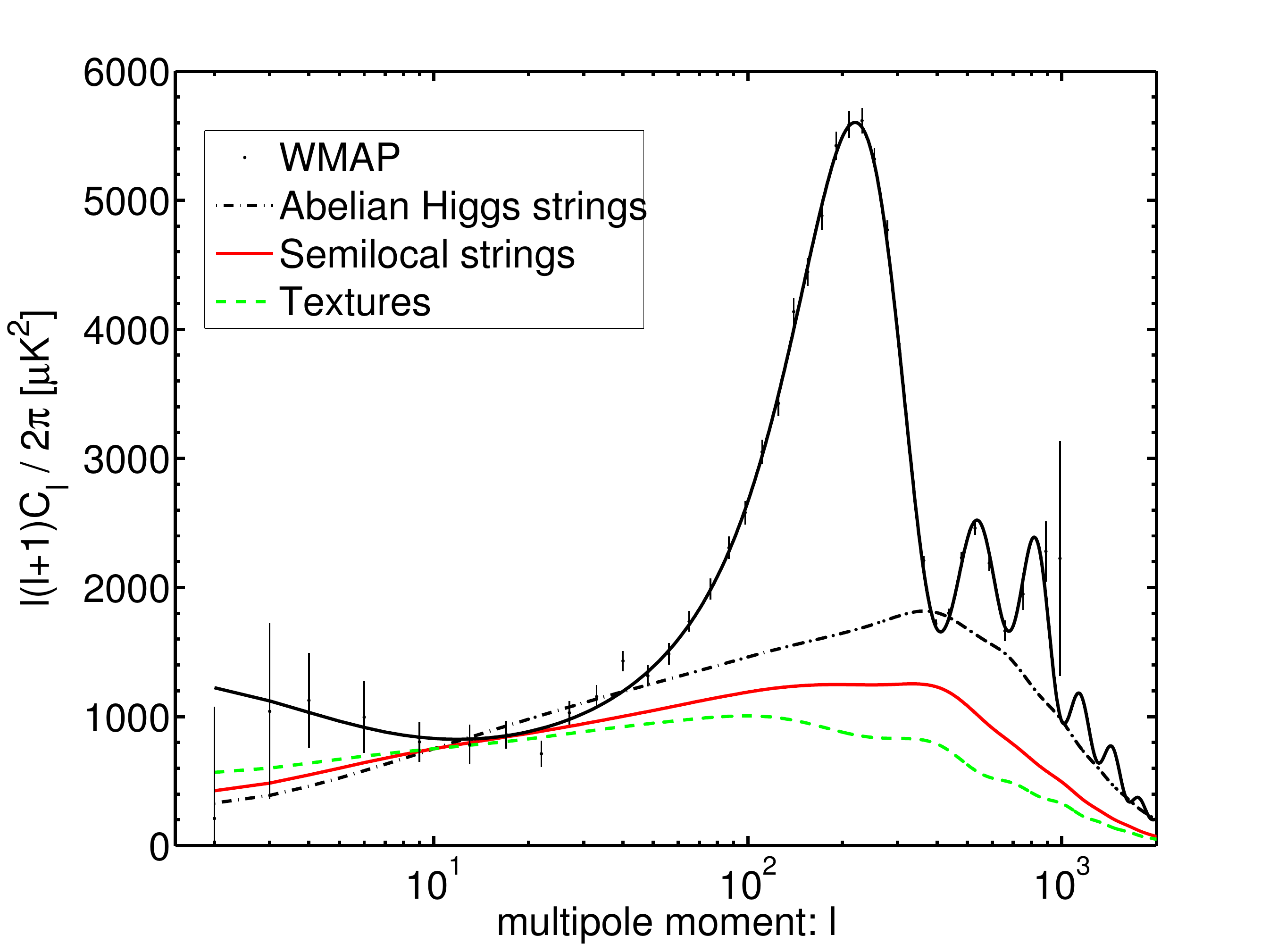}
\caption{The angular spectrum of the CMB temperature
anisotropy in causal models that generate the density perturbations
at the hot stage (non-oscillatory lines) versus data (sketched by
oscillatory line)~\cite{string}.
\label{strings}}
\end{center}
\end{figure}

The fact that the CMB angular spectrum has oscillatory behavior
unambiguously tells us that {\it density perturbations were indeed
superhorizon} at hot cosmological stage. If these perturbations were
generated by some causal mechanism after horizon entry, there would be
no reason for the phase $\psi_0$ in \eqref{may29-2} (better to say,
in the analog of  \eqref{may29-2} for density perturbations)
to take a very
definite value. Instead, one would expect that this phase is a
random function of ${\bf k}$, so there would be no oscillations 
in $l$ in the CMB angular spectrum at all. This is indeed
the case in
concrete causal models aimed at generating the density
perturbations at the hot stage, which make use, e.g., of
topological defects (strings, textures, etc.), see Fig.~\ref{strings}.

Another point to note is that the CMB measurements show that
at recombination, there were density perturbations which were
still superhorizon at that time. These correspond to low multipoles,
$l \lesssim 50$. Perturbations of these wavelengths cannot be
produced at the hot stage before recombination, and, indeed,
causal mechanisms produce small power at low multipoles.
This is also seen in  Fig.~\ref{strings}.

\subsection{Baryon acoustic oscillations}

\begin{figure}[htb!]
\begin{center}
\includegraphics[width=0.6\textwidth]{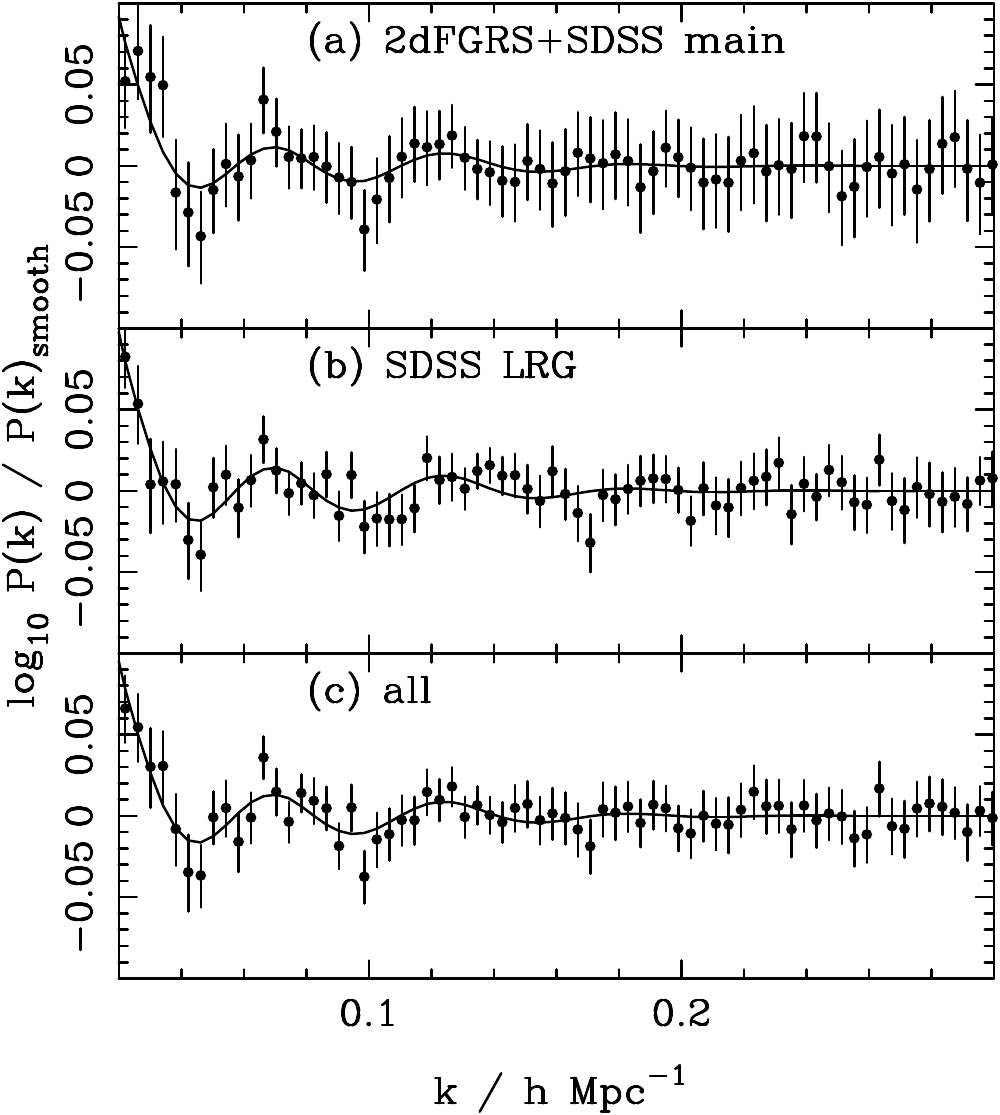}
\caption{Baryon acoustic oscillations in matter power spectrum
detected in galaxy surveys~\cite{BAO}.
\label{BAO}}
\end{center}
\end{figure}

Another manifestation of the well defined phase of
sound waves in baryon-photon medium before recombination is
baryon acoustic oscillations. Right after recombination,
baryons decouple from photons, the sound speed in the 
baryon component
becomes essentially zero, and the spatial distribution of the
baryon 
density freezes out. Since just before recombination
baryons, together with photons, have energy distribution
\eqref{sep21-11-2}
which is oscillatory function of $k$, there is oscillatory
component in the Fourier spectrum of the
total
matter distribution after recombination. This oscillatory component
persists until today, and shows up as oscillations in
the matter power spectrum $P(k)$.
This is a small effect,
since the dominant component at late times is dark matter,
\[
\rho_M ({\bf k})= \rho_{DM} ({\bf k})+ \rho_B ({\bf k}) \; ,
\]
and only $\rho_B$ oscillates as function of $k$
\be
\delta \rho_B ({\bf k}) \approx \rho_B \delta_\gamma ({\bf k})
= \rho_B \cdot   A(k)
\cdot \cos kr_s
\label{BAO-eq}
\ee
(as we already noticed, dark matter has zero pressure
at all times, so there are no sound waves in this component).
Nevertheless, this effect has been observed in
large galaxy surveys, see Fig.~\ref{BAO}. 

\begin{figure}[htb!]
\begin{center}
\includegraphics[width=0.6\textwidth]{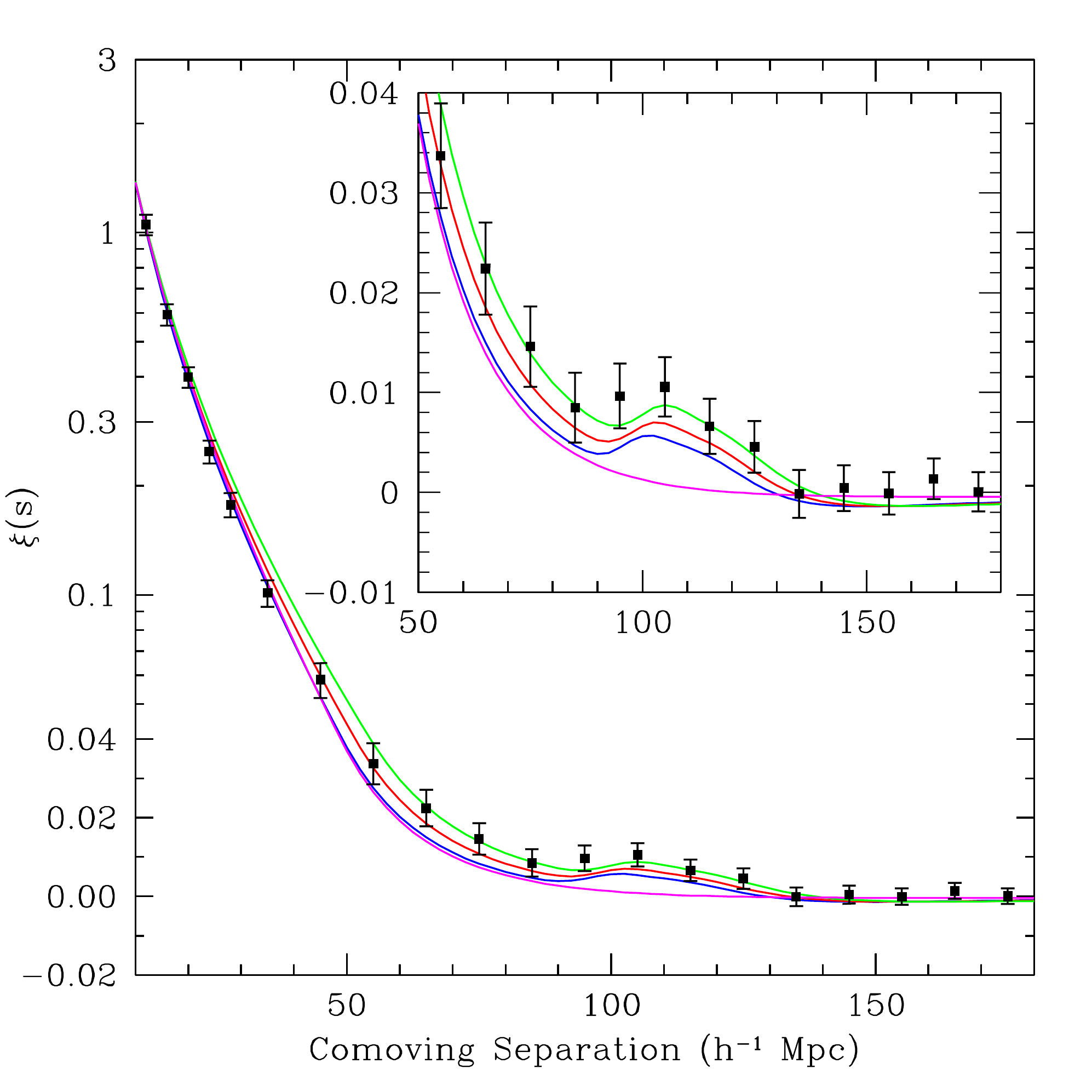}
\caption{Baryon acoustic oscillations in galaxy correlation
function~\cite{BAO-c}.
\label{corfunc}}
\end{center}
\end{figure}

There is a simple interpretation of the effect. As we discuss below,
the overdensities in the
baryon-photon medium and in the dark matter 
are at the same place before horizon entry (adiabatic mode). 
But before recombination the sound speed in baryon-photon plasma is 
of the order of the speed of light, while the sound speed in dark matter is 
basically zero.
So, the overdensity in baryons generates an outgoing
density wave 
after horizon crossing. 
This wave propagates until  recombination, and then freezes out. 
On the other hand, the overdensity in the dark matter remains in its 
original place.
The current distance from the overdensity in dark matter
to the front of the baryon density 
wave equals 150 Mpc. 
Hence, there is an enhanced
correlation between matter perturbations
at this distance scale, which shows up as a feature in the correlation 
function\footnote{Notice that the separation at 
Fig.~\ref{corfunc} is given in $h^{-1}$ Mpc, where $h=H_0/100~
{\rm km/(s\cdot Mpc)} \approx 0.7$.
Hence, 100 $h^{-1}$ Mpc roughly corresponds to 150 Mpc.},
see Fig.~\ref{corfunc}. 
In the Fourier space, this feature produces oscillations~(\ref{BAO-eq}).

\subsection{``Side'' remarks}

Before proceeding to further discussion of primordial
perturbations, let us make a couple of miscellaneous remarks.

\subsubsection{Cosmic variance.}
We can measure only one Universe, and the best one can do is to
{\it define} 
the angular spectrum $C_l$ obtained from the data as
\[
C_l = \frac{1}{2l+1} \sum_{m=-l}^l |a_{lm}|^2 \; . 
\]
This is not the same thing as $C_l$ defined in \eqref{may29-6}, as the
latter definition involves averaging over an ensemble of Universes.
For given $l$, there
are $(2l+1)$ independent coefficients $a_{lm}$ only, so there exists
an irreducible statistical uncertainty of order $\delta C_l/C_l \sim
1/\sqrt{2l+1}$, called cosmic variance. It is particularly pronounced
at small $l$ and, indeed, it is much larger than the experimental
errors in this part of the angular spectrum (as an example,
error bars in the left part of Fig.~\ref{anisotropyspectrum} are
precisely due to the cosmic variance).

\subsubsection{Measuring the cosmological parameters.}

\begin{figure}[htb!]
\begin{center}
\includegraphics[angle=-90,width=0.45\textwidth]{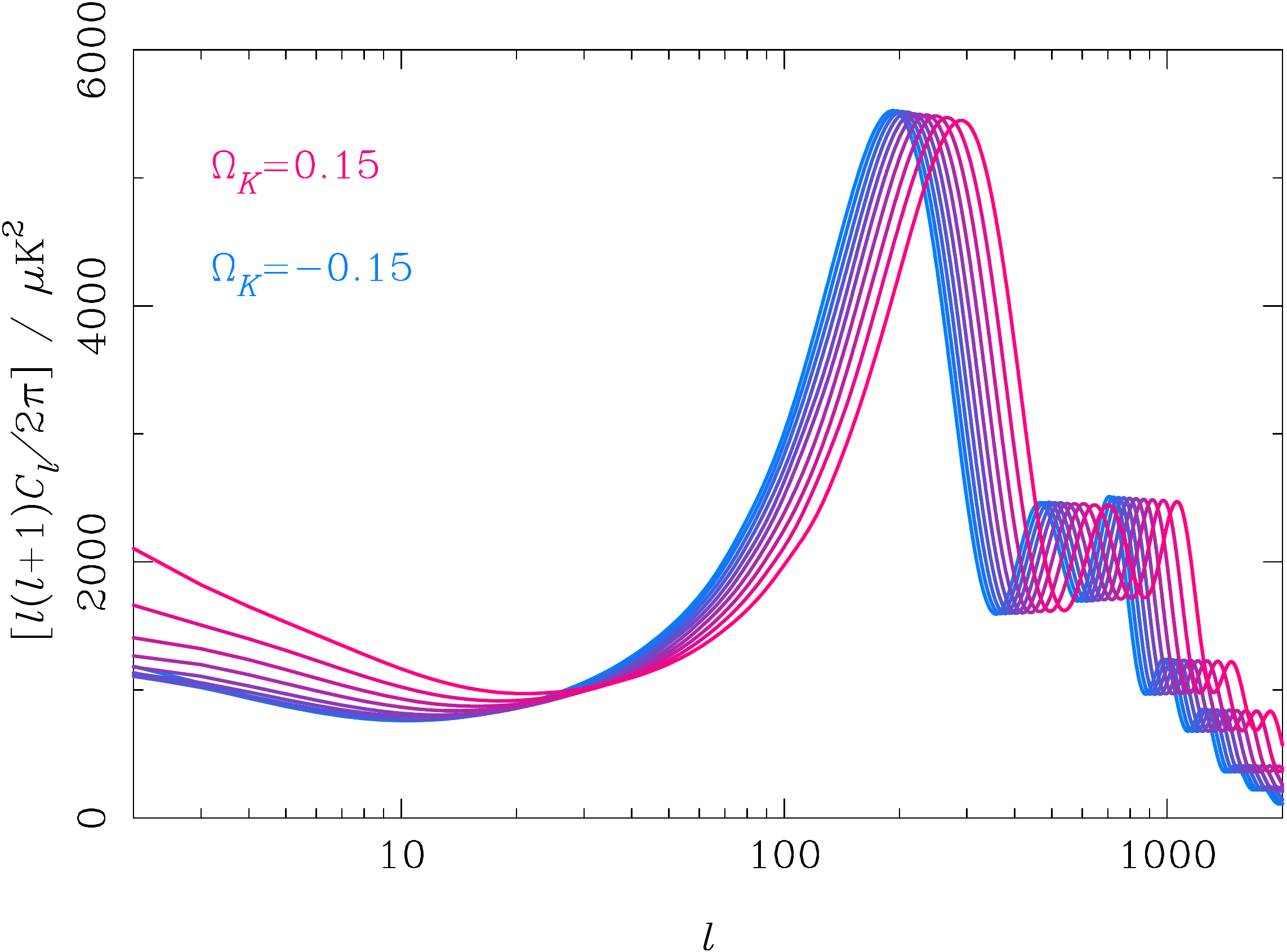}
\hskip 0.03\textwidth
\includegraphics[angle=-90,width=0.45\textwidth]{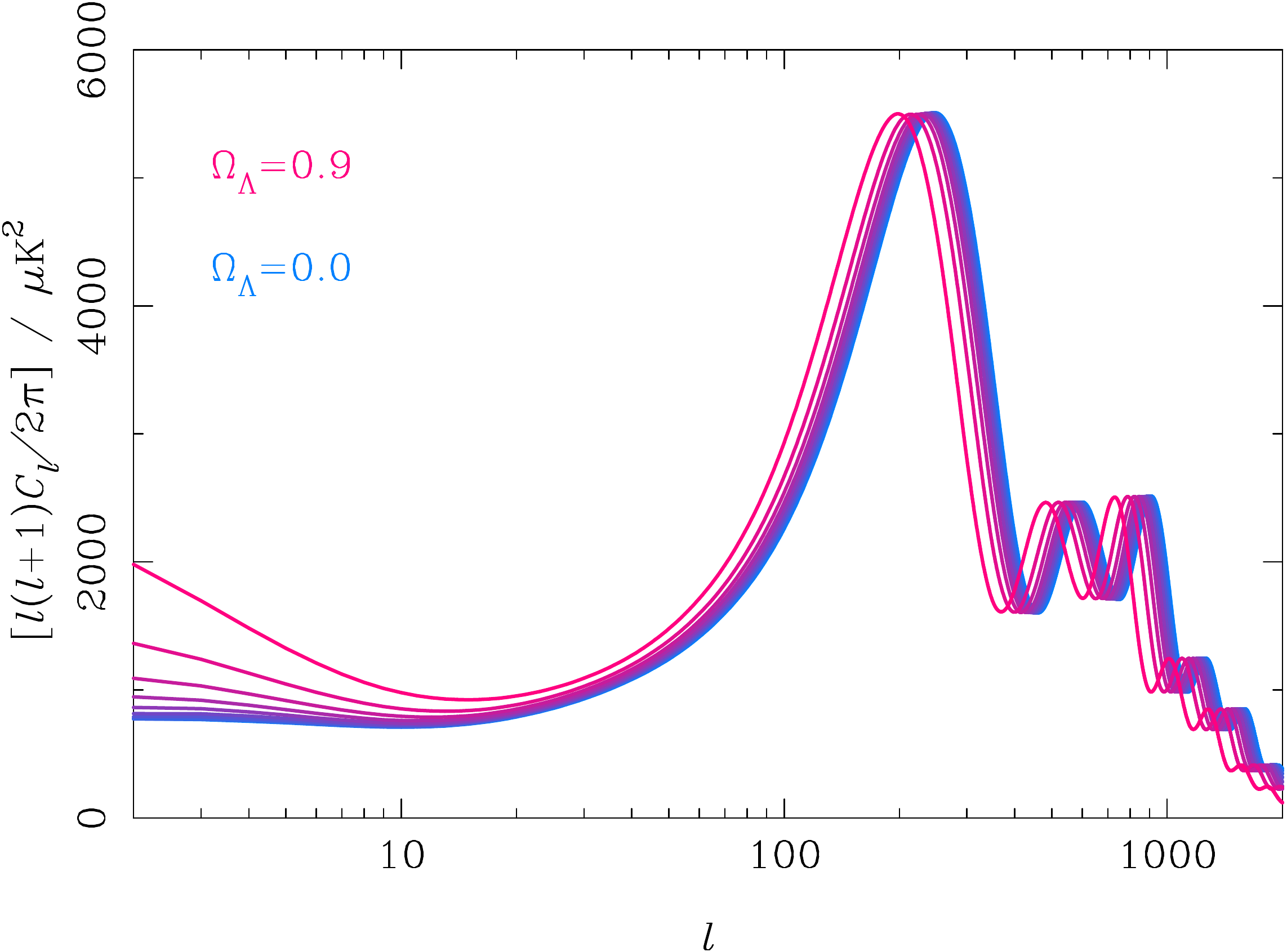}
\caption{Effect of spatial curvature (left) and dark energy
(right) on the CMB temperature angular spectrum~\cite{challinor}.
$\Omega_k = \pm (RH_0)^2$ is the relative contribution of spatial
curvature to the Friedmann equation, with $R$ being the radius of
spatial curvature. Negative sign corresponds to 3-sphere.
As $\Omega_k$ decreases, the curves in the left plot
move left.
Likewise, the curves on the right plot move left as
$\Omega_\Lambda$ increases.
\label{curv-de}}
\end{center}
\end{figure}

The angular spectrum of CMB temperature anisotropy and polarization,
as well as other cosmological data, encodes information on
the cosmological parameters. As an example, the sound horizon at recombination
is a good standard ruler back at that epoch. It is seen at an angle that
depends on the geometry of 3-dimensional space (an interval is seen
at larger angle on a sphere than on a plane) and on the dark energy
density (since dark energy affects the distance to the sphere of
photon last scattering). This is shown in Fig.~\ref{curv-de}.

Likewise, the baryon acoustic oscillations provide a standard ruler
at relatively late times (low redshifts $z \sim 0.2 - 0.4$).
A combination of their measurement with CMB anisotropy
give quite precise determination of both spatial curvature
(which is found to be
zero within error bars) and dark energy density.
Notably, this determination of $\rho_\Lambda$ is in good agreement
with  various independent data, notably, with
the data on SNe~1a, which were the first unambiguous
evidence for dark energy~\cite{Riess:1998cb,Perlmutter:1998np}.

There are many other ways in which the cosmological
parameters, including $\Omega_B$ and $\Omega_{DM}$, affect 
the CMB anisotropies. 
In particular, the hights of the acoustic peaks
in the CMB temperature angular spectrum are very sensitive
to the baryon-to-photon ratio $\eta_B$ (and hence
to $\Omega_B$), the overall shape of the curve
in Fig.~\ref{anisotropyspectrum} stronfly depends on $\Omega_{DM}$,
etc. By fitting the CMB data and combining them with
the results of other cosmological observations, one is able to obtain
quite precise knowledge of our Universe.

\subsection{Properties of primordial density perturbations --- hints 
about the earliest cosmological epoch}

As we emphasized above, the density perturbations were generated
at a very early, pre-hot epoch of the cosmological evolution.
Obviously, it is of fundamental importance to figure out what
precisely that  epoch was. One of its properties is clear right away:
it must be such that the cosmologically relevant wavelengths,
including the wavelengths of the present horizon scale, were
subhorizon early at that epoch. Only in that case the
perturbations of these wavelengths could be generated in a causal
manner at the pre-hot epoch.
Notice that this is another 
manifestation of the horizon problem discussed in 
Section~\ref{sec:regimes}: we know from the observational data
on density perturbtions that our entire visible
Universe was causally connected by the beginning of the hot stage.

An excellent hypothesis on the pre-hot stage is inflation,
the epoch of nearly exponential expansion,
\[
  a(t) = \mbox{e}^{\int H dt} \; , \;\;\;\;\;\;\;\;\;
H \approx \mbox{const} \; .
\]
Originally~\cite{inflation}, inflation was designed
 to solve the problems of the
hot Big Bang cosmology, such as the horizon problem,
as well as the flatness, entropy and other problems. 
It does this job very well: 
the horizon size at inflation
is at least
\[
l_H (t) = a(t) \int_{t_i}^t \frac{dt'}{a(t')} = H^{-1} \mbox{e}^{H (t - t_i)} \; ,
\]
where $t_i$ is the time inflation begins, and
we set $H=\mbox{const}$ for illustrational purposes.
This size is huge for $t - t_i \gg H^{-1}$, so the entire visible Universe
is naturally causally connected. 

\begin{figure}[htb!]
\begin{center}
\includegraphics[width=0.6\textwidth]{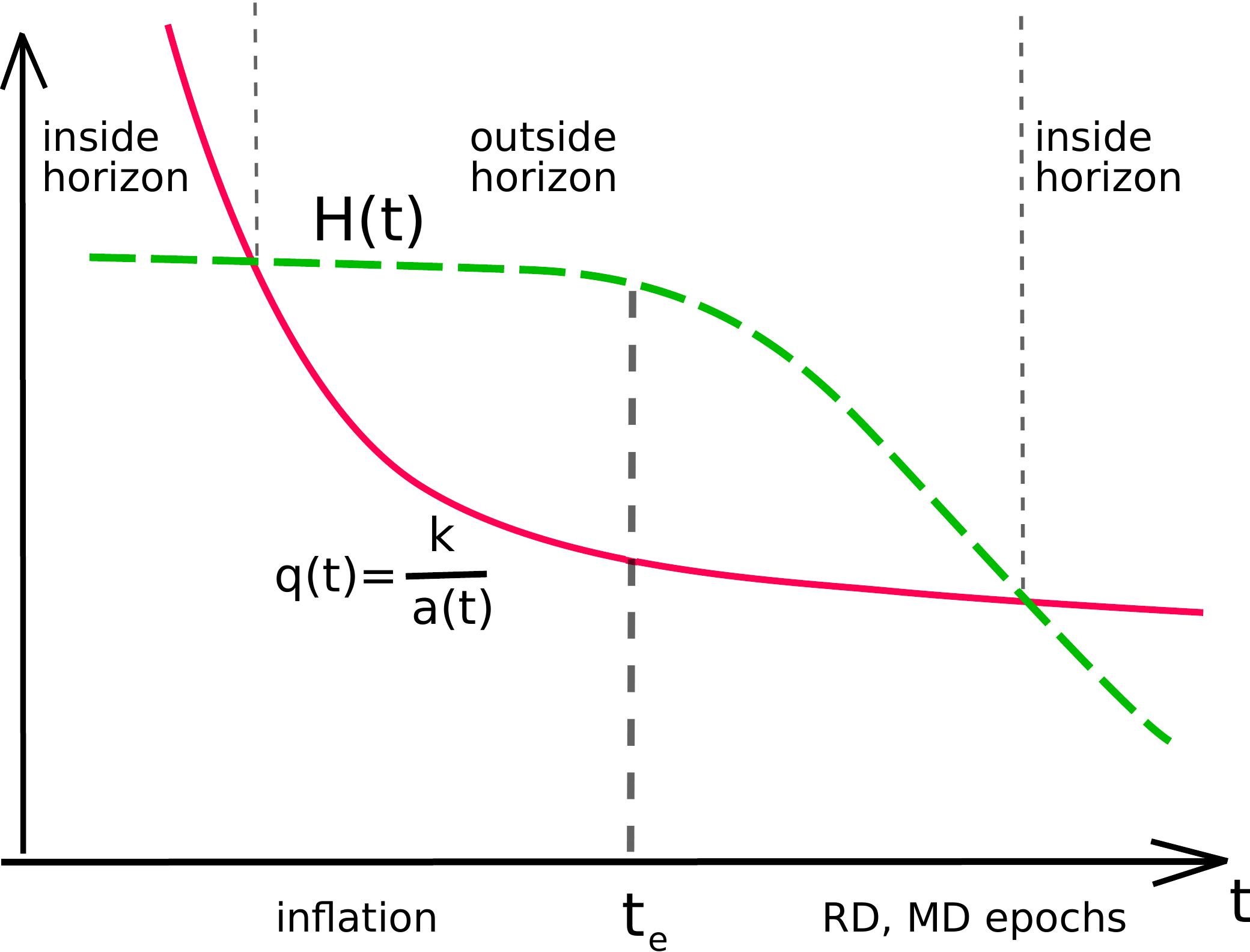}
\caption{Physical momentum and Hubble parameter at inflation and
later. $t_e$ is the time of the inflation end.
\label{infl3}}
\end{center}
\end{figure}

From the viewpoint of perturbations,
the physical momentum $q(t) = k/a(t)$ decreases
(gets redshifted) at inflation, while the Hubble parameter
stays almost constant. So, every mode is first subhorizon
($q(t) \gg H(t)$), and
later superhorizon ($q(t) \ll H(t)$) at inflation. This situation is
opposite to what happens at radiation and matter domination, see
Fig.~\ref{infl3};
this is precisely the pre-requisite for generating the density
perturbations. In fact, inflation does generate primordial
density 
perturbations~\cite{infl-perturbations}, whose properties are consistent
with everything we know about them.

Inflation is not the only hypothesis proposed so far, however.
One option is the bouncing Universe scenario, which assumes
that the cosmological evolution begins from contraction,
then the contracting stage terminates
at some moment of time (bounce) and is followed by expansion.
A version is the cycling Universe scenario with many cycles
of contraction--bounce--expansion.
Another scenario is that the Universe starts out from
nearly flat and static state and then speeds up its expansion.
Theoretical realizations of these scenarios are more difficult
than inflation, but they are not impossible, as became clear recently.
So, one of the major purposes of cosmology is to choose between
various hypotheses on the basis of observational data. The properties
of cosmological perturbations are the key issue in this regard.

There are several things which we already know about the
primordial density perturbations. By ``primordial'' we mean
the perturbations deep in the superhorizon regime at the
radiation-domination epoch. As we already know, perturbations are
time-independent in this regime. They set the initial conditions
for further evolution, and this evolution is well understood, at 
least in the linear regime.
Hence, using observational data, one is able to measure the properties
of primordial perturbations.
Of course, since the properties we know of
are established by observations, thery are valid
within certain error bars. Conversely, deviations from the results
listed below, if observed, would be extremely interesting.

\begin{figure}[htb!]
\begin{center}
\centerline{\includegraphics[angle=-90,width=0.45\textwidth]{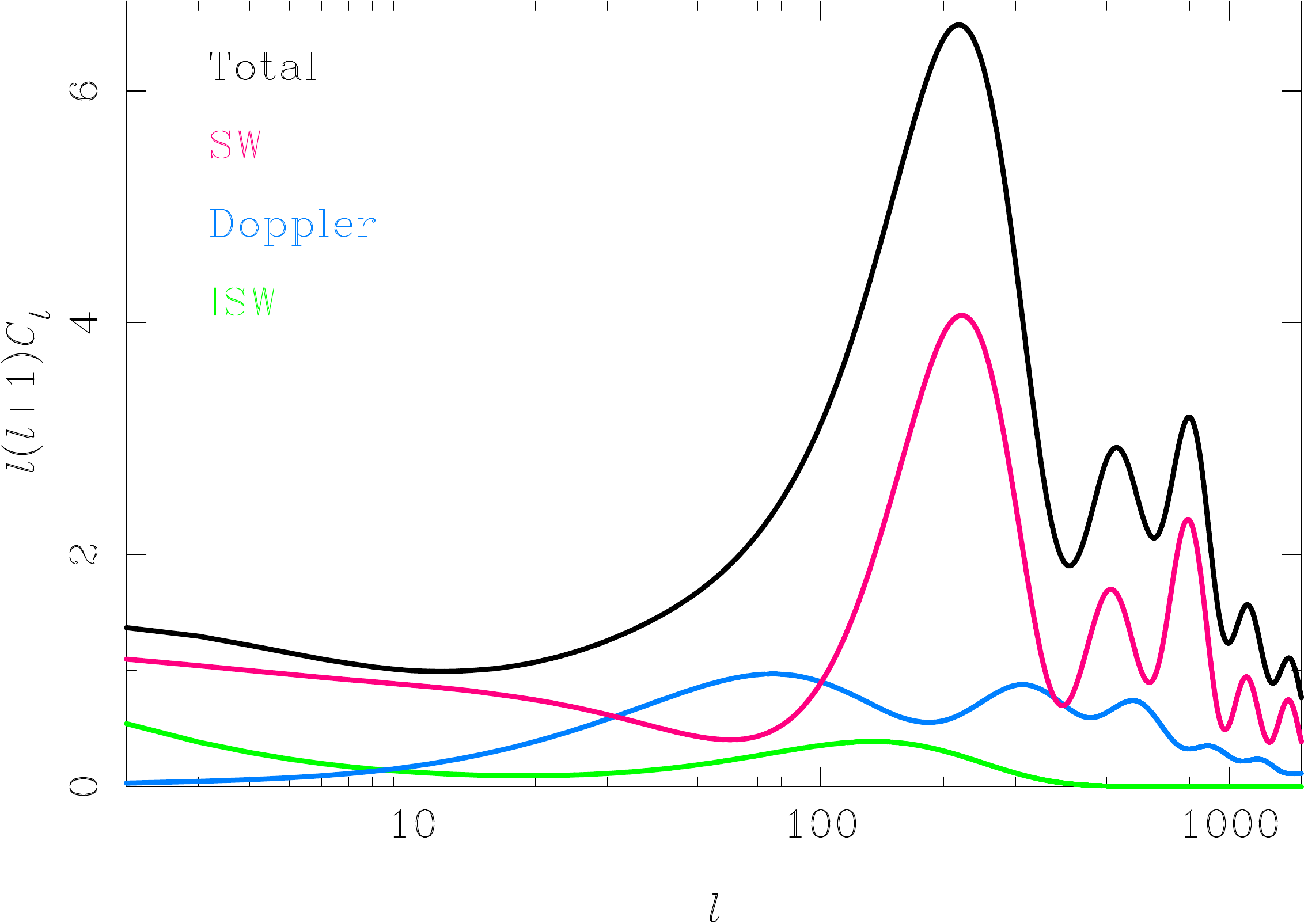}
\includegraphics[angle=-90,width=0.45\textwidth]{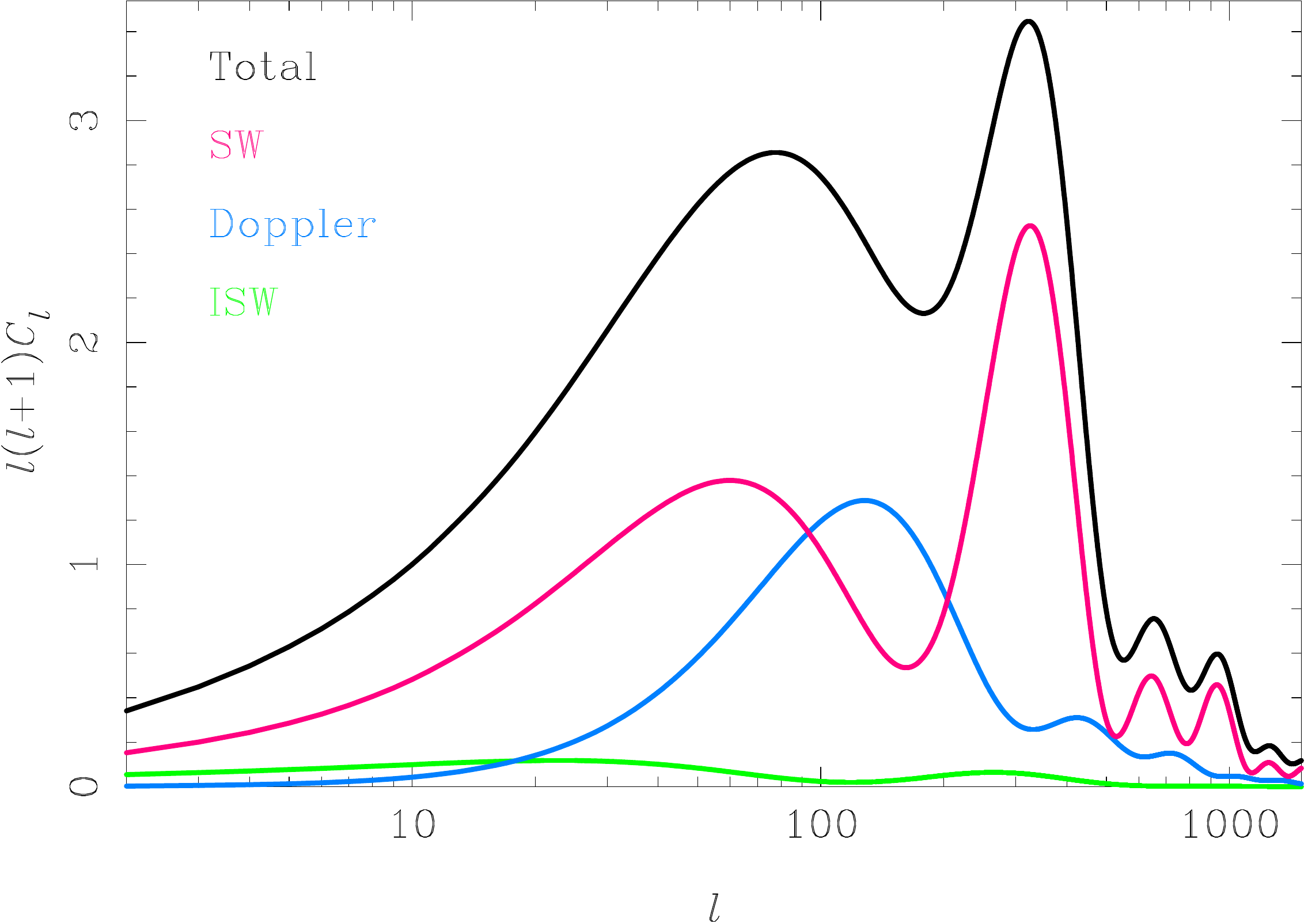}}
\caption{Angular spectrum of the CMB temperature anisotropy
for adiabatic perturbations (left) and entropy perturbations
(right)~\cite{challinor}.
\label{entropy}}
\end{center}
\end{figure}

First, density perturbations are {\bf adiabatic}.
This means that there are perturbations in the energy density,
but {\it not in composition}. More precisely, the baryon-to-entropy
ratio and dark matter-to-entropy ratio are constant in space,
\be
\delta \left(\frac{n_B}{s} \right)= \mbox{const}
\; , \;\;\;\;\;\; \delta \left(\frac{n_{DM}}{s}\right) = \mbox{const} \; .
\label{sep22-11-10}
\ee
This is consistent with the generation of the baryon asymmetry
and dark matter at the hot cosmological epoch: in that case,
all partciles were at thermal equilibrium early at the hot
epoch, the temperature completely characterized the whole
cosmic medium at that time, and as long as physics
behind the baryon asymmetry and dark matter generation is the
same everywhere in the Universe, the baryon and dark matter
abundance (relative to the entropy density) is necessarily
the same everywhere. In principle,
there may exist {\it entropy} (or isocurvature) perturbations,
such that at the early hot epoch energy density (dominated
by relativistic matter) was homogeneous, while the composition was
not. This would give initial conditions for the evolution of
density perturbations, which would be entirely different from
those characteristic of the adiabatic perturbations. As a result,
the angular spectrum of the CMB temperature anisotropy would be entirely
different, see Fig.~\ref{entropy}. No admixture of the
entropy perturbations have been detected so far, but it is worth
emphasizing that even small admixture will show that the most popular
mechanisms for generating dark matter and/or baryon asymmetry
(including those discussed in Sections~\ref{sec:dm} and~\ref{sec:bau})
have nothing to do with reality. One would have to think, instead,
that the baryon asymmetry and/or dark matter were generated
before the beginning of the hot stage.

Second, the primordial density perturbations are {\bf Gaussian random field}.
Gaussianity means that the three-point (and all odd) correlation function
vanishes, while the four-point function and all higher order
even correlation functions are expressed through the two-point
function via  Wick's theorem:
\begin{eqnarray}
\langle \delta({\bf k}_1)  \delta({\bf k}_2)  \delta({\bf k}_3) \rangle &=& 0
\nonumber \\
\langle \delta({\bf k}_1)  \delta({\bf k}_2)  \delta({\bf k}_3) 
\delta ({\bf k}_4)\rangle & =
& \langle \delta({\bf k}_1)  \delta({\bf k}_2)
\rangle \cdot \langle \delta({\bf k}_3)  \delta({\bf k}_4)   \rangle 
\nonumber\\
&~&{\ +}~
{ \mbox{permutations~of~momenta}} \; ,
\nonumber
\end{eqnarray}
while all odd correlation functions vanish.
A technical remark is in order.
As a varibale characterizing the primordial adiabatic
perturbations
we use here 
\[
\delta \equiv \delta \rho_{rad}/\rho_{rad} 
= \delta \rho/\rho
\]
 deep at the 
radiation-dominated epoch. This variable is not gauge-invariant,
so we implicitly have chosen the conformal Newtonian gauge. 
In cosmological literature, other, gauge-invariant quantities
are commonly in use, $\zeta$ and ${\cal R}$. In the conformal
Newtonian gauge, there is a simple relationship, valid in the
superhorizon regime at radiation domination:
\[
{\cal R} = \zeta = \frac{3}{4} \delta \; .
\]
We will continue to use $\delta$ as the basic variable.

Coming back to Gaussianity, we note that this
 property is characteristic of {\it vacuum fluctuations of
non-interacting (linear) quantum fields.} Hence, it is quite likely that
the density perturbations originate from the enhanced vacuum
fluctuations of non-interacting or weakly interacting quantum field(s).
Free quantum field has the general form
\[
\phi ({\bf x}, t) =
\int d^3k e^{-i{\bf kx}} \left(f^{(+)}_{\bf k}(t) a^\dagger_{\bf k}
+  e^{i{\bf kx}} f^{(-)}_{\bf k}(t) a_{\bf k} \right) \; ,
\]
where $a_{\bf k}^\dagger$ and $a_{\bf k}$ are creation and annihilation operators. 
For the field in Minkowski space-time one has  
$f^{(\pm)}_{\bf k} (t) = \mbox{e}^{\pm i\omega_k t}$, while enhancement, e.g.
due to the evolution in time-dependent background, means that
$f^{(\pm)}_{\bf k}$ are large. But in any case,
Wick's theorem is valid, provided that the state of the system is vacuum,
$a_{\bf k} |0\rangle = 0$.

Inflation does the job very well:
fluctuations of all light fields get enhanced greatly
due to the fast expansion of the Universe.
This is true, in particular, for the field that dominates the energy density
at inflation, called inflaton. Enhanced vacuum fluctuations of the
inflaton are nothing but
perturbations in the energy density at inflationary epoch in the
simplest inflationary models,
which are reprocessed into perturbations in the hot medium after
the end of inflation. The generation of the density perturbations
is less automatic in scenarios alternative to inflation,
but there are various examples showing that this is not a particularly
difficult problem.

{\it Non-Gaussianity} is an important topic of current research.
It would show up as a deviation from Wick's theorem.
As an example, the three-point function (bispectrum) may be non-vanishing,
\[\langle
\delta ({\bf k}_1)
\delta ( {\bf k}_2) \delta ({\bf k}_3)
\rangle = \delta( {\bf k}_1 + {\bf k}_2 + {\bf k}_3) ~
G(k_i^2 ;~ {\bf k}_1 {\bf k}_2 ;~
{\bf k}_1  {\bf k}_3) \neq 0 \; .
\]
The shape of  $G(k_i^2 ;~ {\bf k}_1 {\bf k}_2 ;~
{\bf k}_1  {\bf k}_3)$
is different in different models, so this shape is a 
potential discriminator.
In some models the bispectrum vanishes, e.g., due to symmetries. 
In that case the trispectrum
(connected 4-point function) may be measurable instead.
Non-Gaussianity is very small in the simplest inflationary models,
but it can be sizeable in more contrived models of inflation and in
alternatives to inflation. It is worth emphasizing that
non-Gaussianity
has not been detected yet.

Another important property is that the
primordial power spectrum of density perturbations
{\bf is flat} (or almost flat).
A convenient definition of the power spectrum for 
homogeneous and anisotropic  Gaussian random field 
is\footnote{Note that the the definition of the
power spectrum used 
in Figs.~\ref{spec} and~\ref{BAO} is different from
\eqref{sep24-11-1}.}
\be
\langle  \delta  ({\bf k})  
      \delta       ({\bf k}^\prime)  \rangle
= \frac{1}{4 \pi k^3} {\cal P} (k) \delta({\bf k} + {\bf k}^\prime) \; .
\label{sep24-11-1}
\ee
The power spectrum ${\cal P} (k) $ defined in this way
determines the fluctuation in a logarithmic
interval of momenta,
\[
\langle \delta^2 ({\bf x}) \rangle
= \int_0^\infty ~\frac{dk}{k} ~{\cal P}(k) \; .
\]
By definition, the flat spectrum is such that
${\cal P}$ is independent of $k$. It is worth noting that
the flat spectrum was conjectured by E.~Harrison~\cite{Harrison}
and Ya.~Zeldovich~\cite{Zeldovich} in the beginning of
1970's, long before realistic mechanisms of the generation
of density perturbations have been proposed.

In view of the approximate flatness, a natural parametrization is
\be
{\cal P}(k) = A_s \left(\frac{k}{k_*} \right)^{n_s -1} \; ,
\label{sep24-11-6}
\ee
where
$A_s $ is the amplitude, $(n_s-1) $ is the tilt and $k_*$ is a
fiducial momentum, chosen at one's convenience.
The flat spectrum in this parametrization has $n_s=1$.
Cosmological data favor the value $n_s \approx 0.96$ (i.e.,
slightly smaller than 1), see below, but it is fair to say that
$n_s = 1$ is still consistent with observations.

The flatness of the power spectrum calls for some symmetry behind this
property.
In inflationary theory this is the symmetry of the de~Sitter space-time,
which is the space time of constant Hubble rate,
\[
ds^2 = dt^2 - \mbox{e}^{2Ht} d{\bf x}^{2} \; , \;\;\;\;\;\;\;\;
H = \mbox{const} \; .
\]
This metric is invariant under spatial 
dilatations
supplemented by time translations,
\[
{\bf x} \to \lambda {\bf x} \;, \;\;\;
t \to t - \frac{1}{2H} \log \lambda \; .
\]
At inflation, $H$ is almost constant in time, and the
de~Sitter symmetry is an approximate symmetry. For this reason 
inflation automatically generates nearly flat power spectrum.

The de~Sitter symmetry is not the only candidate symmetry
behind the flatness of the power spectrum.
One possible alternative is conformal 
symmetry~\cite{conf1,conf2}. The point is that
the conformal group includes dilatations,
 $x^\mu \to \lambda x^\mu$. This property indicates that
the relevant part of the theory possesses no
scale, and has good chance
for producing the flat spectrum.
Model-building in this direction has begun  recently~\cite{conf2}.

\subsection{What's next?}

Thus, only very basic facts about the primordial denisty perturbations are
observationally established. Even though very suggestive, these facts
by themselves are not sufficient for unambiguously
establishing the properties of the Universe at the pre-hot epoch of
its evolution. In coming years, new properties of cosmological perturbations
will hopefully be discovered, which will shed much more light on this
pre-hot epoch. Let us discuss some of the potential observables.

\subsubsection{Tensor perturbations~=~relic gravity waves}
The simplest, and hence most plausible models of inflation
predict sizeable tensor perturbations, which are perturbations of the
metric independent of perturbations in the energy density. 
After entering the horizon, tensor perturbations are nothing but
gravity waves.
The reason for their generation at inflation is that
the exponential expansion of the Universe  enhances vacuum fluctuations
of all fields, including the gravitational field itself.
In inflationary theory, the primordial tensor perturbations
are Gaussian random field with nearly flat power spectrum
\be
{\cal P}_T = A_T \left( \frac{k}{k_*} \right)^{n_T} \; ,
\label{sep24-11-5}
\ee
where the inflationary prediction is $n_T \approx 0$ (the reason
for different definitions of the tensor spectral index $n_T$ in 
\eqref{sep24-11-5} and scalar spectral index $n_s$ 
in \eqref{sep24-11-6} is purely
historical). 

On the other hand, there seems to be no way of
generating nearly flat tensor power spectrum in alternatives to 
inflation. In fact, most, if not all, alternative scenarios
predict unobservably small amplitude of tensor perturbations.
Thus, the discovery of tensor modes would be the strongest
possible argument in favor of inflation. It is worth noting
that {\it non-observation} of tensor perturbations would not
rule inflation out: there are numerous models of inflation
which predict tensor modes of very small amplitude.

\begin{figure}[htb!]
\begin{center}
\includegraphics[width=0.6\textwidth]{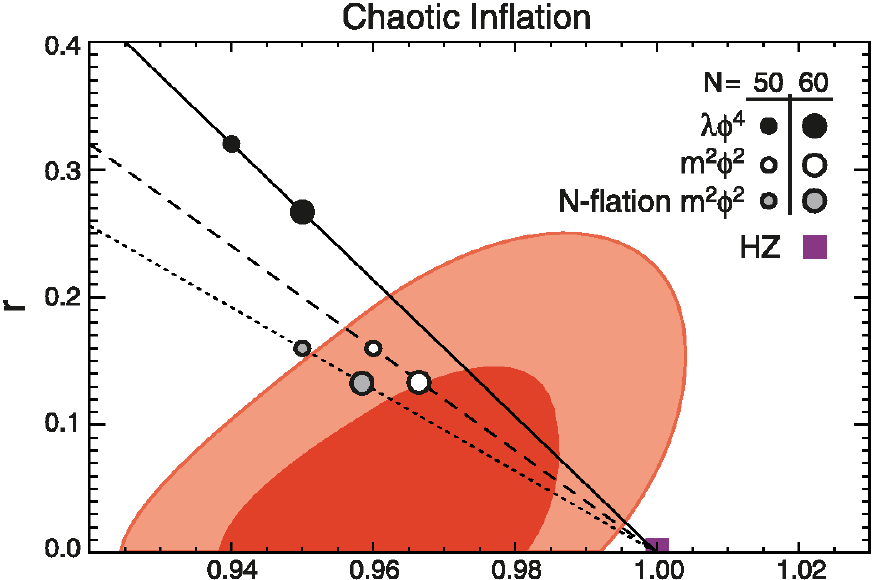}
\caption{Allowed 
regions  (at 68\% and 95\% CL) in the plane $(n_s, r)$, where $n_s$
is the scalar spectral index and $r$ is the tensor-to-scalar
ratio. Rectangle
shows the Harrison--Zeldovich point (flat scalar spectrum, no
tensor modes). Points on the lines show predictions of popular
inflationary models. 
\label{ns-r}}
\end{center}
\end{figure}

The tensor power is usually characterized by the tensor-to-scalar
ratio
\[
r = \frac{A_T}{A_s} \; .
\]
The simplest inflationary models predict, roughly speaking,
$r \sim 0.1 - 0.3$.
The current situation is summarized in Fig.~\ref{ns-r}.
Clearly, there is an indication for the
negative scalar tilt $(n_s-1)$ or non-zero tensor amplitude,
or both, though it is premature to say that the flat scalar spectrum
with no tensor modes (the Harrison--Zeldovich point) is ruled out.
 
For the time being, the most sensitive probe of the tensor perturbations
is the CMB temperature anisotropy. However, the most promising tool
is the CMB polarization. The point is that a certain class of
polarization
patterns (called B-mode) is generated by tensor perturbations,
while scalar perturbations are unable to create it. Hence, the Planck
experiment, and especially dedicated experiments aiming at
measuring the CMB polarization may well discover the tensor perturbations,
i.e., relic gravity waves. Needless to say, this would be a
profound discovery. To avoid confusion, let us note that the CMB 
polarization has been already observed, but it belongs to another class of
patterns (so called E-mode) and is consistent with the 
existence of the
scalar perturbations only.

\subsubsection{Scalar tilt.} 
Inflationary models and their alternatives
will be constrained by the precise determination of the scalar
tilt $(n_s-1)$ and its dependence on momentum $k$. It appears, however,
that the information on $n_s(k)$ that will be obtained in reasonably
near future will be of limited significance from the viewpoint 
of discriminating between different (and even grossly different)
 scenarios.

\subsubsection{Non-Gaussianity.} As we pointed out already, non-Gaussianity
of density perturbations
is very small in the simplest inflationary models. Hence, its discovery
will signalize that either inflation and inflationary generation of
density perturbations
occured in a rather complicated way, or  an alternative
scenario was realized. Once the non-Gaussianity is discovered,
and its shape is revealed even with moderate accuracy,
many concrete models will be ruled out, while at most a few
will get strong support. 

\subsubsection{Statistical anisotropy.} In principle, the power spectrum
of density perturbations may depend on the direction of
momentum, e.g., 
\[
{\cal P}({\bf k}) = {\cal P}_0 (k) \left(1 + 
w_{ij} (k) \frac{k_i k_j}{k^2}
+ \dots \right)
\]
where $w_{ij}$ is a fundamental  tensor in our part
of the Universe (odd powers of $k_i$ would contradict
commutativity of the Gaussian random field $\delta ({\bf k})$,
see Eq.~\eqref{sep24-11-1}). Such a dependence would definitely
imply 
that the Universe was anisotropic at the
pre-hot stage, when the primordial
perturbations were generated. This statistical anisotropy
is rather hard to obtain in inflationary models, though it is
possible in inflation with strong vector fields~\cite{soda}.
On the other hand, statistical anisotropy is 
natural in some other scenarios, including conformal 
models~\cite{mlvr+}

The statistical anisotropy
would show up in correlators~\cite{aniso}
\[
\langle a_{ lm} a_{ l' m'} \rangle \;\;\;\;\;\; \mbox{with}~~{ l'\neq l}
~~
\mbox{and/or}~~{ m' \neq  m}
\]
At the moment, the situation with observational data
is 
controversial~\cite{statanis}, and the new data, notably from the
Planck
experiment, will hopefully clear it up.

\subsubsection{Admixture of entropy perturbations.} 
As we explained above,
even small admixture of entropy perturbations would force us to
abandon the most popular scenarios of the generation of baryon asymmetry
and/or dark matter, which assumed that it happened at the hot epoch.
The WIMP dark matter would no longer be well motivated, 
while other, very weakly interacting dark matter candidates, like axion
or superheavy relic, would be prefered. This would make the direct
searches for dark matter rather problematic.

\section{Conclusion}

We are at the eve of new era not only in particles physics, but
also in cosmology. There is reasonably
well justified expectation that the
LHC will shed light on long-standing cosmological problems
of the origin of the baryon asymmetry and nature of dark
matter in our Universe.
The ideas we discussed in these lectures in this regard
may well 
be not the right ones: we can only hypothesize on physics beyond the
Standard Model and its role in the early Universe.

In fact, the TeV scale
physics may be dramatically different from physics we get used to.
As an example, it is not excluded that TeV is not only electroweak,
but also gravitational scale. This is the case in models with large
extra dimensions, in which the Planck scale is related to
the fundamental gravity scale in  a way that involves the volume
of extra dimensions, and hence the fundamental scale can be much
below $M_{Pl}$ (for a review see, e.g., Ref.~\cite{extra-dim-rev}).  
If the LHC will find that, indeed, the fundamental gravity scale is in
the TeV range, this would have most profound consequences for both
microscopic physics and cosmology. On the microscopic physics
side, this would enable one to study at colliders
quantum gravity and its
high-energy extension --- possibly string theory,
while on the cosmological side,
the entire picture of the early Universe would have to be revised.
Inflation, if any, would have to occur either at low energy density
or in the regime of strong quantum gravity effects. 
The highest temperatures in the usual expansion history would be
at most in the TeV range, so dark matter and baryon
asymmetry would have to be generated either
below TeV temperatures or in quantum gravity regime.
Even more intriguing will be the study of quantum gravity cosmological
epoch, with hints from colliders  
gradually coming. This, probably, is too bright a prospective to
be realistic.

It is more likely that the LHC will find something entirely new,
something theorists have not thought about. Or, conversely, find
so little that one will have to get serious about anthropic principle.
In any case, the LHC results will definitely change the landscape
of fundamental physics, cosmology included.

The observational data unequivocally tell us that the hot
stage of the cosmological evolution was preceeded by some other
epoch, at which the cosmological perturbations were generated.
The best guess for this epoch is inflation, but one should bear
in mind that there are alternative possibilities. It is fascinating
that with new observational data, there is good chance
to learn what precisely that pre-hot epoch was. It may very well be that
 in this way
we will be able to probe physics at the energy, distance and
time scales
well beyond the reach of the LHC.


\begin{thebibliography}{99}

\bibitem{gen-cosmo}
R.~H.~Brandenberger,
{\it Particle physics aspects of modern cosmology,}
arXiv:hep-ph/9701276; 
W.~L.~Freedman and M.~S.~Turner,
Rev.\ Mod.\ Phys.\  {\bf 75} (2003) 1433
[arXiv:astro-ph/0308418]; 
 V.~Rubakov,
  PoS {\bf RTN2005} (2005) 003;
  J.~A.~Peacock,
  {\it Cosmology and particle physics,}
in {\it Proc. 1998 European School of High-Energy Physics, St. Andrews, Scotland}; 
M.~Shaposhnikov,
{\it Cosmology and astrophysics,}
in {\it Proc. 2000 European School of High-Energy Physics, Caramulo, Portugal};
I.~I.~Tkachev,
{\it Astroparticle physics,} 
in {\it Proc. 2003 European School on High-Energy Physics, 
Tsakhkadzor, Armenia},
arXiv:hep-ph/0405168.





\bibitem{dark-rev}
 G.~Jungman, M.~Kamionkowski and K.~Griest,
  Phys.\ Rept.\  {\bf 267} (1996) 195
  [arXiv:hep-ph/9506380];
A.~Bottino and N.~Fornengo,
{\it Dark matter and its particle candidates,}
arXiv:hep-ph/9904469; 
  K.~A.~Olive,
  {\it Dark matter,}
  arXiv:astro-ph/0301505; 
G.~Bertone, D.~Hooper and J.~Silk,
  Phys.\ Rept.\  {\bf 405} (2005) 279.
  [arXiv:hep-ph/0404175].




\bibitem{ew-rev}
A.~D.~Dolgov,
{\it Baryogenesis, 30 years after,}
arXiv:hep-ph/9707419\; 
V.~A.~Rubakov and M.~E.~Shaposhnikov,
  Usp.\ Fiz.\ Nauk {\bf 166} (1996) 493
  [Phys.\ Usp.\  {\bf 39} (1996) 461]
  [arXiv:hep-ph/9603208];
A.~Riotto and M.~Trodden,
  Ann.\ Rev.\ Nucl.\ Part.\ Sci.\  {\bf 49} (1999) 35
  [arXiv:hep-ph/9901362];
M.~Trodden,
  Rev.\ Mod.\ Phys.\  {\bf 71} (1999) 1463
  [arXiv:hep-ph/9803479].

\bibitem{de-rev}
  S.~Weinberg,
  Rev.\ Mod.\ Phys.\  {\bf 61} (1989) 1;
V.~Sahni and A.~A.~Starobinsky,
  Int.\ J.\ Mod.\ Phys.\  D {\bf 9} (2000) 373
  [arXiv:astro-ph/9904398];
P.~J.~E.~Peebles and B.~Ratra,
  Rev.\ Mod.\ Phys.\  {\bf 75} (2003) 559
  [arXiv:astro-ph/0207347]; 
T.~Padmanabhan,
  Phys.\ Rept.\  {\bf 380} (2003) 235
  [arXiv:hep-th/0212290]; 
E.~J.~Copeland, M.~Sami and S.~Tsujikawa,
  Int.\ J.\ Mod.\ Phys.\  D {\bf 15} (2006) 1753
  [arXiv:hep-th/0603057].

\bibitem{pert-rev}
V.~F.~Mukhanov, H.~A.~Feldman and R.~H.~Brandenberger,
  Phys.\ Rept.\  {\bf 215} (1992) 203; 
A.~R.~Liddle and D.~H.~Lyth,
  Phys.\ Rept.\  {\bf 231} (1993) 1
  [arXiv:astro-ph/9303019];
D.~H.~Lyth and A.~Riotto,
  Phys.\ Rept.\  {\bf 314} (1999) 1
  [arXiv:hep-ph/9807278].






\bibitem{topology}
  A.~de Oliveira-Costa, G.~F.~Smoot and A.~A.~Starobinsky,
 {\it Constraining topology with the CMB,}
  arXiv:astro-ph/9705125; 
N.~G.~Phillips and A.~Kogut,
  Astrophys.\ J.\  {\bf 645}, 820 (2006)
  [arXiv:astro-ph/0404400].


\bibitem{black-body} E.~Gawiser and J.~Silk,
Phys.\ Rept.\  {\bf 333}, 245 (2000)
[arXiv:astro-ph/0002044].

\bibitem{WMAP7}
D.~Larson {\it et al.},
  Astrophys.\ J.\ Suppl.\  {\bf 192} (2011) 16
  [arXiv:1001.4635 [astro-ph.CO]].


\bibitem{WMAP7a}
E.~Komatsu {\it et al.}  [WMAP Collaboration],
  Astrophys.\ J.\ Suppl.\  {\bf 192} (2011) 18
  [arXiv:1001.4538 [astro-ph.CO]].

\bibitem{WMAP5}
  E.~Komatsu {\it et al.}  [WMAP Collaboration],
  Astrophys.\ J.\ Suppl.\  {\bf 180} (2009) 330
  [arXiv:0803.0547 [astro-ph]].

\bibitem{Seljak:2004xh}
U.~Seljak {\it et al.}  [SDSS Collaboration],
  Phys.\ Rev.\  D {\bf 71}, 103515 (2005)
  [arXiv:astro-ph/0407372].

\bibitem{'tHooft:1976up}
  G.~'t Hooft,
  Phys.\ Rev.\ Lett.\  {\bf 37} (1976) 8.

\bibitem{Klinkhamer-Manton}
F.~R.~Klinkhamer and N.~S.~Manton,
  Phys.\ Rev.\ D {\bf 30}, 2212 (1984).


\bibitem{Kuzmin:1985mm}
  V.~A.~Kuzmin, V.~A.~Rubakov and M.~E.~Shaposhnikov,
  Phys.\ Lett.\  B {\bf 155} (1985) 36.

\bibitem{Rubakov:2002fi}
  V.~A.~Rubakov,
  {\it Classical theory of gauge fields,}
  Princeton, USA: Univ. Pr. (2002) 444 p.p.



\bibitem{stop}
M.~S.~Carena, M.~Quiros and C.~E.~M.~Wagner,
  Phys.\ Lett.\  B {\bf 380} (1996) 81
  [arXiv:hep-ph/9603420];
M.~S.~Carena, M.~Quiros, A.~Riotto, I.~Vilja and C.~E.~M.~Wagner,
  Nucl.\ Phys.\  B {\bf 503} (1997) 387
  [arXiv:hep-ph/9702409];
M.~S.~Carena, M.~Quiros, M.~Seco and C.~E.~M.~Wagner,
  Nucl.\ Phys.\  B {\bf 650} (2003) 24
  [arXiv:hep-ph/0208043].

\bibitem{Kayser}
B. Kayser, {\it Neutrino Oscillation Physics}, in these proceedings.

\bibitem{prehistory}
V.~A.~Rubakov,
  Phys.\ Rev.\  D {\bf 61}, 061501 (2000)
  [arXiv:hep-ph/9911305]; 
P.~J.~Steinhardt and N.~Turok,
  Science {\bf 312}, 1180 (2006)
  [arXiv:astro-ph/0605173].

\bibitem{Weinberg:1987dv}
  S.~Weinberg,
  Phys.\ Rev.\ Lett.\  {\bf 59}, 2607 (1987).

\bibitem{Linde:1986dq}
  A.~D.~Linde, 
{\it Inflation And Quantum Cosmology,}
in: {\it  Three hundred years of gravitation}.  Cambridge Univ. 
Press, Eds.  
Hawking, S.W.  and Israel, W., 604-630 (1987).


\bibitem{Tegmark}
  M.~Tegmark {\it et al.}  [SDSS Collaboration],
  Astrophys.\ J.\  {\bf 606} (2004) 702
  [arXiv:astro-ph/0310725].

\bibitem{cmbang}
C.~L.~Reichardt {\it et al.},
Astrophys. J. {\bf 694} (2009) 1200
  [arXiv:0801.1491 [astro-ph]].

\bibitem{string}
J.~Urrestilla, N.~Bevis, M.~Hindmarsh, M.~Kunz and A.~R.~Liddle,
  JCAP {\bf 0807}, 010 (2008)
  [arXiv:0711.1842 [astro-ph]].

\bibitem{BAO}
W.~J.~Percival, S.~Cole, D.~J.~Eisenstein, R.~C.~Nichol, 
J.~A.~Peacock, A.~C.~Pope and A.~S.~Szalay,
  Mon.\ Not.\ Roy.\ Astron.\ Soc.\  {\bf 381} (2007) 1053
  [arXiv:0705.3323 [astro-ph]].

\bibitem{BAO-c}
D.~J.~Eisenstein {\it et al.}  [SDSS Collaboration],
  Astrophys.\ J.\  {\bf 633} (2005) 560
  [arXiv:astro-ph/0501171].

\bibitem{challinor}
A.~Challinor,
  Lect. Notes Phys. 2004. {\bf 653}. 71 
  [arXiv:astro-ph/0403344].

\bibitem{Riess:1998cb}
  A.~G.~Riess {\it et al.}  [Supernova Search Team Collaboration],
  Astron.\ J.\  {\bf 116} (1998) 1009
  [arXiv:astro-ph/9805201].


\bibitem{Perlmutter:1998np}
  S.~Perlmutter {\it et al.}  [Supernova Cosmology Project Collaboration],
  Astrophys.\ J.\  {\bf 517} (1999) 565
  [arXiv:astro-ph/9812133].





\bibitem{inflation}
  A.~A.~Starobinsky,
  JETP Lett.\  {\bf 30}  682 (1979)
  [Pisma Zh.\ Eksp.\ Teor.\ Fiz.\  {\bf 30}, 719 (1979)]\;
  Phys.\ Lett.\  B {\bf 91}, 99 (1980);
  A.~H.~Guth,
  Phys.\ Rev.\  D {\bf 23}, 347 (1981);
A.~D.~Linde,
  Phys.\ Lett.\  B {\bf 108}, (1982) 389;
  Phys.\ Lett.\  B {\bf 129} 177 (1983);
A.~Albrecht and P.~J.~Steinhardt,
  Phys.\ Rev.\ Lett.\  {\bf 48}, 1220 (1982).


\bibitem{infl-perturbations}
  V.~F.~Mukhanov and G.~V.~Chibisov,
  JETP Lett.\  {\bf 33}, (1981) 532
  [Pisma Zh.\ Eksp.\ Teor.\ Fiz.\  {\bf 33}, 549 (1981)];
  S.~W.~Hawking,
  Phys.\ Lett.\  B {\bf 115},  295 (1982);
  A.~A.~Starobinsky,
  Phys.\ Lett.\  B {\bf 117}, 175 (1982);
  A.~H.~Guth and S.~Y.~Pi,
  Phys.\ Rev.\ Lett.\  {\bf 49}, 1110 (1982);
  J.~M.~Bardeen, P.~J.~Steinhardt and M.~S.~Turner,
  Phys.\ Rev.\  D {\bf 28}, 679 (1983).

\bibitem{Harrison}
  E.~R.~Harrison,
  Phys.\ Rev.\  D {\bf 1}, 2726 (1970).

\bibitem{Zeldovich}
  Y.~B.~Zeldovich,
  Mon.\ Not.\ Roy.\ Astron.\ Soc.\  {\bf 160}, 1P (1972).

\bibitem{conf1}
  I.~Antoniadis, P.~O.~Mazur and E.~Mottola,
  Phys.\ Rev.\ Lett.\  {\bf 79} (1997) 14
  [arXiv:astro-ph/9611208].

\bibitem{conf2}
  V.~A.~Rubakov,
  {JCAP} {\bf 0909} (2009), 030;
   [arXiv:0906.3693 [hep-th]];
 P.~Creminelli, A.~Nicolis and E.~Trincherini,
  JCAP {\bf 1011}, 021 (2010);
  [arXiv:1007.0027 [hep-th]];
  K.~Hinterbichler and J.~Khoury,
 JCAP {\bf 1204} (2012) 023
  [arXiv:1106.1428 [hep-th]].

\bibitem{soda}
M.~A.~Watanabe, S.~Kanno and J.~Soda,
  Phys.\ Rev.\ Lett.\  {\bf 102} (2009) 191302
  [arXiv:0902.2833 [hep-th]];
T.~R.~Dulaney and M.~I.~Gresham,
  Phys.\ Rev.\  D {\bf 81} (2010) 103532
  [arXiv:1001.2301];
A.~E.~Gumrukcuoglu, B.~Himmetoglu and M.~Peloso,
  Phys.\ Rev.\  D {\bf 81} (2010) 063528
  [arXiv:1001.4088 [astro-ph.CO]].

\bibitem{mlvr+}
M.~Libanov and V.~Rubakov,
  JCAP {\bf 1011} (2010) 045
  [arXiv:1007.4949];
M.~Libanov, S.~Ramazanov and V.~Rubakov,
  JCAP {\bf 1106} (2011) 010
  [arXiv:1102.1390 [hep-th]].

\bibitem{aniso}
L.~Ackerman, S.~M.~Carroll and M.~B.~Wise,
  Phys.\ Rev.\  D {\bf 75} (2007) 083502
  [Erratum-ibid.\  D {\bf 80} (2009) 069901]
  [arXiv:astro-ph/0701357];
A.~R.~Pullen and M.~Kamionkowski,
  Phys.\ Rev.\  D {\bf 76} (2007) 103529
  [arXiv:0709.1144 [astro-ph]].

\bibitem{statanis}
N.~E.~Groeneboom and H.~K.~Eriksen,
  Astrophys.\ J.\  {\bf 690} (2009) 1807
  [arXiv:0807.2242 [astro-ph]];
D.~Hanson and A.~Lewis,
  Phys.\ Rev.\  D {\bf 80} (2009) 063004
  [arXiv:0908.0963 [astro-ph.CO]];
C.~L.~Bennett {\it et al.},
  Astrophys.\ J.\ Suppl.\  {\bf 192} (2011) 17
  [arXiv:1001.4758 [astro-ph.CO]].


\bibitem{extra-dim-rev}
V.~A.~Rubakov,
  Phys.\ Usp.\  {\bf 44} (2001) 871
  [Usp.\ Fiz.\ Nauk {\bf 171} (2001) 913]
  [arXiv:hep-ph/0104152].





\end{thebibliography}
\end{document}